\documentclass[iop]{emulateapj}
\usepackage{hyperref}
\usepackage{amsmath}
\usepackage{algorithm}
\usepackage{amsfonts}
\usepackage{mathrsfs}
\usepackage{bm}
\usepackage{rotating}
\usepackage{color}
\usepackage{graphicx}
\usepackage{subfigure}
\graphicspath{{fig/}}
\usepackage{appendix}
\usepackage{url}

\def\cha{\textit{Chandra}}
\def\XMM{{XMM-{\it Newton}}}
\def\NuSTAR{{\it NuSTAR}}
\def\bat{{{\it Swift}-BAT}}
\def \XSPEC {{\tt XSPEC}}

\def \borus {{\tt borus02}}

\begin{document}
\title{A broadband X-ray study of a sample of AGN\MakeLowercase{s} with [OIII] measured inclinations}
\author{X. Zhao\altaffilmark{1}, S. Marchesi\altaffilmark{1,2}, M. Ajello\altaffilmark{1}, M. Balokovi{\'{c}}\altaffilmark{3,4}, T. Fischer\altaffilmark{5}}
\altaffiltext{1}{Department of Physics \& Astronomy, Clemson University, Clemson, SC 29634, USA}
\altaffiltext{2}{INAF-Osservatorio Astronomico di Bologna, Via Piero Gobetti, 93/3, I-40129, Bologna, Italy}
\altaffiltext{3}{Center for Astrophysics $\vert$ Harvard \& Smithsonian, 60 Garden Street, Cambridge, MA 02138, USA}
\altaffiltext{4}{Black Hole Initiative at Harvard University, 20 Garden Street, Cambridge, MA 02138, USA}
\altaffiltext{5}{Astrometry Department, United States Naval Observatory, 3450 Massachusetts Ave., NW, Washington, DC 20392, USA}
\begin{abstract}
In modeling the X-ray spectra of active galactic nuclei (AGNs), the inclination angle is a parameter that can play an important role in analyzing the X-ray spectra of AGN, but it has never been studied in detail. We present a broadband X-ray spectral analysis of the joint \NuSTAR-\XMM\ observations of 13 sources with [OIII] measured inclinations determined by \citet{Fischer_2013}. By freezing the inclination angles at the [OIII] measured values when modeling the observations, the spectra are well fitted and the geometrical properties of the obscuring structure of the AGNs are slightly better constrained than those fitted when the inclination angles are left free to vary. We also test if one could freeze the inclinations at other specific angles in fitting the AGN X-ray spectra as commonly did in the literatures. We find that one should always let the inclination angle free to vary in modeling the X-ray spectra of AGNs, while fixing the inclination angle at [OIII] measured values and fixing the inclination angle at 60$^\circ$ also present correct fits of the sources in our sample. Correlations between the covering factor and the average column density of the obscuring torus with respect to the Eddington ratio are also measured, suggesting that the distribution of the material in the obscuring torus is regulated by the Eddington ratio, which is in agreement with previous studies. In addition, no geometrical correlation is found between the narrow line region of the AGN and the obscuring torus, suggesting that the geometry might be more complex than what is assumed in the simplistic unified model.
\end{abstract}


%
%
\section{Introduction}
\label{sec:intro} 
It is commonly accepted that the main structure of an active galactic nucleus (AGN) is composed of a supermassive black hole (SMBH; $M_{\rm BH}\approx10^{6-9.5}M_{\sun}$) at the center of the AGN, an accretion disk surrounding the SMBH, a subparsec-scale dust-free region known as the broad line region (BLR), where broad lines with full-width at half-maximum (FWHM) $>$2000\,km\,s$^{-1}$ are observed in optical, a parsec-scale toroidal structure composed of gas and dust obscuring the emission from the center engine of the AGN, and a broaden structure ($\sim$10\,pc to $\sim$1\,kpc) namely the narrow line region \citep[NLR; FWHM $<$1000\,km\,s$^{–1}$; see, e.g.,][for recent reviews]{Netzer15,Naure_astro2017,Hickox18}. AGNs are optically classified as type 1 or type 2 AGNs, if the broad-emission lines can be observed in their optical spectra or not. According to the AGN unified model \citep{Antonucci1993,Urry_1995}, type 2 AGNs are the AGNs whose line-of-sight between the central engine and the observer passes the dusty toroidal structure and type 1 AGNs are those whose line-of-sight does not intercept the torus. Furthermore, the torus is also thought to play an important role in the co-evolution of the SMBH and the host galaxy \citep[see, e.g.,][]{Kormendy13,Heckman14}. Therefore, putting strong constraints on the physical and geometrical properties of the toroidal structure is essential to understand the basics of AGNs.

Observing the X-ray emission from AGNs is a powerful method to probe their obscuring toroidal structure. The intrinsic X-ray emission produced by the central engine of the AGN is reprocessed by the obscuring torus: studying this reprocessed X-ray emission can then provide abundant of information about the properties of the torus. One of the ubiquitous signatures of this reprocessed component is the fluorescent Fe K$\alpha$ line at 6.4\,keV, originating from the outer side of the accretion disk or the inner edge of the torus \citep[see, e.g.,][for reviews]{Fabian_2000,Reynolds03,Yaqoob_2004}, which could provide significant information on both the physics and dynamics of the circumnuclear materials \citep{Leahy1993,Reynolds1999,Giorgio02,Shu10,EW1}. \XMM\ is the best instrument to study such a signature in terms of both effective area between 0.3\,keV to 10\,keV and spectral resolution. Indeed, many studies have been done on the properties of the torus utilizing \XMM\ \citep[e.g.,][]{Georgantopoulos2013,LaMassa_2014}. Another spectral signature of the reprocessed component, which is particularly prominent in heavily obscured AGN (i.e., sources with column density N$_{\rm H}\ge10^{24}$\,cm$^{-2}$), is the so called ``Compton hump'' peaked at 10--40\,keV \citep[see, e.g.,][]{Ghisellini94,Krolik1994}. Thus the proper characterization of heavily obscured AGNs, which are thought to be $\sim$20-30\% of all AGNs according to different CXB synthesis models predictions \citep{Alexander03,Gandhi03,gilli07,Treister09,Ueda14,Tasnim_Ananna_2019}, requires an X-ray telescope sensitive above 10\,keV. The launch of the \textit{Nuclear Spectroscopic Telescope Array} \citep[hereafter, \NuSTAR,][]{harrison}, which is the first instrument to focus on X-ray at energy $>$10\,keV and provides a two orders of magnitude better sensitivity than previous telescopes \citep[e.g., \textit{INTEGRAL} and \bat;][]{Winkler2003,Barthelmy2005} at $\sim$10--50\,keV, allowed us to characterize the physical properties of heavily obscured AGNs with unprecedented accuracy \citep[e.g.,][]{Balokovic14,puccetti14,Annuar15,Ursini15,koss2016,Stefano2017,Ursini18}. Therefore, the combination of \NuSTAR\ and \XMM\ is one of the best methods to study the properties of an heterogeneous AGN population in the local universe \citep[see, e.g.,][]{Marinucci14,F_rst_2016,Ursini16,Caria19,Marchesi_2019,Walton19,Zhao_2019_2,Zhao_2019_1}. 

In recent years, several tori models based on Monte Carlo simulation have been developed to characterize the X-ray spectra of AGNs \citep{Matt1994,Shinya09,MYTorus2009,BNtorus,Paltani17,Borus,Tanimoto_2019}. Different models adopt different assumptions on the geometry of the torus, e.g., \citet{MYTorus2009,Liu14,Furui16} assume a half-opening angle of the torus fixed at $\theta_{\rm h.o.}$ = 60$^\circ$ but variable inclination angle of the torus ($\theta_{\rm obs}$); \citet{Shinya09,BNtorus,Paltani17,Borus,Tanimoto_2019} assume a flexible half-opening angle of the torus and variable inclination angle. Given the intrinsic complexity of these models and the multiple free parameters involved, applying them in full capability is still difficult especially with low-quality X-ray spectra: in particular, the inclination angle of the AGN is hard to constrain. Thus it is common to freeze $\theta_{\rm obs}$ in the spectral analysis process \citep[see, e.g.,][]{MYTorus2012,Kawamuro_2013,Brightman_2015,MYTorus2015,NuSTAR6,koss2016,Ricci_2016,Gandhi2017,Marchesi2018}.
However, the validity of the method of freezing the inclination angle has not yet been studied in a systematic way. The studies on the NLRs of AGNs can provide us with a method to overcome this issue by measuring the inclinations of the AGNs by mapping the kinematics of their NLRs. \citet{Fischer_2013} successfully measured the inclinations of the NRLs and thus the torus with respect to our line-of-sight in 17 AGNs by fitting the radial outflow dominated NLR kinematics resolved by \textit{Hubble Space Telescope} (HST) [OIII] imaging and Space Telescope Imaging Spectrograph (STIS) with a biconical outflow model.

In this work, we study the role of inclination angle in fitting the AGN X-ray spectra by comparing the best-fit results obtained when the broad X-ray spectra of the sources in the sample of \citet{Fischer_2013} are fitted with the inclination angle being (i) left free to vary, (ii) fixed at [OIII] measured values, (iii) fixed at 60$^\circ$ and (iv) fixed at 87$^\circ$. The paper is organized as follows: in Section \ref{sec:Observation} , we report the sample selection rules and the \NuSTAR, \XMM\ and \cha\ data reduction process; in Section \ref{sec:spectral}, we describe the model used to fit the broadband X-ray spectra, the fitting procedure and best-fit results of each source in our sample; in Section \ref{discussion}, we discuss how fixing the inclination angle affects the broadband X-ray spectral analysis of AGNs and study the geometrical properties of the AGNs in both X-ray and optical.
All reported uncertainties on spectral parameters are at 90\% confidence level. Standard cosmological parameters are adopted as follows: $<H_0>$ = 70 km s$^{-1}$ Mpc$^{-1}$, $<q_0>$ = 0.0 and $<\Omega_\Lambda>$ = 0.73.

%
\section{Sample selection and data reduction}
\label{sec:Observation}
%

\begingroup
\renewcommand*{\arraystretch}{1.3}
\begin{table*}
\scriptsize
\centering
\caption{Inclination angle of 15 sources in \citet{Fischer_2013}}
\label{Table:Information}
\vspace{.1cm}
  \begin{tabular}{ccccccccccccc}
       \hline
       \hline     
       Source&$z$&log(M$\rm_{BH}$)\footnote{Logarithm of the mass of the black hole at the center of the AGN in solar mass. Circinus: \citet{Beifiori12}; Mrk~3, Mrk~78, Mrk~573, Mrk~1066: \citet{Nelson1995}; Mrk~34: \citet{Gandhi_2014}; NGC~1068:\citet{Merritt01}; NGC~3227: \citet{Onken_2003}; NGC~3783: \citet{Onken_2002}; NGC~4051: \citet{Denney_2009}, NGC~4151: \citet{Onken_2007}; NGC~4507: \citet{Nicastro_2003}; NGC~5506: \citet{Nikoajuk09}; NGC~5643: \citet{Goulding10}; NGC~7674: \citet{Woo02}.}&$i_{AGN}$\footnote{AGN inclination angle reported in \citet{Fischer_2013}: 0$^\circ$ corresponds to an ``face-on'' orientation. The typical error is 5$^\circ$.}&$\theta_{\rm max}$\footnote{The opening angle between the bicone axis and the outer edge of the narrow line region, assuming a typical error of 5$^\circ$.}&\NuSTAR&\NuSTAR\footnote{Total effective exposure time after data cleaning of \NuSTAR\ FPMA and FPMB.}&XMM&XMM\footnote{Total effective exposure time after data cleaning of \XMM\ MOS1, MOS2 and pn.}&\cha&\cha\footnote{Effective exposure time after data cleaning for ACIS-S of \cha.}\\
       &&&(deg)&(deg)&date&ks&date&ks&date&ks\\
       \hline
	Circinus&0.00145&6.23&65&41&2013 Jan 25&108&2013 Feb 03&131\\
	Mrk~3&0.01351&8.65&85&51&2015 Apr 08&50&2015 Apr 08&7\\
	Mrk~34&0.05050&7.12&65&40&2013 Sept 19&48&2005 Apr 04&31&2017 Jan 30&100\\
	Mrk~78&0.03715&7.87&60&35&2018 Nov 19&48&2006 Mar 11&16&2017 Jan 01&50\\
	...&...&...&...&...&...&...&...&...&2017 Jan 07&50\\
	Mrk~573&0.01718&7.28&60&53&2018 Jan 06&64&2004 Jan 15&33&&&\\
	Mrk~1066&0.01202&7.01&80&25&2014 Dec 06&60&2005 Feb 20&33&2003 July 14&20\\
	NGC~1068&0.00379&7.20&85&40&2015 Feb 05&108&2015 Feb 03&89\\
	NGC~3227&0.00386&7.56&15&55&2016 Dec 01&84&2016 Dec 01&176\\
	NGC~3783&0.00973&6.94&15&55&2016 Dec 11&52&2016 Dec 11&126\\
	NGC~4051&0.00234&6.24&12&25&2013 Oct 09&100&2009 June 10&71\\
	NGC~4151&0.00332&7.66&45&33&2012 Nov 14&124&2012 Nov 14&16\\
	NGC~4507&0.01180&8.26&47&50&2015 June 10&68&2010 Aug 03&51\\
	NGC~5506&0.00618&7.94&80&40&2014 Apr 01&114&2015 July 08&322\\
	NGC~5643&0.00400&6.44&65&55&2014 May 24&42&2009 July 25&92\\
	...&...&...&...&...&2014 June 30&40&...&...\\
	NGC~7674&0.02892&7.56&60&40&2014 Sept 30&104&2004 June 02&22\\
       \hline
	\vspace{0.06cm}
\end{tabular}
\end{table*}
\endgroup
\subsection{Selection Rule}
To better constrain and properly study the physical and geometrical properties of AGNs, we utilize the sample reported in \citet{Fischer_2013}, who measured the nuclear inclinations of 17 nearby AGNs ($z<\,$0.1) in optical. In these 17 AGNs, 15 sources have high-quality \NuSTAR\ archival data (the 2 sources without \NuSTAR\ archival data are Mrk~279 and NGC~1667). All 15 sources also have \XMM\ archival observations: for 6 of these 15 sources, the \NuSTAR\ and \XMM\ observations were took simultaneously. We also supplement \cha\ data for three sources, i.e., Mrk~34, Mrk~78 and Mrk~1066, of which the \XMM\ spectra is not in high-quality. The summary of the observations is reported in Table~\ref{Table:Information}.

It is worth mentioning that 2 sources in the \citet{Fischer_2013} sample are excluded in our analysis, which are Circinus and NGC~1068. The Circinus AGN X-ray spectra has been shown to be contaminated by two bright off-nuclear X-ray sources, the X-ray binary CGX1 and the supernova remnant CGX2 \citep{Bauer_2001}. Furthermore, \citet{Ar_valo_2014} showed that the contamination from CGX1 and CGX2 contributes to 18\% of the nuclear flux in the Iron line region and becomes comparable to the nuclear flux at energy $>$30\,keV. The off-nuclear sources can be resolved by \XMM, but not by \NuSTAR. Therefore, Circinus is excluded from our sample due to the fact we do not have the ability to extract the AGN broadband X-ray spectrum of Circinus without any contamination. Furthermore, NGC~1068 is also excluded from our final sample, since we find that it is difficult to fit both \NuSTAR\ and \XMM\ spectra properly with the standard model presented in Section~\ref{section:model}. Indeed, \citet{Bauer15} suggests a best-fit model of three reprocessed components with distinct column densities, rather than the single reprocessed component used in our analysis. Therefore, 13 sources are analyzed as our finalized sample in the rest of the work.

\subsection{Data Reduction}
\subsubsection{\NuSTAR}
For \NuSTAR\ data, the raw files are calibrated, cleaned and screened using the \NuSTAR\ \texttt{nupipeline} script version 0.4.6 and calibration database (CALDB) version 20181030. The sources spectra, ancillary response files (ARF) and response matrix files (RMF) are obtained using the \texttt{nuproducts} script version 0.3.0. The sources spectra are extracted from a 75$^{\prime\prime}$ circular region, unless otherwise indicated, corresponding to $\approx$80\% of the encircled energy fraction (EEF) at 10\,keV, centered on the source. The background spectra are extracted using a 75$^{\prime\prime}$ circular region near the source but avoiding contamination from it.

\subsubsection{\XMM} 
The \XMM\ observations are taken with two MOS cameras \citep[][]{MOS} and the EPIC CCD cameras \citep[pn;][]{pn}. The \XMM\ data are reduced using the Science Analysis System \citep[SAS;][]{SAS} version 17.0.0 following standard procedures. The source spectra are extracted from a circular region with radius of 15$^{\prime\prime}$ (corresponding to $\approx$70\% of the EEF at 1.5\,keV) or 30$^{\prime\prime}$ (corresponding to $\approx$85\% of the EEF at 1.5\,keV), based on which spectra has higher signal to noise ratio (SNR); the background spectra are extracted from a circle nearby the source with the same radius as the source spectra but avoiding contamination from sources. ARF and RMF files are produced using the tasks \texttt{arfgen} and \texttt{rmfgen}. 

\subsubsection{\cha}
Archived \cha\ ACIS-S observations are used for three sources (Mrk~34, Mrk~78 and Mrk~1066) which have low-quality \XMM\ data because of their short exposure time and low observing luminosity in soft X-ray band. We reduced the \cha\ data using the \cha's data analysis system, CIAO software package \citep{CIAO} version 4.11 and \cha\ CALDB version 4.8.2. The level = 1 data are reprocessed as suggested to apply updated calibrations as suggested using the CIAO \texttt{chandra\_repro} script. The source spectrum is extracted from a circular region centered at the source with a radius of 5$^{\prime\prime}$; background spectrum is extracted from a circular region near the source with a radius of 10$^{\prime\prime}$. The CIAO \texttt{specextract} tool is used to extract both source and background spectra, ARF and RMF files following standard procedures. 

The \NuSTAR, \XMM\ and \cha\ spectra are rebinned with a minimum of 20 counts per bin using the HEAsoft task \texttt{grppha}.

%
\section{Spectral Analysis and Results} 
\label{sec:spectral}
%
The spectral are fitted using the \texttt{XSPEC} software \citep{Arnaud1996} version 12.10.0c. The photoelectric cross section for the absorption component is from \cite{Verner1996}; the element abundance is from \citet{Anders1989} and the metal abundance is fixed to Solar; the Galactic absorption column density is obtained using the \texttt{nh} task \citep{Kalberla05} in \texttt{HEAsoft} for each source. The redshift of each source is adopted from NED\footnote{\url{https://ned.ipac.caltech.edu}}. In this work, the spectra are analyzed using the self-consistent \borus\ model \citep{Borus}, which is suitable to characterize AGNs with high-quality broadband X-ray spectra.  

%
\subsection{Spectral Modeling}
\label{section:model}
%
The recently published Monte Carlo radiative transfer code {\texttt {BORUS}} \citep{Balokovic_17} has already widely used to model the reprocessed component of AGN spectra \citep[e.g.,][]{Boorman18,Caria19,Kammoun_2019,Li_2019,Masini_2019,Marchesi_2019}; see also the \borus\ website\footnote{\url{http://www.astro.caltech.edu/~mislavb/download/}} for more details. The complete model used in fitting the spectra is composed of four parts:
\begin{enumerate}
\item An absorbed intrinsic continuum, described by a cut-off power-law, denoted by \texttt{cutoffpl} in \XSPEC, multiplied by a obscuring component, considering both the photoelectric absorption (\texttt{zphabs}) and the Compton scattering (\texttt{cabs}) effects.

\item A reprocessed component produced by the obscuring material near the center of the AGN, including the  scattered component and fluorescent lines, characterized by \borus.

\item A second, leaked unabsorbed intrinsic continuum, modeling the fractional AGN emission which is deflected, rather than absorbed by the obscuring material.

\item A thermal component, namely \texttt{mekal} \citep{mekal}, modeling the soft excess observed below 1\,keV, potentially describing the emission caused by the processes other than AGN, such as star-formation and/or diffuse gas emission.
\end{enumerate}

The reprocessed component, \borus\footnote{The energy coverage of \borus\ model is 1\,keV $<E$ $<$1000\,keV. The model cut-off at 1\,keV does not affect the fit of the sources in our sample since their spectra in soft energy band ($E<$ 3\,keV) are dominated by the leaked component.} assumes a sphere with conical cutouts at both poles \citep{Borus}, approximating a torus with an opening angle which can vary in the range of $\theta_{\rm Tor}$ = [0--84]$^\circ$, corresponding to a torus covering factor, $c_f$ = cos($\theta_{\rm Tor}$) = [1--0.1]. Another parameter in the reprocessed component is the inclination angle, which is the angle between the axis of the AGN and the observer line-of-sight, $\theta_{\rm inc}$ = [18--87]$^\circ$, where $\theta\rm_{obs}$ = 0$^\circ$ is when the AGN is observed ``face-on'' and $\theta\rm_{obs}$ = 87$^\circ$ is observed ``edge-on". Another parameter, the relative iron abundance of the reprocessed component, A$\rm _{Fe}$, is fixed to 1 (i.e., the iron abundance in solar, A$\rm _{Fe,\odot}$), unless a much better result is obtained leaving the parameter free to vary. We plot the spectra of the \borus\ \citep{Borus} model prediction when varying different parameters in Appendix~A to illustrate how the spectra of the reprocessed component vary with different parameters, i.e., $\theta\rm_{obs}$, $\theta_{\rm Tor}$, $\theta_{\rm Tor}$ and A$\rm _{Fe}$.  Evidence in Infrared and X-ray observations have shown that the torus is clumpy rather than having an uniform density \citep[e.g.,][]{Krolik1988,Risaliti_2002,Nenkova08,Ramos_Almeida_2009,Markowitz14}. Therefore, the column density of the obscuring torus in the reprocessed component is decoupled from the one in the absorbed intrinsic continuum in our modeling to approximate the clumpy nature of the obscuring torus. In this scenario, the column density of the reprocessed component is an average property of the clumpy torus while the column density of the absorbed intrinsic continuum represents a line-of-sight quantity.

In the process of modeling the spectra, the photon index, $\Gamma$, the cut-off energy, $E_{\rm cut}$ and the normalization, $norm$, of the intrinsic continuum, the reprocessed component and the fractional unabsorbed continuum are tied together, assuming that the three component have the same origin. The cut-off energy is fixed at $E_{\rm cut}$ = 500\,keV, unless a much better result is obtained leaving the parameter free to vary. The fractional unabsorbed continuum is usually less than 5--10\% of the intrinsic continuum \citep[see, e.g.,][]{Noguchi_2010,Marchesi2018}. We denote this fraction as $f_s$, and we model it with a constant (\texttt{constant$_2$}). Finally, the temperature and the relative metal abundance in \texttt{mekal} are both left free to vary. Lines are added if strong emission lines are found in the spectra using \texttt{zgauss} model in \XSPEC.

The \borus\ model is used in the following \XSPEC\ configuration:
\begin{equation*}\label{eq:Borus}
\begin{aligned}
Model =&constant_1*phabs*(borus02+zphabs*cabs\\
&*cutoffpl+constant_2*cutoffpl+mekal) 
\end{aligned}
\end{equation*}
where \texttt{constant$_1$} is the cross-calibration between \NuSTAR\ and \XMM\ (separate cross-calibration constants are applied if \cha\ data are used); \texttt{phabs} models the Galactic absorption.

%
\subsection{Results}
\label{section:results}
%
\begingroup
\renewcommand*{\arraystretch}{1.5}
\begin{table*}
\footnotesize
\centering
\caption{Best-fit results of 13 sources}
\label{Table:results}
\vspace{.1cm}
  \begin{tabular}{ccccccccccc}
       \hline
       \hline     
         Model&$\chi^2$/d.o.f.&$\Gamma$&N$\rm _{H,l.o.s}$\footnote{Logarithm of line-of-sight column density in cm$^{-2}$; for sources which variability has been obsvered between the \NuSTAR\ and \XMM, we report here the line-of-sight column density of the \XMM\ observation}&N$\rm _{H,tor}$\footnote{Logarithm of average torus column density in cm$^{-2}$}&cos($\theta\rm _{inc}$)\footnote{Inclination angle, i.e., the angle between the axis of the torus and the edge of the torus}&$c_{\rm f,tor}$\footnote{Effective covering factor of the torus}&norm\footnote{normalization of the main cut-off power-law component at 1\,keV in 10$^{-2}$ photons keV$^{-1}$\,cm$^{-2}$\,s$^{-1}$ of \XMM\ observations}&$f_s$\footnote{Fraction of scattering component in 10$^{-2}$}&F$_{2-10}$\footnote{Flux between 2--10\,keV in $10^{-12}$\,erg\,cm$^{-2}$\,s$^{-1}$ of \XMM\ observation}&L$_{\rm int}$\footnote{Intrinsic luminosity between 2--10\,keV in $10^{42}$\,erg\,s$^{-1}$ of \XMM\ observation}\\
       \hline
       \hline
\vspace{.1cm}
&&&&&Mrk~3&&&&&\\
\hline
free&1056/1073&1.48$_{-u}^{+0.11}$&23.94$_{-0.04}^{+0.06}$&23.30$_{-0.15}^{+0.24}$&0.47$_{-0.07}^{+0.16}$&0.50$_{-0.22}^{+0.06}$&1.42$_{-0.22}^{+0.54}$&0.98$_{-0.89}^{+0.46}$&8$_{-4}^{+1}$&31\\
       \hline
       [OIII]&1061/1074&1.40$_{-u}^{+0.05}$&23.91$_{-0.05}^{+0.01}$&23.04$_{-0.26}^{+0.24}$&0.09$^f$&0.62$_{-0.05}^{+0.28}$&1.12$_{-0.11}^{+0.01}$&1.36$_{-0.14}^{+0.55}$&8$_{-4}^{+2}$&28\\
       \hline
       60$^\circ$&1057/1074&1.46$_{-u}^{+0.18}$&23.95$_{-0.05}^{+0.06}$&23.32$_{-0.24}^{+0.44}$&0.50$^{f}$&0.50$_{-0.17}^{+0.06}$&1.35$_{-0.17}^{+0.82}$&0.81$_{-0.50}^{+0.36}$&8$_{-5}^{+1}$&31\\
       \hline
       87$^\circ$&1061/1074&1.40$_{-u}^{+0.12}$&23.90$_{-0.04}^{+0.02}$&23.07$_{-0.31}^{+0.19}$&0.05$^f$&0.60$_{-0.04}^{+0.29}$&1.12$_{-0.11}^{+0.01}$&1.41$_{-0.14}^{+0.44}$&8$_{-3}^{+3}$&28\\
       \hline
\vspace{.1cm}
&&&&&Mrk~34&&&&&\\
\hline
      free&74/82&1.45$_{-u}^{+0.67}$&24.74$_{-0.46}^{+u}$&25.04$_{-0.73}^{+u}$&0.42$_{-u}^{+u}$&0.40$_{-0.27}^{+0.44}$&0.06$_{-0.03}^{+0.35}$&0.85$_{-0.65}^{+2.02}$&0.2$_{-0.2}^{+18}$&21\\
       \hline
       [OIII]&74/83&1.49$_{-u}^{+0.59}$&24.73$_{-0.34}^{+u}$&24.98$_{-0.64}^{+u}$&0.42$^f$&0.41$_{-0.31}^{+0.06}$&0.07$_{-0.02}^{+0.56}$&0.74$_{-0.53}^{+0.53}$&0.2$_{-0.2}^{+3.7}$&24\\
       \hline
       60$^\circ$&74/83&1.46$_{-u}^{+0.55}$&24.67$_{-0.36}^{+u}$&25.00$_{-0.71}^{+u}$&0.50$^{f}$&0.43$_{-u}^{+0.09}$&0.05$_{-0.10}^{+0.47}$&1.12$_{-0.82}^{+0.65}$&0.2$_{-0.2}^{+2.2}$&16\\
       \hline
       87$^\circ$&76/83&1.41$_{-u}^{+0.33}$&24.59$_{-0.04}^{+0.08}$&23.66$_{-0.09}^{+0.17}$&0.05$^f$&0.10$_{-u}^{+0.13}$&0.25$_{-0.02}^{+0.07}$&0.29$_{-0.05}^{+0.06}$&0.2$_{-0.2}^{+0.5}$&96\\
       \hline
\vspace{.1cm}
&&&&&Mrk~78&&&&&\\
       \hline
   free&276/271&1.40$_{-u}^{+0.21}$&23.91$_{-0.02}^{+0.13}$&24.21$_{-0.33}^{+0.18}$&0.43$_{-0.18}^{+0.45}$&0.45$_{-0.23}^{+u}$&0.06$_{-0.03}^{+0.00}$&1.80$_{-0.20}^{+2.00}$&0.5$_{-0.5}^{+0.3}$&12\\
       \hline
       [OIII]&276/272&1.40$_{-u}^{+0.22}$&23.89$_{-0.07}^{+0.07}$&24.15$_{-0.25}^{+0.15}$&0.50$^f$&0.54$_{-0.12}^{+u}$&0.05$_{-0.01}^{+0.00}$&2.19$_{-0.23}^{+1.56}$&0.5$_{-0.5}^{+0.2}$&10\\
       \hline
       60$^\circ$&276/272&1.40$_{-u}^{+0.22}$&23.89$_{-0.07}^{+0.07}$&24.15$_{-0.25}^{+0.15}$&0.50$^f$&0.54$_{-0.12}^{+u}$&0.05$_{-0.01}^{+0.00}$&2.19$_{-0.23}^{+1.56}$&0.5$_{-0.5}^{+0.2}$&10\\
       \hline
       87$^\circ$&278/272&1.40$_{-u}^{+0.16}$&23.82$_{-0.02}^{+0.06}$&23.95$_{-0.06}^{+0.07}$&0.05$^f$&0.99$_{-0.34}^{+u}$&0.04$_{-0.01}^{+0.00}$&3.32$_{-0.29}^{+0.90}$&0.5$_{-0.3}^{+0.1}$&8.0\\
       \hline
\vspace{.1cm}
&&&&&Mrk~573&&&&&\\
       \hline
       free&152/194&2.35$_{-0.65}^{+u}$&24.52$_{-0.30}^{+u}$&24.91$_{-0.68}^{+u}$&0.60$_{-u}^{+u}$&0.61$_{-u}^{+0.37}$&0.56$_{-0.51}^{+2.25}$&0.17$_{-0.19}^{+1.91}$&0.3$_{-0.3}^{+10}$&5.7\\
       \hline
       [OIII]&152/195&2.36$_{-0.62}^{+u}$&24.63$_{-0.32}^{+u}$&25.00$_{-0.73}^{+u}$&0.50$^f$&0.54$_{-u}^{+0.04}$&0.96$_{-0.87}^{+2.58}$&0.10$_{-0.10}^{+1.13}$&0.3$_{-0.3}^{+12}$&9.7\\
       \hline
       60$^\circ$&152/195&2.36$_{-0.62}^{+u}$&24.63$_{-0.32}^{+u}$&25.00$_{-0.73}^{+u}$&0.50$^f$&0.54$_{-u}^{+0.04}$&0.96$_{-0.87}^{+2.58}$&0.10$_{-0.10}^{+1.13}$&0.3$_{-0.3}^{+12}$&9.7\\
       \hline
       87$^\circ$&152/195&2.60$_{-0.02}^{+u}$&24.94$_{-0.05}^{+u}$&24.99$_{-0.02}^{+0.04}$&0.05$^f$&0.12$_{-u}^{+0.06}$&110$_{-8}^{+6}$&0.00$_{-0.00}^{+0.01}$&0.3$_{-0.3}^{+1.1}$&798\\
       \hline
\vspace{.1cm}
&&&&&Mrk~1066&&&&&\\
       \hline
      free&142/147&1.52$_{-u}^{+0.02}$&23.97$_{-0.05}^{+0.03}$&24.16$_{-0.10}^{+0.31}$&0.65$_{-u}^{+u}$&1.00$_{-0.36}^{+u}$&0.08$_{-0.01}^{+0.11}$&4.05$_{-2.36}^{+2.38}$&0.3$_{-0.3}^{+0.1}$&1.1\\
       \hline
       [OIII]&142/148&1.53$_{-0.06}^{+0.06}$&23.98$_{-0.07}^{+0.03}$&24.17$_{-0.05}^{+0.06}$&0.17$^f$&1.00$_{-0.37}^{+u}$&0.08$_{-0.01}^{+0.02}$&4.04$_{-0.47}^{+1.51}$&0.3$_{-0.3}^{+0.1}$&1.1\\
       \hline
       60$^\circ$&142/148&1.54$_{-0.04}^{+0.02}$&23.98$_{-0.02}^{+0.00}$&24.17$_{-0.02}^{+0.01}$&0.50$^{f}$&1.00$_{-0.36}^{+u}$&0.09$_{-0.00}^{+0.01}$&3.78$_{-0.08}^{+0.63}$&0.3$_{-0.3}^{+0.1}$&1.2\\
       \hline
       87$^\circ$&142/148&1.53$_{-0.03}^{+0.06}$&23.97$_{-0.06}^{+0.04}$&24.17$_{-0.07}^{+0.04}$&0.05$^f$&1.00$_{-0.36}^{+u}$&0.09$_{-0.03}^{+0.10}$&3.78$_{-0.70}^{+2.32}$&0.3$_{-0.3}^{+0.1}$&1.2\\
        \hline
\vspace{.1cm}
&&&&&NGC~3227&&&&&\\
       \hline
       free&4684/4008&1.68$_{-0.01}^{+0.01}$&21.39$_{-0.01}^{+0.03}$&23.14$_{-0.02}^{+0.03}$&0.15$_{-u}^{+0.14}$&1.00$_{-0.07}^{+u}$&0.86$_{-0.01}^{+0.01}$&0$^{f}$&37.7$_{-0.2}^{+0.3}$&1.2\\
       \hline
       [OIII]&4686/4009&1.68$_{-0.01}^{+0.01}$&21.39$_{-0.04}^{+0.05}$&23.14$_{-0.04}^{+0.03}$&0.95$^f$&1.00$_{-0.05}^{+u}$&0.86$_{-0.01}^{+0.01}$&0$^{f}$&37.7$_{-0.1}^{+0.2}$&1.2\\
       \hline
       60$^\circ$&4686/4009&1.68$_{-0.01}^{+0.01}$&21.39$_{-0.06}^{+0.05}$&23.14$_{-0.04}^{+0.03}$&0.95$^f$&1.00$_{-0.13}^{+u}$&0.86$_{-0.01}^{+0.01}$&0$^{f}$&37.7$_{-0.1}^{+0.1}$&1.2\\
       \hline
       87$^\circ$&4686/4009&1.68$_{-0.01}^{+0.01}$&21.39$_{-0.06}^{+0.05}$&23.15$_{-0.04}^{+0.03}$&0.05$^f$&1.00$_{-0.11}^{+u}$&0.87$_{-0.02}^{+0.01}$&0$^{f}$&37.7$_{-0.3}^{+0.3}$&1.2\\
        \hline
\vspace{.1cm}
&&&&&NGC~3783&&&&&\\
       \hline
       free&3349/2929&1.51$_{-0.04}^{+0.02}$&22.85$_{-0.01}^{+0.01}$&25.00$_{-0.22}^{+0.11}$&0.54$_{-0.02}^{+0.02}$&0.41$_{-0.04}^{+0.08}$&0.68$_{-0.03}^{+0.03}$&10.14$_{-0.28}^{+0.30}$&26$_{-1}^{+1}$&6.9\\
       \hline
       [OIII]&3367/2930&1.49$_{-0.02}^{+0.04}$&22.84$_{-0.01}^{+0.02}$&25.11$_{-0.23}^{+u}$&0.95$^f$&0.26$_{-0.01}^{+0.01}$&0.66$_{-0.03}^{+0.05}$&10.55$_{-0.70}^{+0.39}$&26$_{-2}^{+1}$&6.9\\
       \hline
       60$^\circ$&3356/2930&1.49$_{-0.05}^{+0.01}$&22.85$_{-0.01}^{+0.01}$&24.98$_{-0.27}^{+0.12}$&0.50$^f$&0.40$_{-0.02}^{+0.01}$&0.67$_{-0.05}^{+0.02}$&10.42$_{-0.33}^{+0.11}$&26$_{-2}^{+1}$&6.9\\
       \hline
       87$^\circ$&3439/2930&1.55$_{-0.01}^{+0.01}$&22.82$_{-0.01}^{+0.01}$&23.58$_{-0.08}^{+0.06}$&0.05$^f$&1.00$_{-0.03}^{+u}$&0.66$_{-0.04}^{+0.01}$&10.47$_{-0.18}^{+0.69}$&27$_{-1}^{+0}$&6.8\\
       \hline
        \hline
\end{tabular}
	\vspace{0.2cm}
\end{table*}
\endgroup

\begingroup
\renewcommand*{\arraystretch}{1.5}
\begin{table*}
\small
\centering
\caption{Best-fit results of 13 sources}
\label{Table:results2}
\vspace{.1cm}
\begin{tabular}{ccccccccccc}
       \hline
       \hline     
         Model&$\chi^2$/d.o.f.&$\Gamma$&N$\rm _{H,l.o.s}$\footnote{Logarithm of line-of-sight column density in cm$^{-2}$; for sources which variability has been obsvered between the \NuSTAR\ and \XMM, we report here the line-of-sight column density of the \XMM\ observation}&N$\rm _{H,tor}$\footnote{Logarithm of average torus column density in cm$^{-2}$}&cos($\theta\rm _{inc}$)\footnote{Inclination angle, i.e., the angle between the axis of the torus and the edge of the torus}&$c_{\rm f,tor}$\footnote{Effective covering factor of the torus}&norm\footnote{normalization of the main cut-off power-law component at 1\,keV in 10$^{-2}$ photons keV$^{-1}$\,cm$^{-2}$\,s$^{-1}$ of \XMM\ observations}&$f_s$\footnote{Fraction of scattering component in 10$^{-2}$}&F$_{2-10}$\footnote{Flux between 2--10\,keV in $10^{-12}$\,erg\,cm$^{-2}$\,s$^{-1}$ of \XMM\ observation}&L$_{\rm int}$\footnote{Intrinsic luminosity between 2--10\,keV in $10^{42}$\,erg\,s$^{-1}$ of \XMM\ observation}\\
       \hline
       \hline
        \vspace{0.1cm}    
&&&&&NGC~4051&&&&&\\
       \hline
       free&2686/2390&1.72$_{-0.01}^{+0.01}$&22.53$_{-0.01}^{+0.01}$&24.45$_{-0.09}^{+0.07}$&0.95$_{-0.01}^{+u}$&0.95$_{-0.01}^{+0.01}$&0.45$_{-0.01}^{+0.01}$&0$^{f}$&14$_{-1}^{+1}$&0.2\\
       \hline
       [OIII]&2686/2391&1.72$_{-0.01}^{+0.01}$&22.53$_{-0.01}^{+0.01}$&24.45$_{-0.06}^{+0.07}$&0.95$^f$&0.95$_{-0.01}^{+0.01}$&0.45$_{-0.01}^{+0.01}$&0$^{f}$&14$_{-1}^{+1}$&0.2\\
       \hline
       60$^\circ$&2739/2391&1.70$_{-0.05}^{+0.03}$&22.53$_{-0.02}^{+0.03}$&24.16$_{-0.04}^{+0.03}$&0.50$^f$&1.00$_{-0.04}^{+u}$&0.45$_{-0.02}^{+0.01}$&0$^{f}$&14$_{-2}^{+0}$&0.2\\
       \hline
       87$^\circ$&2738/2391&1.70$_{-0.05}^{+0.03}$&22.53$_{-0.01}^{+0.03}$&24.16$_{-0.05}^{+0.03}$&0.05$^f$&1.00$_{-0.02}^{+u}$&0.45$_{-0.02}^{+0.01}$&0$^{f}$&14$_{-2}^{+0}$&0.2\\
        \hline
        \vspace{0.1cm}    
&&&&&NGC~4151&&&&&\\
       \hline
       free&5200/4664&1.67$_{-0.04}^{+0.02}$&23.00$_{-0.01}^{+0.01}$&23.94$_{-0.02}^{+0.02}$&0.05$_{-u}^{+0.03}$&0.80$_{-0.07}^{+0.04}$&3.64$_{-0.23}^{+0.16}$&2.75$_{-0.13}^{+0.20}$&88$_{-1}^{+1}$&3.6\\
       \hline
       [OIII]&5208/4665&1.74$_{-0.01}^{+0.02}$&23.01$_{-0.00}^{+0.01}$&24.01$_{-0.03}^{+0.04}$&0.71$^f$&0.88$_{-0.06}^{+u}$&4.09$_{-0.12}^{+0.19}$&2.49$_{-0.14}^{+0.11}$&88$_{-1}^{+1}$&3.7\\
       \hline
       60$^\circ$&5201/4665&1.69$_{-0.04}^{+0.03}$&23.00$_{-0.01}^{+0.02}$&23.98$_{-0.02}^{+0.02}$&0.50$^f$&0.78$_{-0.09}^{+0.08}$&3.75$_{-0.27}^{+0.22}$&2.68$_{-0.16}^{+0.23}$&88$_{-1}^{+1}$&3.6\\
       \hline
       87$^\circ$&5200/4665&1.67$_{-0.04}^{+0.02}$&23.00$_{-0.01}^{+0.01}$&23.94$_{-0.03}^{+0.04}$&0.05$^f$&0.80$_{-0.06}^{+0.04}$&3.64$_{-0.24}^{+0.16}$&2.75$_{-0.13}^{+0.20}$&88$_{-1}^{+1}$&3.6\\
\hline
        \vspace{0.1cm}    
&&&&&NGC~4507&&&&&\\
       \hline
       free&1601/1614&1.71$_{-0.03}^{+0.05}$&23.98$_{-0.05}^{+0.03}$&23.40$_{-0.09}^{+0.09}$&0.43$_{-0.13}^{+0.10}$&0.55$_{-0.06}^{+0.09}$&2.03$_{-0.46}^{+0.40}$&0.27$_{-0.19}^{+0.18}$&7.1$_{-0.3}^{+0.3}$&25\\
       \hline
       [OIII]&1603/1615&1.85$_{-0.02}^{+0.04}$&23.87$_{-0.02}^{+0.02}$&25.50$_{-0.19}^{+u}$&0.68$^f$&0.40$_{-0.06}^{+0.06}$&1.67$_{-0.10}^{+0.24}$&0.18$_{-0.02}^{+0.15}$&7.1$_{-0.3}^{+0.2}$&17\\
       \hline
       60$^\circ$&1603/1615&1.68$_{-0.05}^{+0.02}$&23.98$_{-0.06}^{+0.02}$&23.44$_{-0.11}^{+0.04}$&0.50$^f$&0.59$_{-0.04}^{+0.07}$&1.81$_{-0.34}^{+0.14}$&0.23$_{-0.18}^{+0.17}$&7.1$_{-0.3}^{+0.2}$&23\\
       \hline
       87$^\circ$&1607/1615&1.69$_{-0.04}^{+0.06}$&23.97$_{-0.03}^{+0.05}$&23.34$_{-0.05}^{+0.06}$&0.05$^f$&0.63$_{-0.05}^{+0.08}$&1.89$_{-0.29}^{+0.52}$&0.43$_{-0.09}^{+0.10}$&7.1$_{-0.3}^{+0.1}$&24\\
        \hline    
        \vspace{0.1cm}    
&&&&&NGC~5506&&&&&\\
       \hline
       free&5378/4543&1.72$_{-0.01}^{+0.01}$&22.50$_{-0.00}^{+0.01}$&23.91$_{-0.02}^{+0.02}$&0.55$_{-0.25}^{+0.09}$&1.00$_{-0.02}^{+u}$&1.90$_{-0.01}^{+0.03}$&1.33$_{-0.04}^{+0.04}$&61$_{-0}^{+4}$&4.9\\
       \hline
       [OIII]&5381/4544&1.72$_{-0.01}^{+0.02}$&22.50$_{-0.00}^{+0.01}$&23.91$_{-0.01}^{+0.01}$&0.17$^f$&1.00$_{-0.02}^{+u}$&1.90$_{-0.03}^{+0.04}$&1.33$_{-0.05}^{+0.04}$&61$_{-1}^{+4}$&4.9\\
       \hline
       60$^\circ$&5379/4544&1.72$_{-0.01}^{+0.01}$&22.50$_{-0.00}^{+0.01}$&23.91$_{-0.02}^{+0.02}$&0.50$^f$&1.00$_{-0.02}^{+u}$&1.90$_{-0.02}^{+0.03}$&1.33$_{-0.05}^{+0.04}$&61$_{-0}^{+4}$&4.9\\
       \hline
       87$^\circ$&5380/4544&1.72$_{-0.01}^{+0.02}$&22.50$_{-0.00}^{+0.01}$&23.91$_{-0.01}^{+0.01}$&0.05$^f$&1.00$_{-0.02}^{+u}$&1.90$_{-0.03}^{+0.04}$&1.33$_{-0.05}^{+0.04}$&61$_{-1}^{+5}$&4.9\\
        \hline
                \vspace{0.1cm}    
&&&&&NGC~5643&&&&&\\
       \hline
       free&198/170&1.77$_{-0.37}^{+0.27}$&24.65$_{-0.32}^{+u}$&24.15$_{-0.35}^{+0.22}$&0.51$_{-0.24}^{+u}$&0.50$_{-0.23}^{+0.38}$&0.41$_{-0.27}^{+1.23}$&0.10$_{-0.10}^{+0.39}$&0.8$_{-0.6}^{+15}$&0.5\\
       \hline
       [OIII]&199/171&1.77$_{-0.34}^{+0.32}$&24.65$_{-0.25}^{+0.23}$&24.13$_{-0.28}^{+0.23}$&0.42$^f$&0.42$_{-0.13}^{+0.04}$&0.51$_{-0.28}^{+0.63}$&0.08$_{-0.08}^{+0.28}$&0.8$_{-0.3}^{+19}$&0.6\\
       \hline
       60$^\circ$&198/171&1.80$_{-0.32}^{+0.27}$&24.65$_{-0.28}^{+u}$&24.15$_{-0.31}^{+0.24}$&0.50$^{f}$&0.49$_{-0.12}^{+0.04}$&0.47$_{-0.25}^{+0.54}$&0.10$_{-0.10}^{+0.31}$&0.8$_{-0.5}^{+17}$&0.6\\
       \hline
       87$^\circ$&220/171&1.40$_{-u}^{+0.12}$&24.35$_{-0.05}^{+0.06}$&23.44$_{-0.08}^{+0.04}$&0.05$^f$&0.43$_{-0.07}^{+0.03}$&0.12$_{-0.01}^{+0.01}$&1.59$_{-0.21}^{+0.21}$&0.8$_{-0.6}^{+0.4}$&0.3\\
        \hline
        \vspace{0.1cm}    
&&&&&NGC~7674&&&&&\\
       \hline
       free&240/251&2.21$_{-0.18}^{+0.12}$&24.15$_{-0.05}^{+0.08}$&23.65$_{-0.05}^{+0.14}$&0.25$_{-0.04}^{+0.52}$&0.10$_{-u}^{+0.15}$&1.49$_{-0.06}^{+2.87}$&0$^f$&0.6$_{-0.6}^{+0.6}$&54\\
       \hline
       [OIII]&241/252&2.20$_{-0.19}^{+0.17}$&24.13$_{-0.06}^{+0.06}$&23.73$_{-0.05}^{+0.11}$&0.50$^f$&0.10$_{-u}^{+0.07}$&1.26$_{-0.06}^{+0.36}$&0$^f$&0.7$_{-0.7}^{+0.2}$&46\\
       \hline
       60$^\circ$&241/252&2.20$_{-0.19}^{+0.17}$&24.13$_{-0.06}^{+0.06}$&23.73$_{-0.05}^{+0.11}$&0.50$^f$&0.10$_{-u}^{+0.07}$&1.26$_{-0.06}^{+0.36}$&0$^f$&0.7$_{-0.7}^{+0.2}$&46\\
       \hline
       87$^\circ$&245/251&2.00$_{-0.31}^{+0.40}$&24.10$_{-0.17}^{+0.15}$&23.27$_{-0.18}^{+0.13}$&0.05$^f$&0.30$_{-u}^{+0.39}$&0.66$_{-0.49}^{+1.65}$&0.60$_{-0.04}^{+1.47}$&0.7$_{-0.7}^{+0.1}$&32\\
       \hline
\end{tabular}
	\vspace{0.2cm}
\end{table*}
\endgroup

We fit the spectra twice at first: letting the inclination angle free to vary when fitting the spectra and fixing the inclination angle at the values reported in \citet{Fischer_2013}. The best-fit results are reported in Table~\ref{Table:results} and Table~\ref{Table:results2}. Furthermore, to extend our analysis on the role of inclination angle in the spectral analysis, we also fit the spectra by fixing the inclination at some specific angles, i.e., $\theta_{\rm inc}$ = 60$^\circ$ or cos($\theta_{\rm inc}$) = 0.5 \citep[i.e., the opening angle of the torus in MYTorus model][]{MYTorus2009}, and fixing inclination at $\theta_{\rm inc}$ = 87$^\circ$ (the maximum angle in \borus\ model) representing an ``edge-on'' scenario, which is commonly used when analyzing heavily obscured AGN spectra \citep[see, e.g.,][]{BNtorus,NuSTAR6,Ricci_2016,Marchesi2018,Zappacosta_2018}, although not all of the sources in our sample are heavily obscured. The best-fit results of the two scenarios are also reported in Table~\ref{Table:results} and Table~\ref{Table:results2}. The details of fitting procedure of each source and best-fit results of the 13 sources fitted when the inclination angles are left free to vary are reported in Appendix~C. The unfolded spectra and the model predictions of each source when fitted leaving the inclination angle free to vary are plotted in Appendix~C as well.

6 out of 13 sources in our sample have been observed displaying strong variability between their soft X-ray observations and \NuSTAR\ observations. The observed variability is commonly explained by either the variation of the accretion rate of the SMBH or the variation in the so-called corona when the fluctuation of intrinsic emission of the AGN is observed, i.e., the shape of the spectrum does not change while the normalization of the intrinsic power-law varies \citep{NANDRA2001295}, or the change in the absorption column density along the line-of-sight when the shape of the spectra varies \citep[see, e.g.,][]{Risaliti_2002,Bianchi12}, or both. To properly characterize the spectra of these sources, we fit them three times: 1. disentangling the normalization of the intrinsic cut-off power law, {\tt norm}, of the soft X-ray observatories observations and the \NuSTAR\ observations, modeling the flux variability caused by the intrinsic emission variation; 2. disentangling the line-of-sight column densities of the soft X-ray observations, N$_{\rm H,los,soft}$ and \NuSTAR\ observations, N$_{\rm H,los,NuS}$, modeling the flux variability results from the line-of-sight column density variability; 3. disentangling both the {\texttt norm} and N$_{\rm H,los}$ between the soft X-ray observations and the \NuSTAR\ observations, assuming the flux variability is caused by both the intrinsic emission variation and line-of-sight column density variability. Here, we treat the reprocessed emission as an invariable component during the two observations, assuming a stable structure and constant global properties of the obscuring torus.

\begin{figure*} 
\begin{minipage}[b]{.5\textwidth}
\centering
\includegraphics[width=1\textwidth]{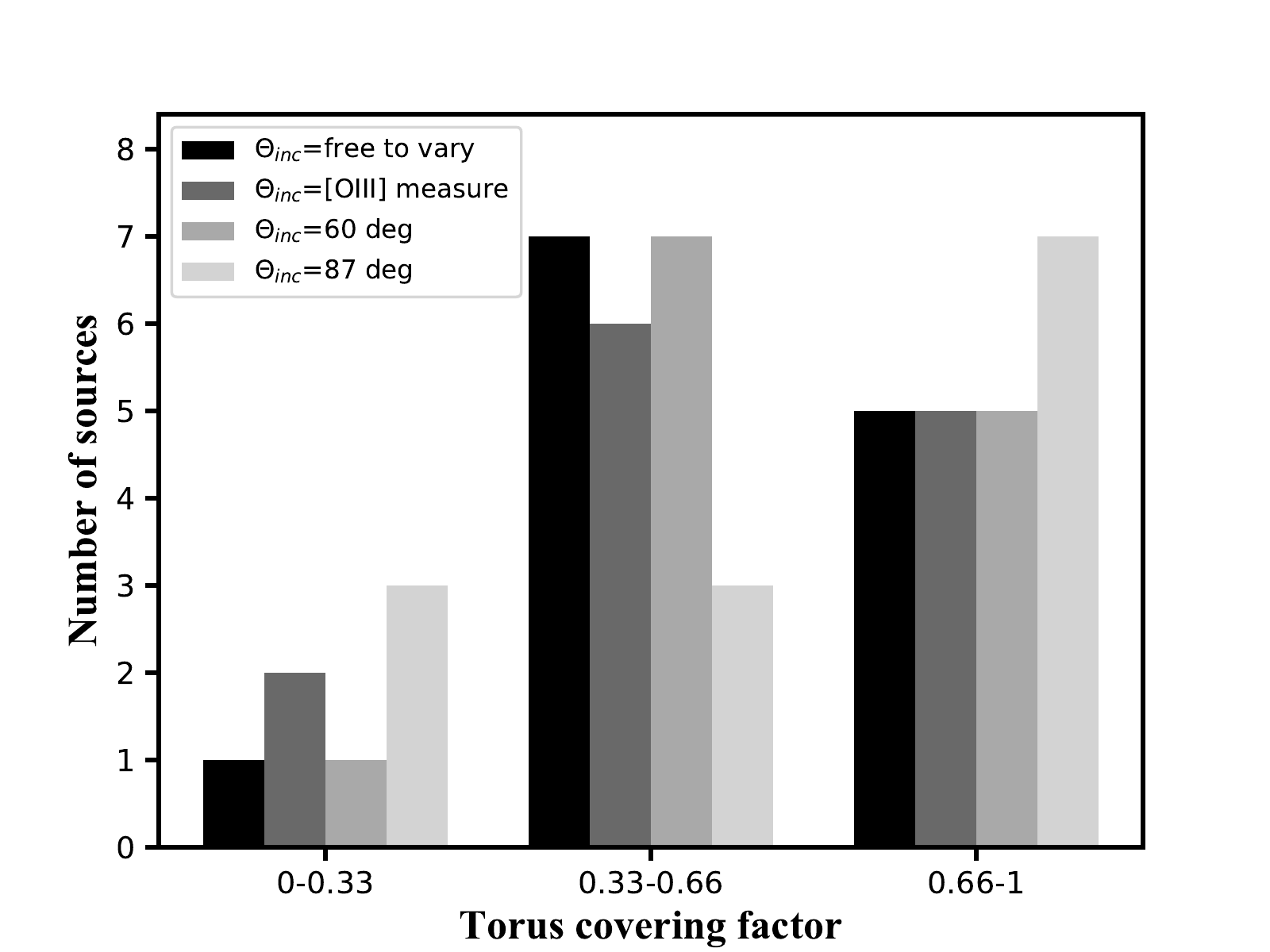}
\end{minipage}
\begin{minipage}[b]{.5\textwidth}
\centering
\includegraphics[width=1\textwidth]{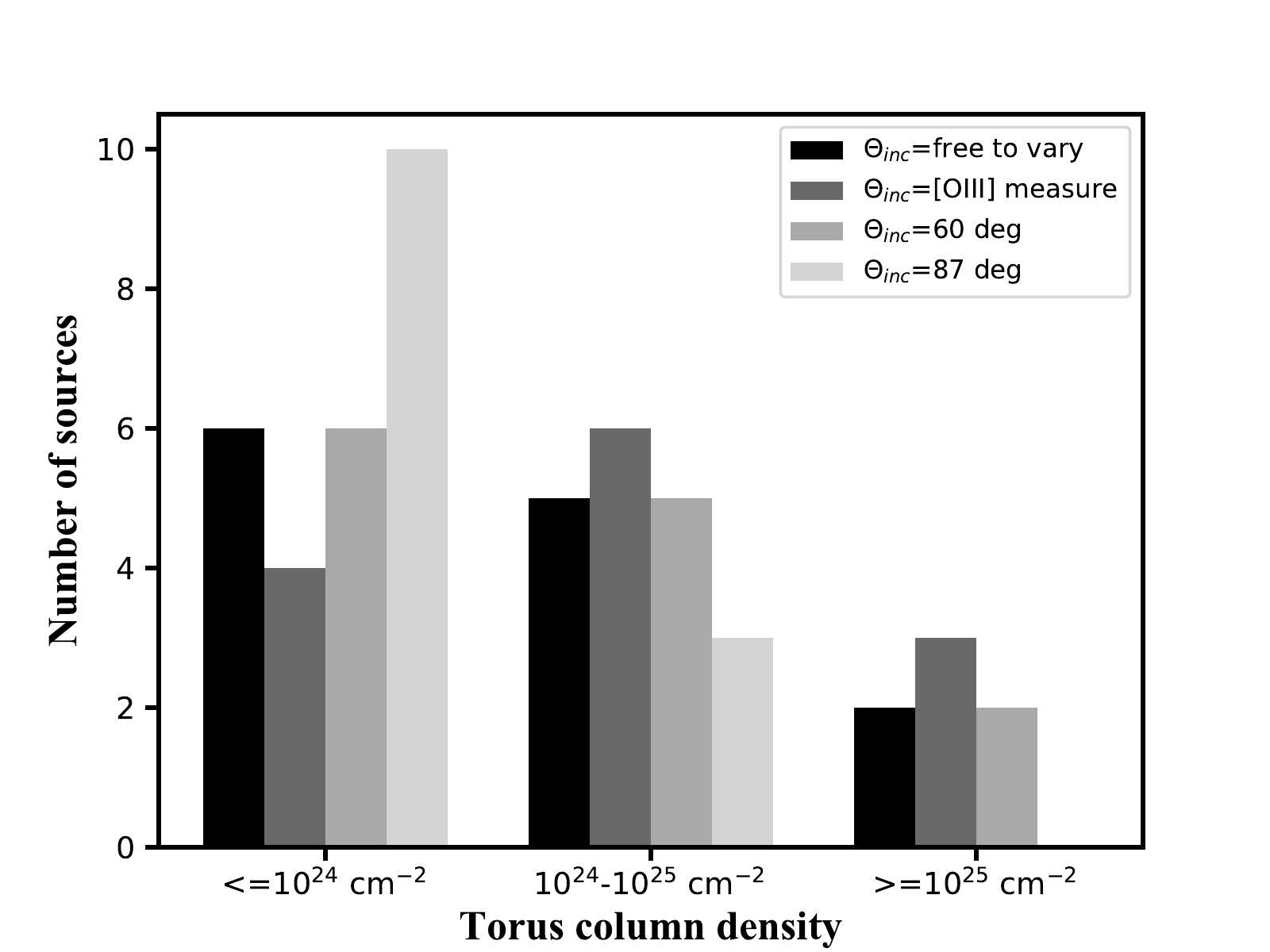}
\end{minipage}
\caption{The figures from the left to right in the first row are the number of sources with specific best-fit torus covering factor and torus column density when $\rm \theta_{inc}$ is left free to vary (black histogram), when $\rm \theta_{inc}$ is fixed at [OIII] measured values (dark gray histogram), when $\rm \theta_{inc}$ is fixed at $\rm \theta_{inc}$ = 60$^\circ$ (light gray histogram) and $\rm \theta_{inc}$ = 87$^\circ$ (silver histogram).}
\label{fig:bar}
\end{figure*}   
The decoupling line-of-sight column density and the torus average column density applied to fit the spectra in this work is commonly used to approximate the non-uniform (clumpy) torus. To interpret the obtained results, we separate the sources in three categories. 
\begin{enumerate}
\item Obscured AGNs with log(N$\rm _{H,l.o.s}$) $\ge$22 where the line-of-sight does not intercept the torus (cos($\theta\rm _{inc}$) $>$$c_{\rm f,tor}$), i.e., Mrk 34, NGC 3783, NGC 5643. The interpretation of this result is that an obscured clump is above the torus along our line-of-sight. However, we cannot exclude the possibility that the inclination angle smaller than the half-opening angle of the torus (cos($\theta\rm _{inc}$) $>$$c_{\rm f,tor}$) may correspond to a clumpy torus seen edge-on \citep[see, Fig.~3 and Section~2.3 in][]{Borus} and \citep[Fig.~6.1 in][]{Balokovic_17}. 
\item Unobscured AGN log(N$\rm _{H,l.o.s}$) $<$22 where the line-of-sight intercept the torus (cos($\theta\rm _{inc}$) $\le$$c_{\rm f,tor}$), i.e., NGC 3227. This result may suggest that the source is observed through an underdense region of a clumpy obscured torus. 
\item Obscured AGNs with their line-of-sight intercepting the torus but the line-of-sight column density is significantly different from the torus average column density ($|\Delta \rm log(N_H)|\gtrsim$1), e.g., NGC 3227, NGC 4051, NGC 4151 and NGC 5506. This result shows that we are currently looking through either an overdense or underdense region in their non-uniform tori, and this could be changing with time \citep[see, e.g.,][]{Risaliti_2005}. 
\end{enumerate}

\begin{figure*} 
\begin{minipage}[b]{.5\textwidth}
\centering
\includegraphics[width=1\textwidth]{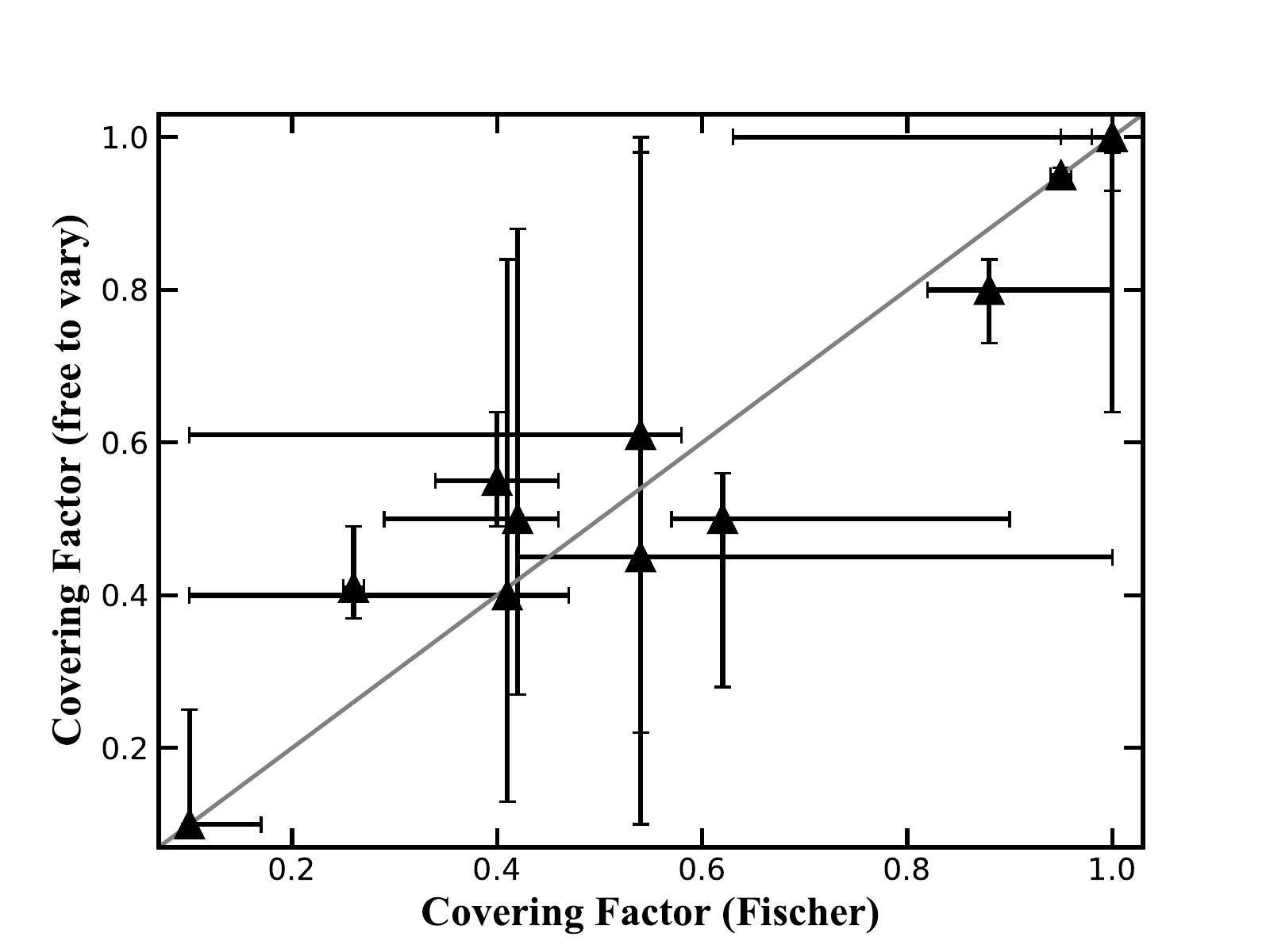}
\includegraphics[width=1\textwidth]{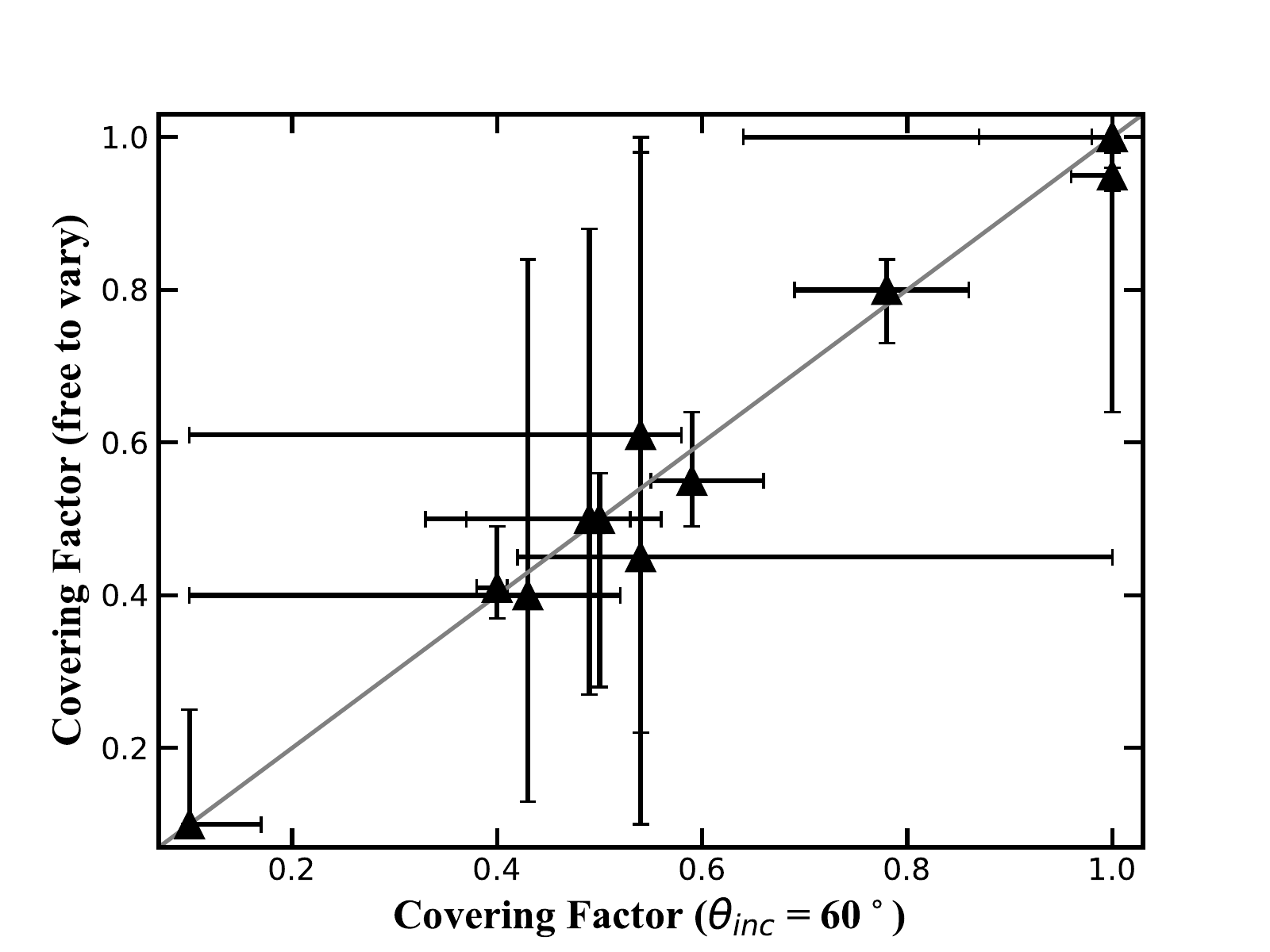}
\includegraphics[width=1\textwidth]{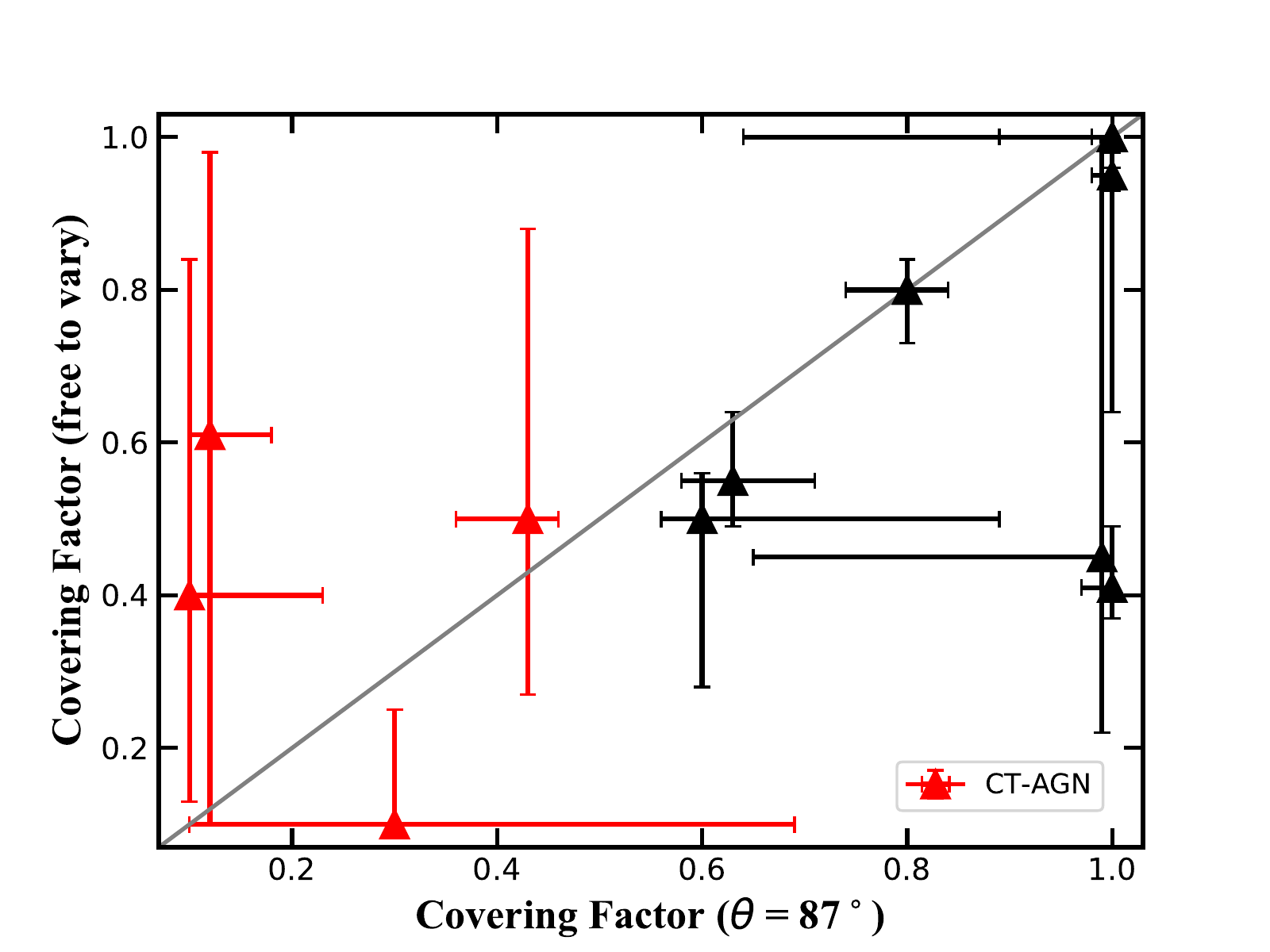}
\end{minipage}
\begin{minipage}[b]{.5\textwidth}
\centering
\includegraphics[width=1\textwidth]{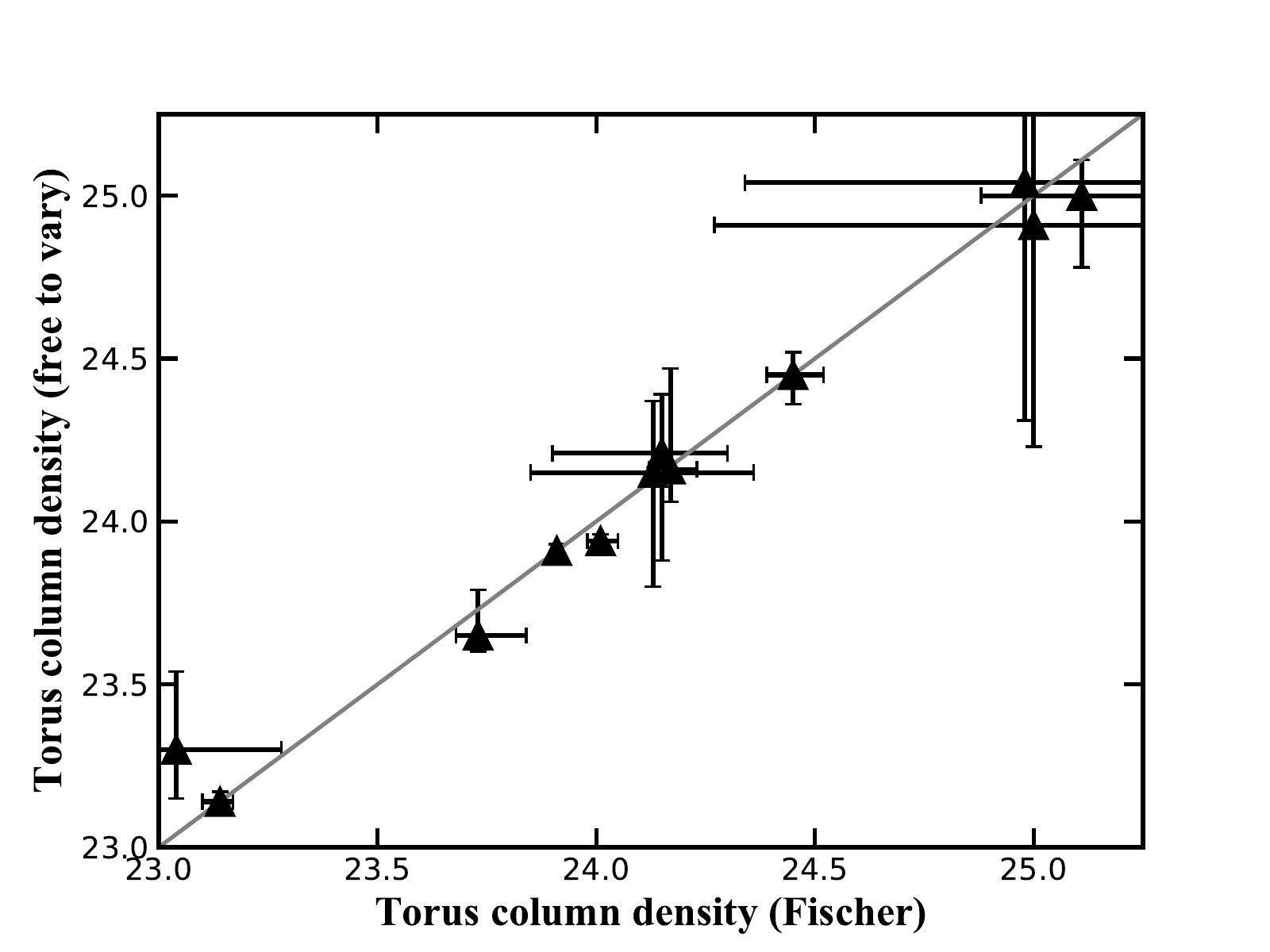}
\includegraphics[width=1\textwidth]{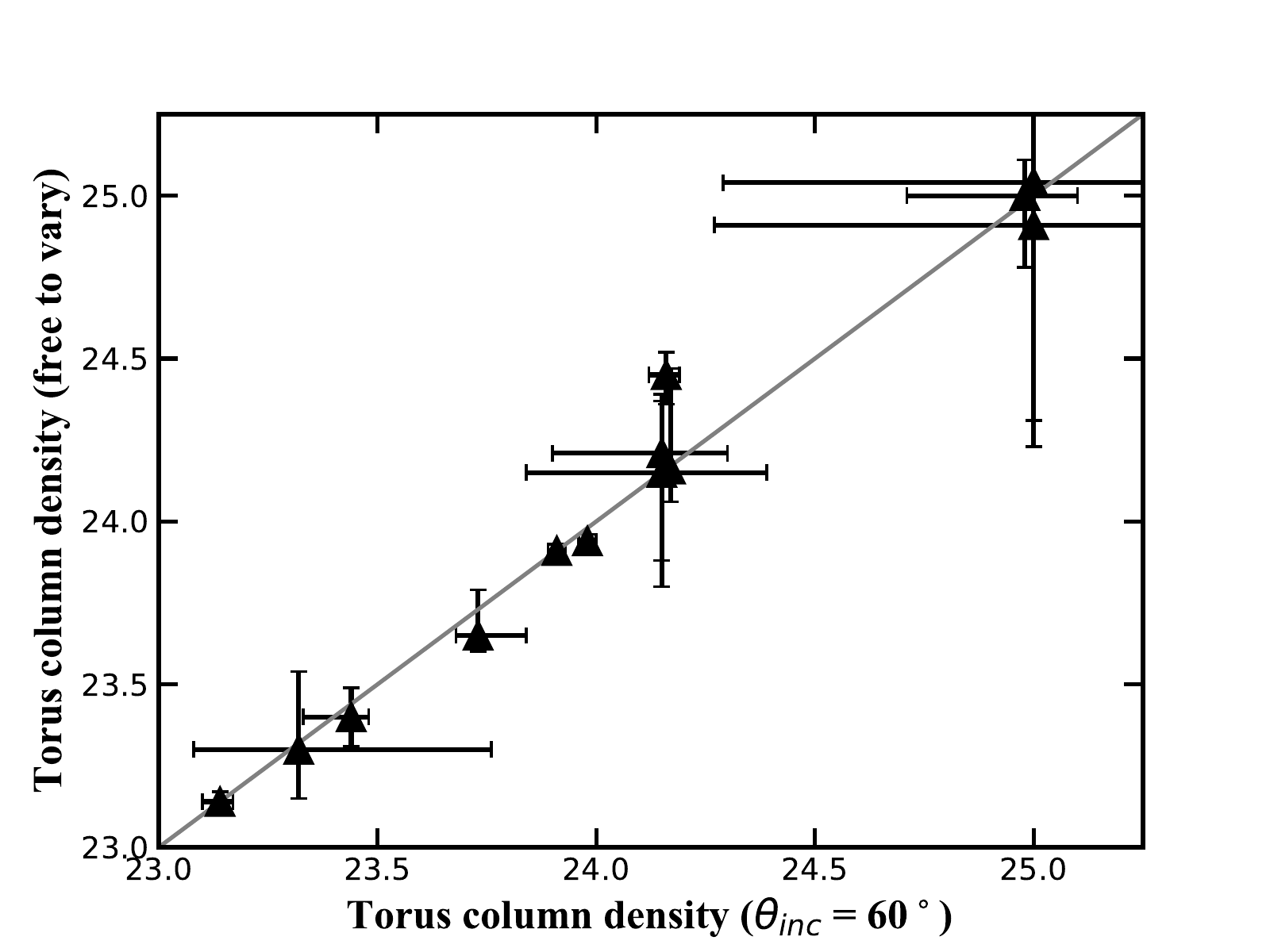}
\includegraphics[width=1\textwidth]{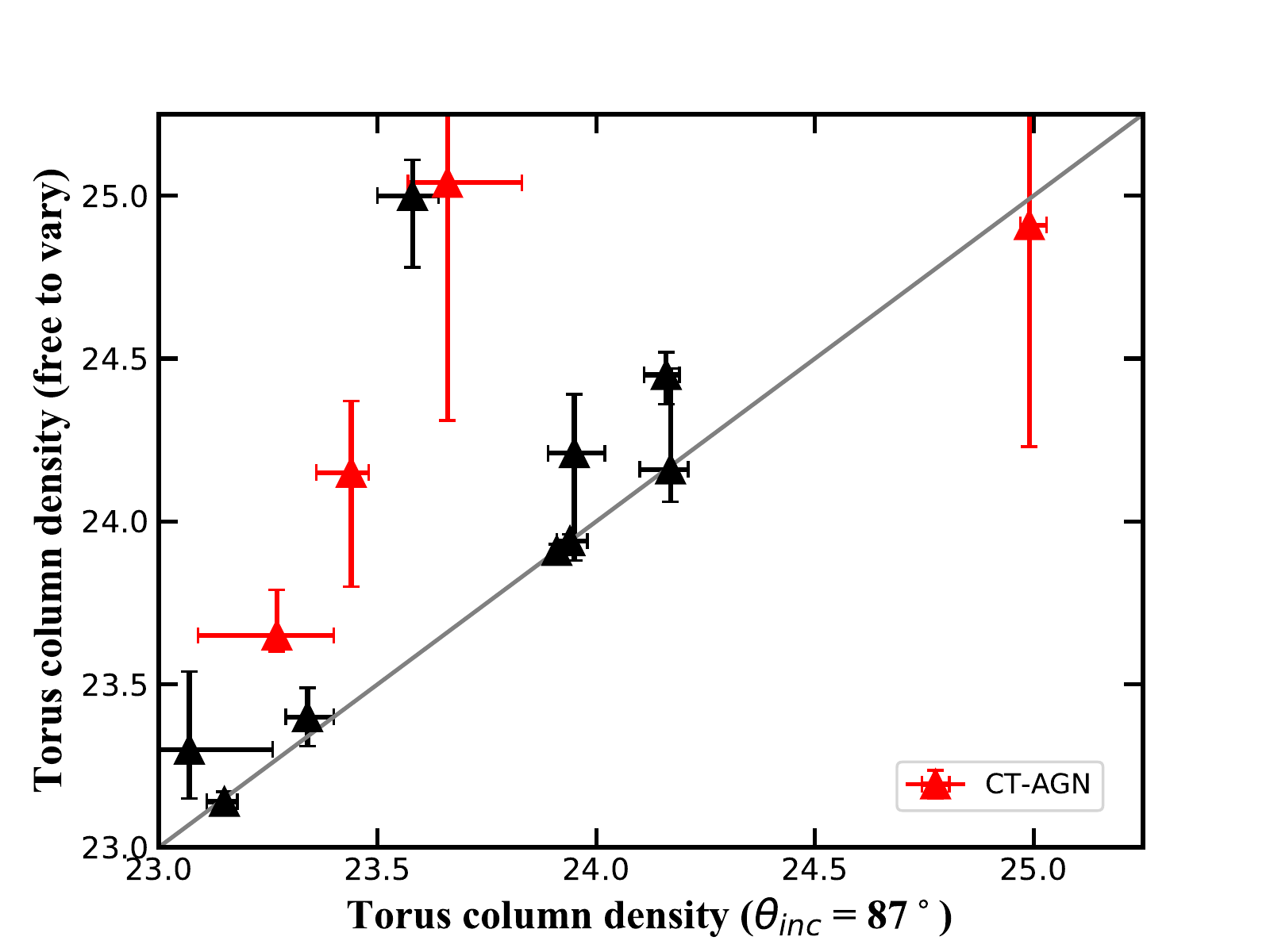}
\end{minipage}
\caption{The figures from the left to right in different rows are the $c_{\rm f,tor}$ and N$\rm_{H,tor}$ best-fit values obtained when $\rm \theta_{inc}$ is left free to vary with respect to those obtained when $\rm \theta_{inc}$ is fixed at [OIII] measured values (first row), when $\rm \theta_{inc}$ is fixed at $\rm \theta_{inc}$ = 60$^\circ$ (second row) and $\rm \theta_{inc}$ = 87$^\circ$ (third row), where gray solid line represents the 1:1 result. CT-AGNs are plotted in red in third row.}
\label{fig:compare}
\end{figure*}   

%
\section{Discussion}\label{discussion}
%
Thanks to the \borus\ model and the [OIII] measured inclination, we are able to properly study the role of inclination angle in analyzing the AGN spectra: the best-fit results of 13 sources when fitted with different scenarios of applying the inclination angle are reported in Section~\ref{sec:spectral}. In the rest of the work, we discuss how varying the inclination angle affects the measurement of the other spectral parameters (Section~\ref{sec:comp}, Section~\ref{sec:validity_60} and Section~\ref{sec:validity_87}), the correlations between the covering factor of the obscuring torus and other AGN properties (Section~\ref{sec:cf}), and study the geometrical properties of the AGNs in both X-ray and optical (Section~\ref{sec:X-ray_Optical}).

%
\subsection{Fixing the Torus Inclination Angle at $\theta_{\rm inc,[OIII]}$} 
\label{sec:comp}
%
We compare the best-fit results of different spectral parameters (i.e., $\chi^2$/d.o.f., $\Gamma$, N$_{\rm H,l.o.s}$, $\rm N_{H,tor}$, $c_{\rm f,tor}$ and $\theta_{\rm inc}$), computed either leaving the inclination angle free to vary or using the [OIII] measured values, $\theta_{\rm inc,[OIII]}$, reported in \citet{Fischer_2013}. The best-fit results obtained in two scenarios are reported in Table~\ref{Table:results} and Table~\ref{Table:results2}.  We compare the best-fit $c_{\rm f,tor}$ and $\rm N_{H,tor}$ computed in two methods in Fig.~\ref{fig:compare}. The above comparison is also plotted as an histogram in Figure~\ref{fig:bar} for better readability. 

\begin{itemize}
\item The inclination angles of 3 sources in our sample (Mrk~34, Mrk~573 and Mrk~1066) are fully unconstrained when inclination angle is left free to vary in fitting the spectra due to the poor quality (d.o.f $\le$200) of the data. 

The best-fit inclination angles measured in X-ray do not always match the inclination angles measured in optical using [OIII], e.g., we found 6 sources of which the differences between the two inclination angles is $\Delta\theta_{\rm inc}$ $>$20$^\circ$.

\item In spite of large $\Delta\theta_{\rm inc}$ found in some sources, the best-fit results of the other key parameters, e.g., $\rm N_{H,l.o.s}$, $\rm N_{H,tor}$ and $c_{\rm f,tor}$ are in good agreement with each other within the uncertainties.

\item The goodness of the spectral fits shows no improvement or only a marginal improvement when inclination angle is left free to vary in fitting the spectra. The sources with most improved fit statitics in our sample is NGC~3783, which improves from $\chi^2$/d.o.f = 3367/2930 when fixing the inclination angle at the [OIII] measured inclination to $\chi^2$/d.o.f = 3349/2929 when letting the inclination free to vary.

\item A minor improvement on constraining the spectral parameters are found when fixing the $\theta_{\rm inc}$ at [OIII] measured value. The average uncertainties on each parameters are reported in Table~\ref{Table:constraints}.
\end{itemize}

The above results suggest that: 1. the inclination angle measured in the optical band using [OIII] can be used in the X-ray spectral analysis of AGNs, since it provides similar best-fit results to those obtained when inclination angle is left free to vary when fitting the spectra considering uncertainties; 2. in some sources, significant different inclination angles measured in the optical compared to those derived from the X-ray spectra are found, but the other fitted parameters, as well as the best-fit statistic, are only marginally affected by this variation, suggesting that we do not have enough power to constrain the inclination angle of these sources even with high-quality broadband X-ray data. 

\begingroup
\renewcommand*{\arraystretch}{1.5}
\begin{table}
\scriptsize
\centering
\caption{Average uncertainties of different parameters assuming different torus inclination angles}
\label{Table:constraints}
\vspace{.1cm}
  \begin{tabular}{cccccc}
       \hline
       \hline     
         Paramter&Free to vary&[OIII]&60$^\circ$&87$^\circ$\\
       \hline
       $\Gamma$&10\%&9\%&10\%&6\%\\
       \hline
       N$\rm_{H,l.o.s}$&19\%&18\%&18\%&10\%\\
       \hline
       N$\rm_{H,tor}$&38\%&36\%&37\%&17\%\\
       \hline
       $c_{\rm f,tor}$\footnote{Absolute average uncertainty}&0.17&0.11&0.11&0.10\\
       \hline
       $\chi^2_\nu$&1.04&1.04&1.04&1.05\\
       \hline
	\vspace{0.03cm}
\end{tabular}

\end{table}
\endgroup

%
\subsection{Fixing the Torus Inclination Angle at $\theta_{\rm inc}$ = 60$^\circ$} 
\label{sec:validity_60}
%
\begin{figure*} 
\begin{minipage}[b]{.5\textwidth}
\centering
\includegraphics[width=1\textwidth]{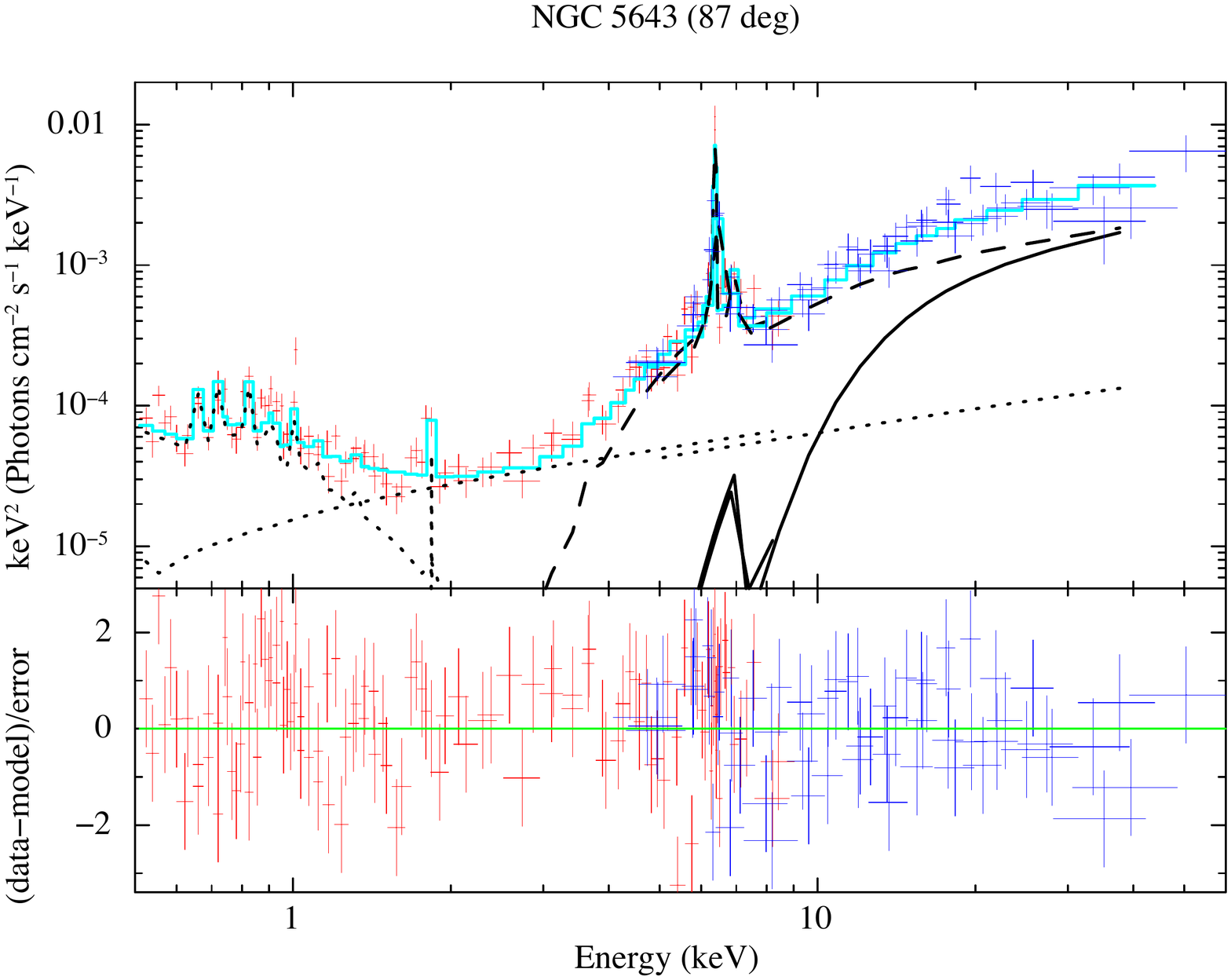}
\includegraphics[width=1\textwidth]{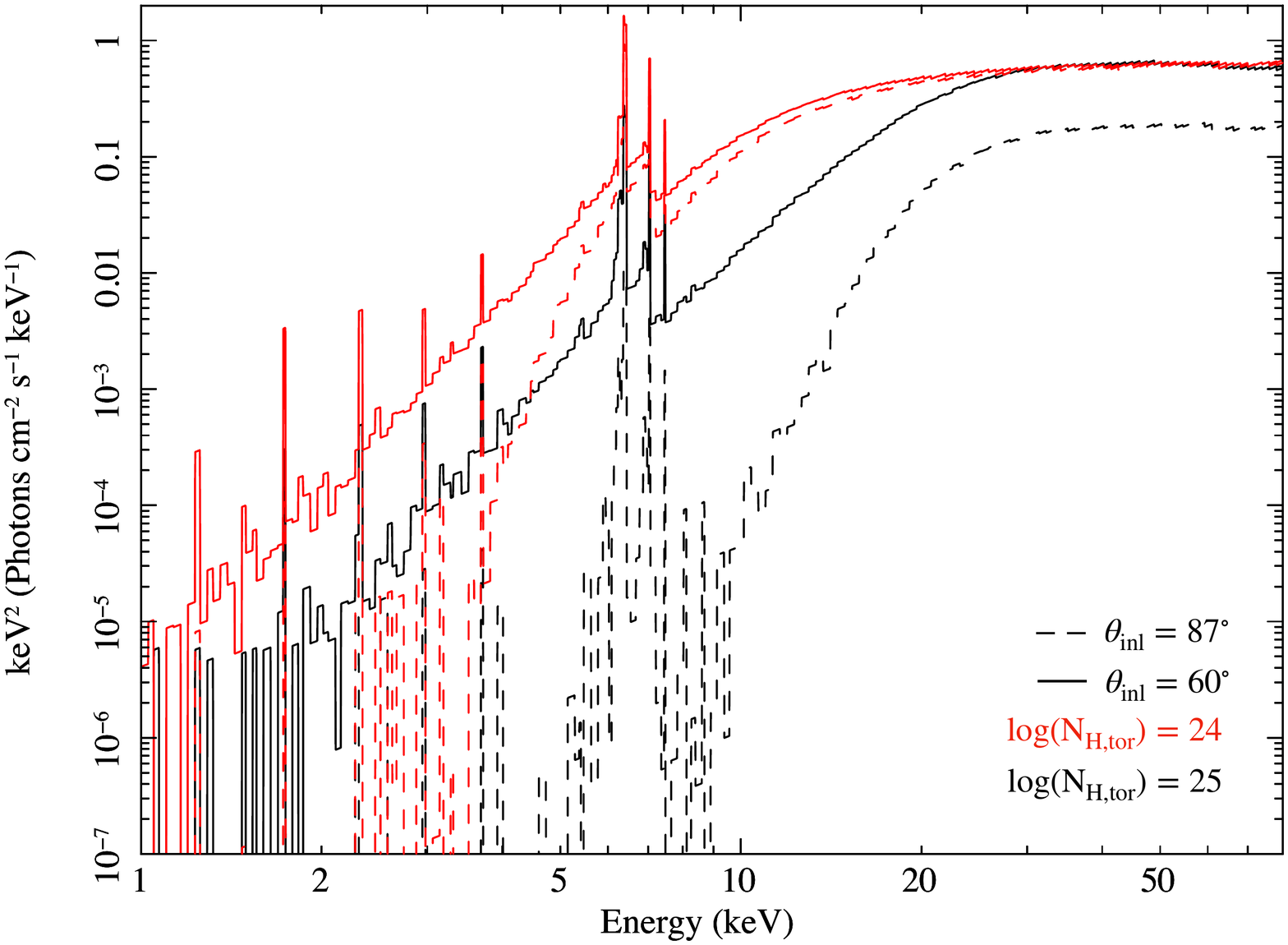}
\end{minipage}
\begin{minipage}[b]{.5\textwidth}
\centering
\includegraphics[width=1\textwidth]{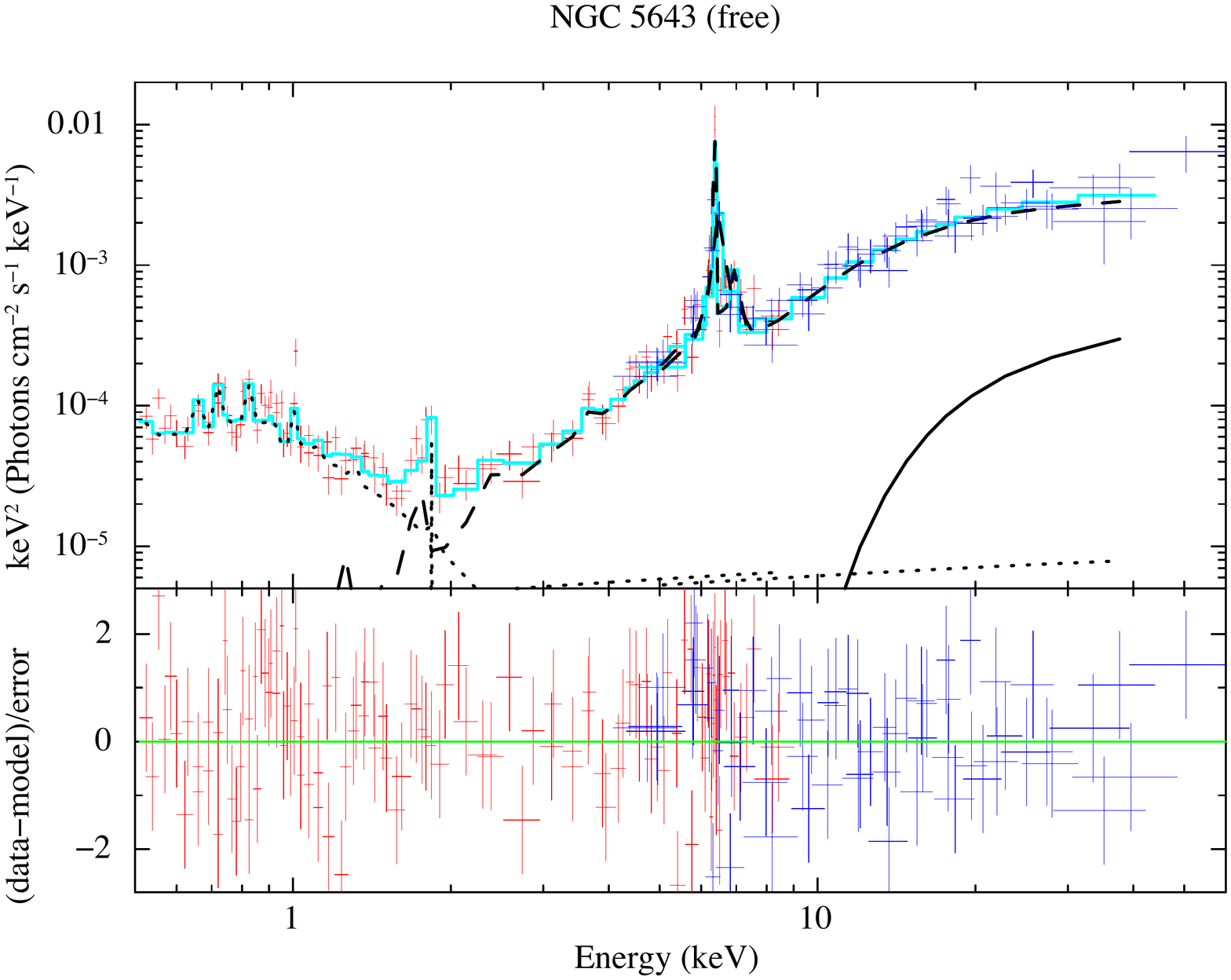}
\includegraphics[width=1\textwidth]{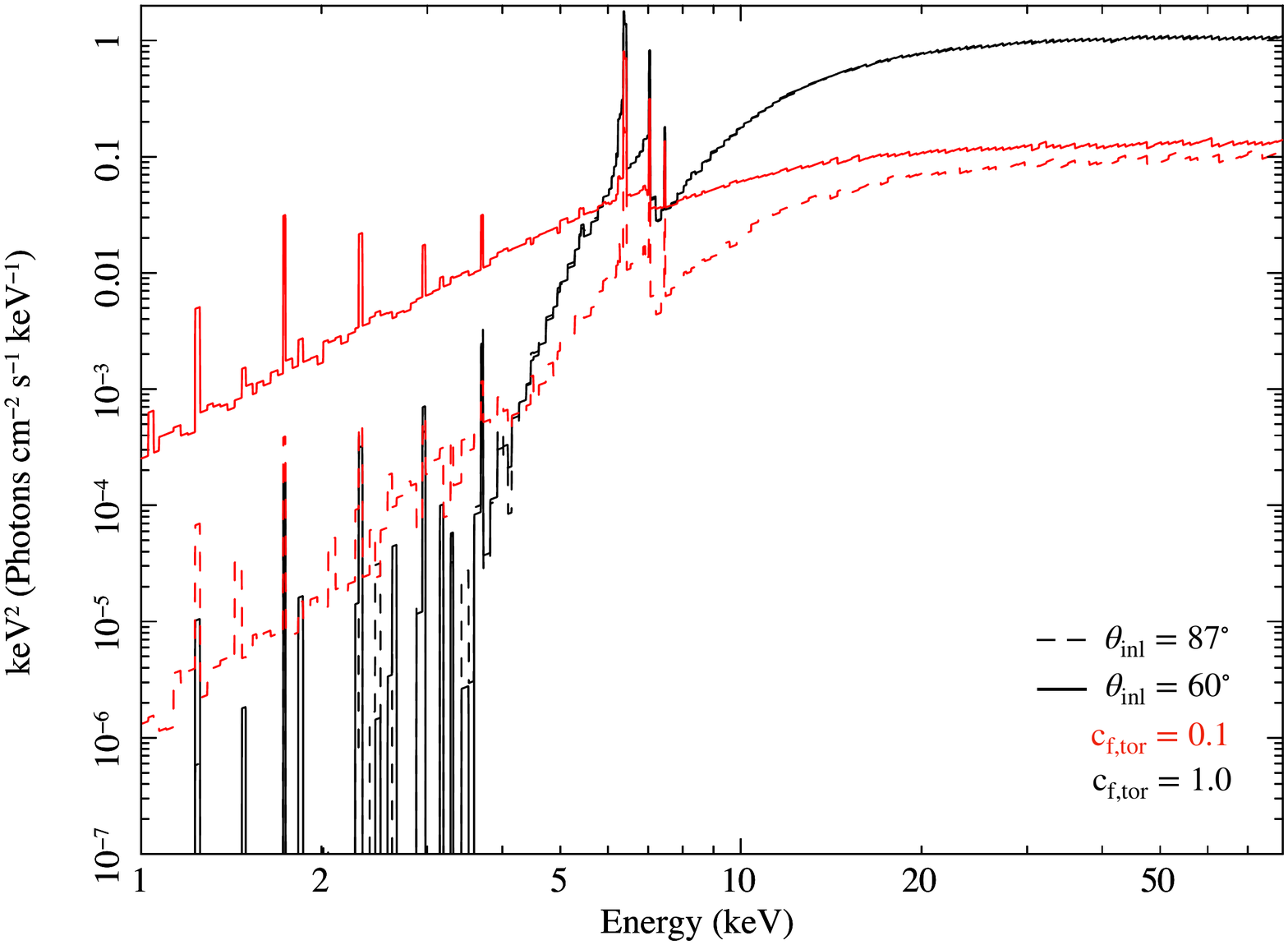}
\end{minipage}
\caption{ The figure is an illustrative example of how fixing the inclination angle at $\theta_{\rm inc}$ = 87$^\circ$ gives very different best-fit results compared to when the inclination angle is left free to vary in fitting the spectra and when the inclination angle at $\theta_{\rm inc}$ = 60$^\circ$. Top: unfolded spectra of NGC~5643 with inclination angle being fixed at $\theta_{\rm inc}$ = 87$^\circ$ and being left free to vary. The \NuSTAR\ data are plotted in blue and the \XMM\ data are plotted in red. The best-fit model prediction is plotted as cyan solid lines. The single components of the model are plotted in black with different line styles, i.e., the absorbed intrinsic continuum with solid lines, the reflection component with dashed lines, the scattered component, the {\tt mekal} component and emission lines with dotted lines.
Bottom: \borus\ model predictions of the reprocessed component when inclination angle are fixed at $\rm \theta_{inc}$ = 87$^\circ$ (dashed line) and $\rm \theta_{inc}$ = 60$^\circ$ (solid line). Left: the model predictions for log($\rm N_{H,tor}$) = 24 (red) and log($\rm N_{H,tor}$) = 25 (black): a photon index with $\Gamma$ = 1.8 and torus covering factor with $c_{f,\rm Tor}$ = 0.6 are assumed. Right: the model predictions for $c_{f,\rm Tor}$ = 0.1 (red) and  $c_{f,\rm Tor}$ = 1.0 (black): a photon index with $\Gamma$ = 1.8 and torus column density with log($\rm N_{H,tor}$) = 24 are assumed.} 
\label{fig:NHS_model}
\end{figure*}  

Following the method in Section~\ref{sec:comp}, we compare the best-fit results obtained when leaving the inclination angle free to vary with the best-fit results computed when fixing the inclination at some specific angles, e.g., $\theta_{\rm inc}$ = 60$^\circ$. The best-fit results when the sources are fitted with $\theta_{\rm inc}$ = 60$^\circ$ is reported in Table~\ref{Table:results} and Table~\ref{Table:results2}. The comparison of $c_{\rm f,tor}$ and $\rm N_{H,tor}$ between the two scenarios are plotted in Figure~\ref{fig:compare}. The above comparison is also plotted as a histogram in Figure~\ref{fig:bar} for better readability.

We find that the differences of the best-fit results of different parameters between when $\theta_{\rm inc}$ = 60$^\circ$ and when leaving $\theta_{\rm inc}$ free to vary are marginal, e.g., the average differences are $\sim$1\% for $\Gamma$, $\sim$2\% for N$\rm_{H,l.o.s}$, $\sim$5\% for N$\rm_{H,tor}$ and 2\% for $c_{\rm f,tor}$. We measure a similar average goodness of fitting in the two scenarios as reported in Table~\ref{Table:constraints}. Only marginal improvements are found in the fits of most sources when inclination angle is left free to vary in fitting the spectra than the inclination angle is fixed at $\theta_{\rm inc}$ = 60$^\circ$, except for NGC~4051, whose fit improves from $\chi^2$/d.o.f = 2739/2391 to $\chi^2$/d.o.f = 2686/2390 after letting the inclination free to vary. Fixing the inclination angle at $\rm \theta_{inc}$ = 60$^\circ$ provides similar constraints on fitting which are consistent with those obtained when fixing $\rm \theta_{inc}$ at [OIII] measured values. 

Although, according to our spectral fits, fixing the inclination angle at $\theta_{\rm inc}$ = 60$^\circ$ provides similar goodness of fits and similar other key properties to those when $\rm \theta_{inc}$ is left free to vary, this results may be biased by the sample that we have selected, since 9 out of 13 sources in our sample have a $\theta_{\rm inc}$ $\sim$60$^\circ$. Indeed, we find that fixing the inclination angle at $\theta_{\rm inc}$ = 60$^\circ$ does not reproduce the N$_{\rm H,tor}$ and other key parameters measured when $\rm \theta_{inc}$ is left free to vary in some heavily obscured sources out of our sample, e.g., we reanalyze the spectra of a Compton thick (CT-) AGN, NGC~1358, and the best-fit N$\rm _{H,tor}$ and $c_{f,tor}$ obtained when fixing the inclination angle at $\rm \theta_{inc}$ = 60$^\circ$ is about 5 times lager than those obtained when the inclination is left free to vary as reported in \citet{Zhao_2019_1}, which measured a best-fit inclination angle of $\rm \theta_{inc,NGC1358}$ $\approx$87$^\circ$.

%
\subsection{Fixing the Torus Inclination Angle at $\theta_{\rm inc}$ = 87$^\circ$} 
\label{sec:validity_87}
%
We also fit the spectra when fixing the inclination angle at $\theta_{\rm inc}$ = 87$^\circ$. The differences of the best-fit results of $\Gamma$ and N$\rm_{H,l.o.s}$ between obtained when $\theta_{\rm inc}$ = 87$^\circ$ and when leaving $\theta_{\rm inc}$ free to vary are marginal, e.g., the average differences are $\sim$2\% for $\Gamma$, $\sim$9\% for N$\rm_{H,l.o.s}$. However, the measurement of N$\rm_{H,tor}$ when $\theta_{\rm inc}$ = 87$^\circ$, especially for some sources with CT torus (i.e., log($\rm N_{H,tor}$)$\gg$24), are considerably different (the average difference is $\sim$30\%) from those obtained when $\theta_{\rm inc}$ is left free to vary. The discrepancy of $c_{\rm f,tor}$ between the two cases is large as well, e.g., the average difference is 22\%. However, unlike the $\rm N_{H,tor}$ case where the measurements of higher $\rm N_{H,tor}$ tend to a lower value, such trend is not found in $c_{\rm f,tor}$. Notably, $\sim$30\% of the sources in our sample are measured with best-fit photon indices stuck at $\Gamma$ $\sim$1.4, which is the lower limit of the parameter in \borus\ model. Such a result can be explained by the fact that a flatter $\Gamma$ is needed to compensate the change of the spectral shape caused by the unrealistic measurement of $c_{\rm f,tor}$ and $\rm N_{H,tor}$. It is worth noting that fixing the inclination angle at $\theta_{\rm inc}$ = 87$^\circ$ leads to significant different best-fit torus covering factor and torus column density than when the inclination angle is left free to vary even when fitting the CT-AGNs, which are plot in red in Fig.~\ref{fig:compare}.

To illustrate the torus column density bias mentioned above, we take NGC~5643 as an example. The source is a CT-AGN and the spectrum of the source is dominated by reprocessed component. We plot the spectra and different components of the model predictions of NGC~5643 with inclination angle being fixed at $\theta_{\rm inc}$ = 87$^\circ$ and being left free to vary in Fig.~\ref{fig:NHS_model}. The best-fit inclination angle measured for NGC~5643 is $\theta_{\rm inc}$ $\sim$59$^\circ$. When the inclination is left free to vary, the spectra above 2\,keV are dominated by the reprocessed component and the measured best-fit torus column density is log($\rm N_{H,tor,free}$) $\sim$24.15. However, when the inclination angle is fixed at $\theta_{\rm inc}$ = 87$^\circ$, the best-fit torus column density is measured as log($\rm N_{H,tor,free}$) $\sim$23.44. Such result can be understood by looking at the bottom left panel of Fig.~\ref{fig:NHS_model}, where we plot the \borus\ model prediction of the reprocessed component when the inclination angles are $\theta_{\rm inc}$ = 60$^\circ$ and $\theta_{\rm inc}$ = 87$^\circ$ and the torus column densities are log($\rm N_{H,tor,free}$) = 24 and log($\rm N_{H,tor,free}$) = 25. When the inclination moves from $\theta_{\rm inc}$ = 60$^\circ$ to $\theta_{\rm inc}$ = 87$^\circ$, the spectra are suppressed significantly and nonlinearly: the spectra with energy below $\sim$20\,keV are much affected. Therefore, the torus column density decreases to compensate this reduction, which better describes the energy between $\sim$5\,keV to $\sim$20\,keV. The discrepancy of the spectra in other energy bands thus needs other components to make up, e.g., the line-of-sight component contributes more to the spectra at energy $>$20\,keV by decreasing the line-of-sight column density and the scattering component dominates the spectra at energy below $\sim$4\,keV by artificially increasing $f_{\rm s}$.

The discrepancy of $c_{\rm f,tor}$, however, is more complex due to the fact that the reprocessed component is energy dependent with respect to $c_{\rm f,tor}$, as plotted in the bottom right panel of Figure~\ref{fig:NHS_model}, which plots the spectra of reprocessed component with different combinations of $\theta_{\rm inc}$ and $c_{\rm f,tor}$. For more information about the reprocessed component of the \borus\ model, we plot the \borus\ model predictions with varying different parameters in Appendix~A.

Fixing the inclination angle at $\rm \theta_{inc}$ = 87$^\circ$ puts the strongest constraint on the parameters, e.g., $\rm N_{H,l.o.s}$ and $\rm N_{H,tor}$, among other cases. Such a results is caused by the fact that by fixing the inclination angle at $\rm \theta_{inc}$ = 87$^\circ$, the change in other parameters will lead to large variation on the spectrum: for example, in figure~\ref{fig:NHS_model}, we find that the spectral shape variation is much larger for $\rm \theta_{inc}$ = 87$^\circ$ than for $\rm \theta_{inc}$ = 60$^\circ$ when $\rm N_{H,tor}$ varies, thus, the uncertainty of $\rm N_{H,tor}$ is much less when using $\rm \theta_{inc}$ = 87$^\circ$ than using $\rm \theta_{inc}$ = 60$^\circ$. However, the $c_{\rm f,tor}$-related spectral variation is energy dependent in both $\rm \theta_{inc}$ = 87$^\circ$ and $\rm \theta_{inc}$ = 60$^\circ$ cases, therefore the average uncertainties of $c_{\rm f,tor}$ are similar in the two cases.

\subsection{Distribution of the obscuring material}
\label{sec:cf}
\begin{figure*} 
\centering
\includegraphics[width=\textwidth]{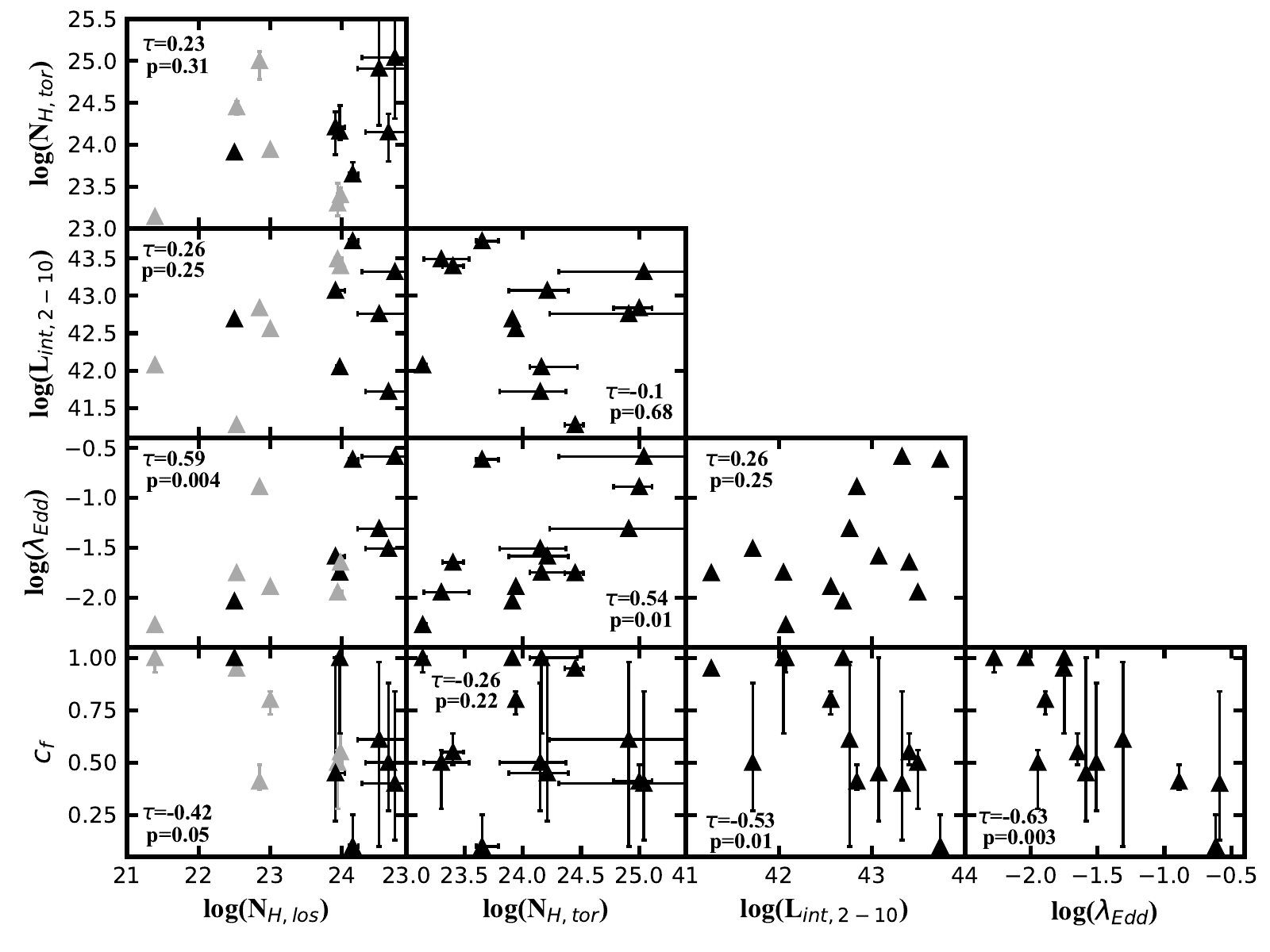}
\caption{Best-fit results of different parameters, i.e., line-of-sight column density, N$_{\rm H,los}$, torus column density, N$_{\rm H,tor}$, torus covering factor, $c_{\rm f}$, 2--10\,keV intrinsic luminosity, L$_{\rm int,2-10}$ and Eddington ratio, $\lambda_{\rm Edd}$, as a function of each other. Sources with known line-of-sight column density variability are plotted in gray. The tau values and $p$ values in the Kendall’s tau test are computed for each pair of parameters and are also labeled in each sub-plot.}
\label{fig:correlation}
\end{figure*}   


In Section~\ref{sec:comp}, Section~\ref{sec:validity_60} and Section~\ref{sec:validity_87}, we discussed how varying the inclination angle affects the measurement of the other spectral parameters of AGNs. Thanks to the flexible and powerful \borus\ model, we could explore a larger parameter space in modeling the spectrum of AGN, e.g., we could now directly measure the average column density and the covering factor of the obscuring torus in AGN. Therefore, in this section, we focus on the correlations among different physical and geometrical properties of the sources in our sample.

A corner plot is drawn in Fig.~\ref{fig:correlation} to explore the correlation among different parameters, i.e., line-of-sight column density, log(N$_{\rm H,l.o.s}$), torus column density, log(N$_{\rm H,tor}$), torus covering factor, $c_{\rm f}$, 2--10\,keV intrinsic luminosity, L$_{\rm int,2-10}$, and Eddington ratio\footnote{The Eddington ratio is calculated by $\lambda_{\rm Edd}$ $\equiv \rm L_{bol}/L_{Edd}$, where $\rm L_{bol}$ is the bolometric luminosity. Bolometric luminosity is calculated by $\rm L_{bol}$ = $\kappa\,\rm L_{int,2-10}$, assuming a bolometric correction $\kappa$ = 20 \citep{Vasudevan10}. Eddington luminosity is calculated as $L_{\rm Edd}$ = 4$\pi$GM$_{\rm BH}$m$_p$c/$\sigma_T$, where M$_{\rm BH}$ is the mass of SMBH and m$_p$ is the mass of proton.}, $\lambda_{\rm Edd}$. Kendall’s tau tests are performed for each pair of parameters and are labeled in each subplot. The best-fit values used are obtained when the inclination angle is left free to vary in fitting the spectra. We also plot the inclination angle as functions of other properties in Appendix~B to explore the correlation between the inclination angle and other properties of the sources.

\begin{figure*} 
\begin{minipage}[b]{.5\textwidth}
\centering
\includegraphics[width=1\textwidth]{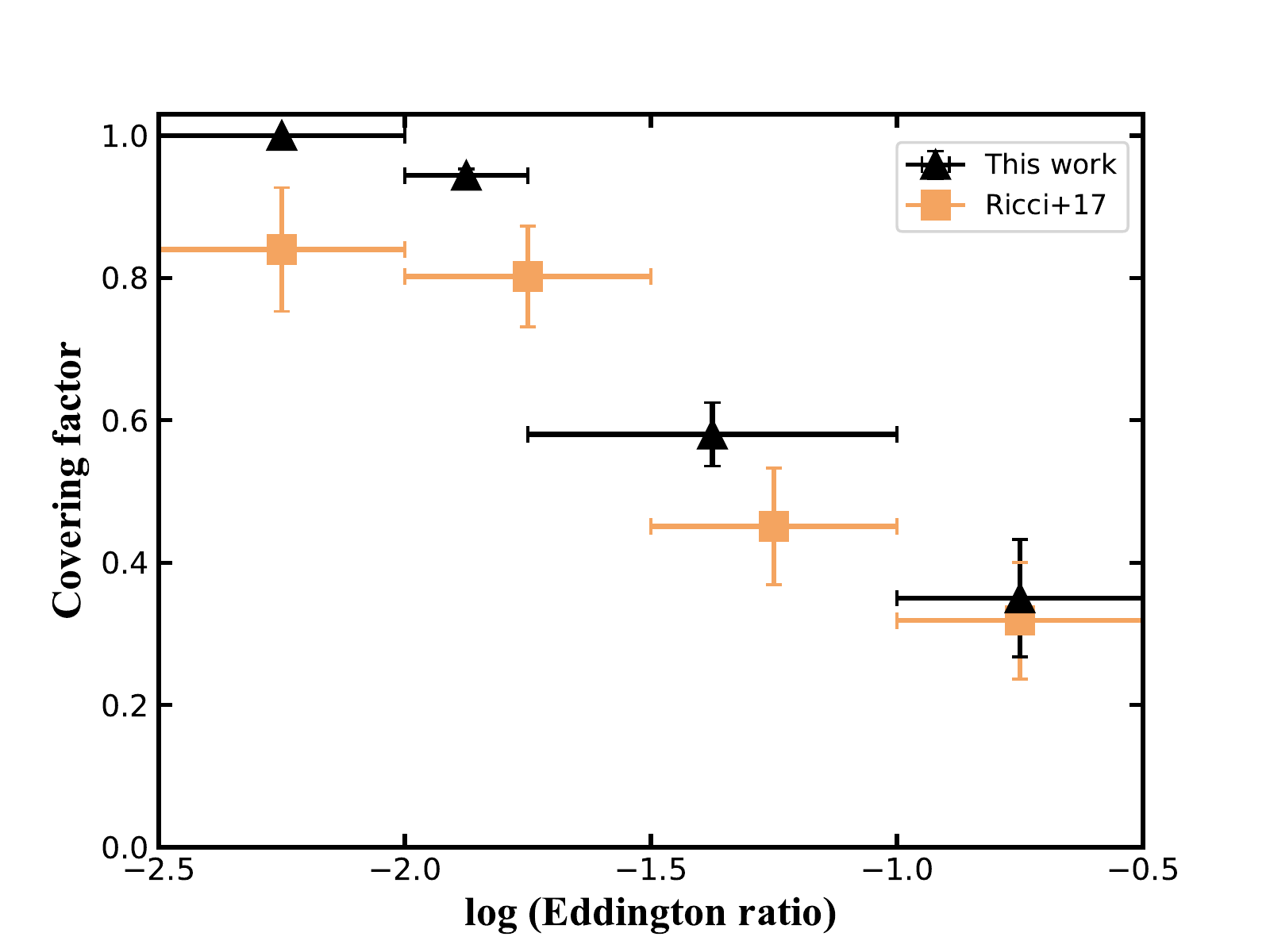}
\end{minipage}
\begin{minipage}[b]{.5\textwidth}
\centering
\includegraphics[width=1\textwidth]{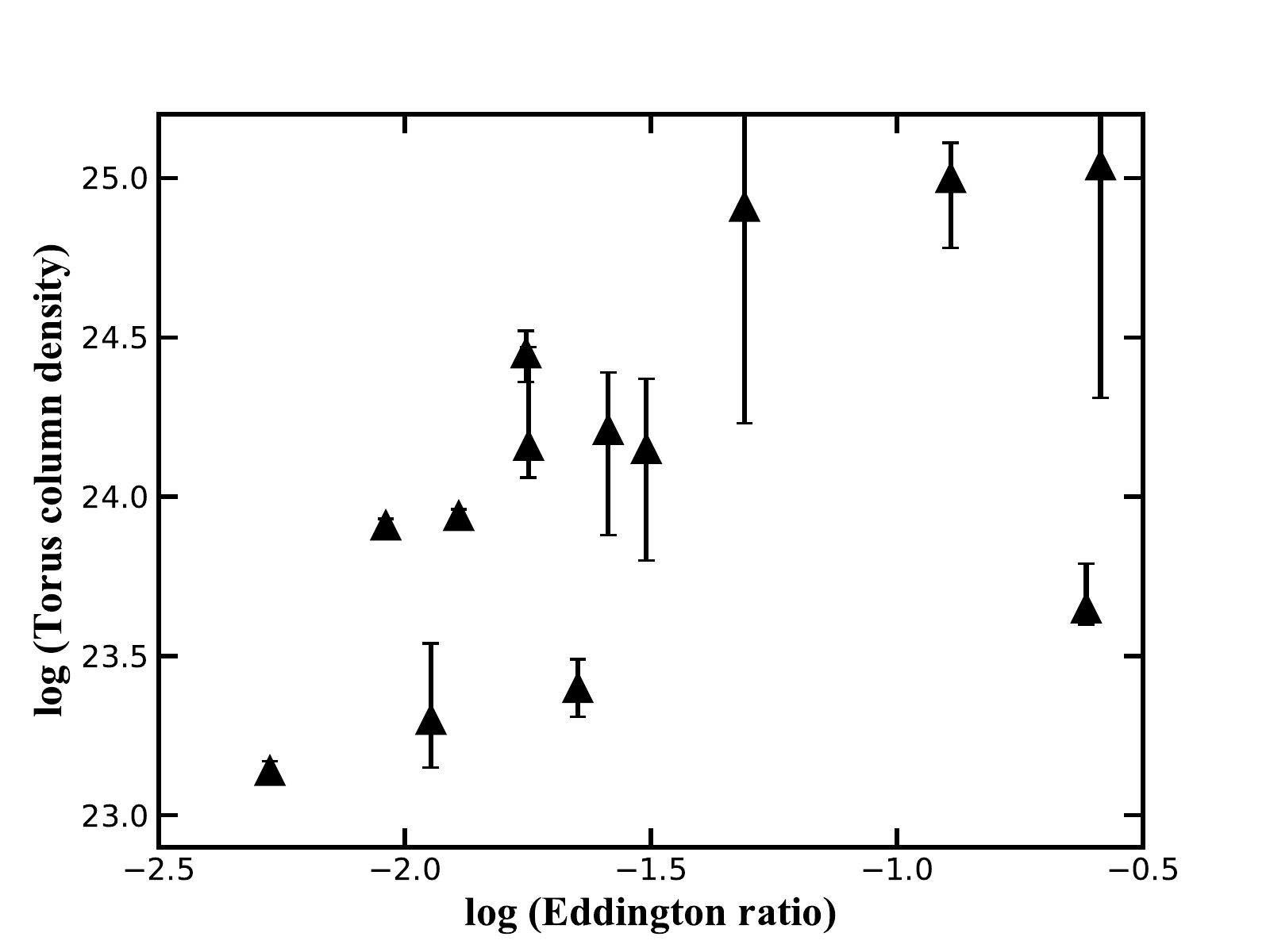}
\end{minipage}
\caption{Left: torus covering factors as a function of the Eddington ratio of the 13 sources, which are rebinned to compare with the torus covering factor and Eddington ratio relationship measured in \citet{Ricci:2017aa} (orange squares). The results are rebinned to make sure that each bin has similar number of sources. Right: torus column densities as a function of Eddington ratio of the 13 sources.}
\label{fig:edd}
\end{figure*}

\begin{itemize}
\item We find no correlation between the measured AGN inclination angle and the other physical and geometrical properties of the AGN as show in Fig.~\ref{fig:inc}. Such a result is reasonable since the sources are randomly observed and the properties of the sources should not be related to the angle at which they are observed.

\item We find no correlations between intrinsic luminosity and torus column density ($p$ = 0.68), torus column density and line-of-sight column density ($p$ = 0.31), intrinsic luminosity and line-of-sight column density ($p$ = 0.25), Eddington ratio and 2--10\,keV intrinsic luminosity ($p$ = 0.25) and torus covering factor and torus column density ($p$ = 0.22).

\item We find a correlation with confidence level $\sim2.9\,\sigma$ between the line-of-sight column density and Eddington ratio ($p$ = 0.004), i.e., as the line-of-sight column density increases, the Eddington ratio also increases. We also find an inverse correlation with confidence level $\sim2\,\sigma$ between the line-of-sight column density and torus covering factor ($p$ = 0.05), i.e., as the line-of-sight column density increases, the torus covering factor decreases. However, such trends are less evident if we exclude the sources which have been observed to be variable due to the shift of line-of-sight column density, which are marked as gray in Figure~\ref{fig:correlation} from our analysis: the $p$ values become $p$ = 0.07 for line-of-sight column density and Eddington ratio and $p$ = 0.22 for line-of-sight column density and torus covering factor. Therefore, we are not able confirm the correlation found between line-of-sight column density and Eddington ratio and the inverse correlation between the line-of-sight column density and torus covering factor.

\item We find a correlation with confidence level $\sim2.6\,\sigma$ between the 2--10\,keV intrinsic luminosity and the torus covering factor ($p$ = 0.01), i.e., as the intrinsic luminosity increases, the torus covering factor decreases. Such a trend has been reported in many previous works with larger samples and higher statistical accuracy in different redshift range \citep[e.g.,][]{Lawrence_1982,Hasinger08,Ueda14}. The covering factor of the torus in these works is derived from the X-ray hardness ratio or the fraction of the obscured Compton thin sources (22$\le$log(N$_{\rm H}$)$\le$24) in these works. Similar trend has also been found by \citet{Balokovic_17}, which measured the individual torus covering factors and their intrinsic luminosities as in this work. 

\item We find an inverse correlation with confidence level $\sim3\,\sigma$ between the Eddington ratio and torus covering factor ($p$ = 0.003), i.e., as the Eddington ratio of the AGN increases, the covering factor of the obscuring torus decreases. We also find a correlation with confidence level $\sim2.6\,\sigma$ between the Eddington ratio and torus column density ($p$ = 0.01), i.e., as the Eddington ratio of the AGN increases, the average column density of the obscuring torus decreases. Such results are in good agreement with those reported in \citet{Ricci:2017aa}, who found that the torus covering factor and the torus average column density strongly depends on the Eddington ratio of the AGN using a larger BAT selected sample of 392 AGNs. The dependence of torus covering factor and torus average column density on the Eddington ratio can be explained assuming that the distribution of the circumnuclear material around the SMBH is mainly regulated by the radiative feedback: as the accretion rate increases, the radiation pressure from the accretion disk blows the less dense (log(N$_{\rm H}$) $\le$24) materials away and leaves only the CT materials, thus decreasing the torus covering factor and increasing the torus average column density \citep{Fabian_2006,Fabian_2009,Ricci:2017aa}. We point out that in \citet{Ricci:2017aa}, the covering factors are indirectly measured, using are the fraction of obscured (22 $\le$log(N$_{\rm H}$)) AGNs with respect to all AGNs with 20 $\le$log(N$_{\rm H}$) in their sample. To better visualize the above correlations, we display the best-fit torus covering factors as a function of their measured Eddington ratio and the best-fit torus column density as a function of their measured Eddington ratio separately in Fig.~\ref{fig:edd}. To compare with the results obtained in \citet{Ricci:2017aa}, we rebin our results in plotting the torus covering factors as a function of Eddington ratio. We find that our average torus covering factor are in good agreement with what \citet{Ricci:2017aa} obtain, especially at large Eddington ratio.  
\end{itemize}

The fact that no correlation is found between 2--10\,keV intrinsic luminosity and torus average column density, together with the correlations found between torus covering factor and torus average column density with respect to the Eddington ratio, suggests that the distribution of the obscuring materials surrounding the SMBH of the AGNs in our sample is mainly regulated by the Eddington ratio rather than the intrinsic luminosity, which is in agreement with what is found in \citet{Ricci:2017aa}. Nevertheless, the exploration of the distribution of the materials in the obscuring torus in AGN needs to be further studied in a larger unbiased sample with high-quality spectra.

\subsection{Geometrical properties of torus and NLR}
\label{sec:X-ray_Optical}
According to the unified model, the dusty torus obscures the radiation from the center engine of the AGN, and is therefore thought to form the biconical shape of the NLR or the ionization cone \citep{Malkan_1998}. \citet{Fischer_2013} report the opening angle of the outer edge of the NLR obtained by modeling the kinematics of the sources' NLR observed with the \textit{Hubble Space Telescope} (HST) and the Space Telescope Imaging Spectrograph (STIS). In this section, we explore the relation between the geometrical properties of the AGNs in our sample measured in optical and measured in X-ray. 

\begin{figure} 
\centering
\includegraphics[width=.5\textwidth]{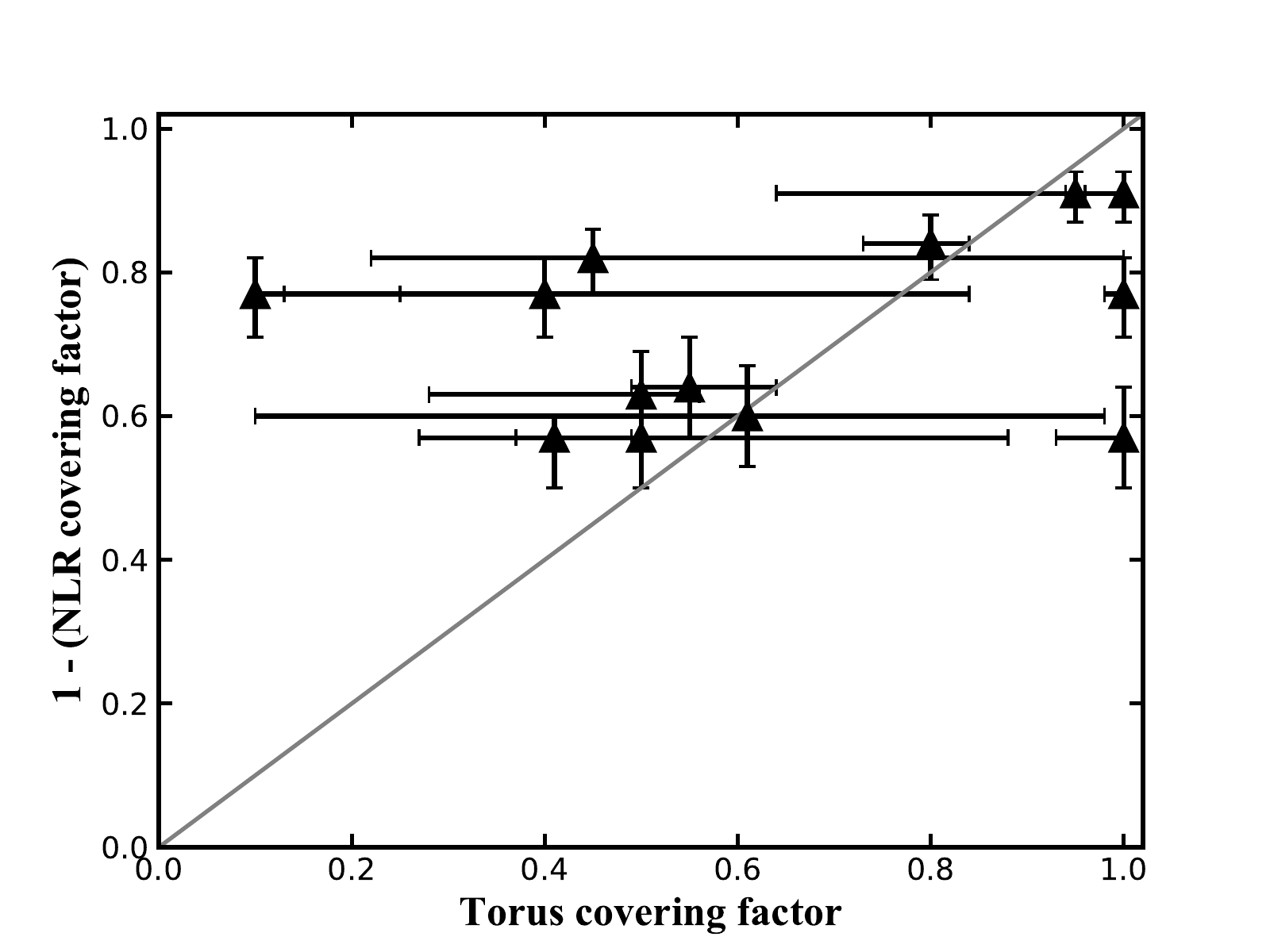}
\caption{Relationship between the covering factor of the region excluding the NLR, i.e., $1-c_{f,\rm NLR}$, with respect to the covering factor of the torus, i.e., $c_{f,\rm Tor}$. The gray solid line represents the 1:1 result.}
\label{fig:cf_NLR}
\end{figure}

In Figure~\ref{fig:cf_NLR}, we plot the covering factor of the region excluding the NLR measured in optical, i.e., 1--$c_{f,\rm NLR}$, as a function of the covering factor of the torus measured in X-ray, i.e., $c_{f,\rm Tor}$. We find that: 1. there is no correlation between $c_{f,\rm tor}$ and 1-$c_{f,\rm NLR}$ ($\tau$ = 0.18 and $p$ = 0.42); 2. while $c_{f,\rm Tor}$ span all values from 0.1 to 1, $c_{f,\rm NLR}$ does not (i.e., $c_{f,\rm NLR}<0.5$). However, our results may be biased by the fact that: \borus\ assumes a uniformly distributed obscuring material scenario, therefore the $c_{f,\rm Tor}$ measured in \borus\ is the effective fraction of the sky that is covered by the obscuring material, which thus gives the lower limit of the realistic clumpy $c_{f,\rm Tor}$; the optical emission associated with the NLR measured by \citet{Fischer_2013} may contain the emission from star formation process, which might lead to inaccurate measurement of the geometry of the NLR.
Therefore, further studies with larger sample of AGNs with multi-wavelength datasets are needed to understand the geometrical properties of different components of the AGN.

%
\section{Conclusion}\label{conclusion}
%
We performed a broadband X-ray spectral analysis on a sample of AGNs selected from \citet{Fischer_2013} with [OIII] measured inclination angle, using high-quality \NuSTAR, \XMM\ and \cha\ archival data. To model the spectra, we utilized the recently published self-consistent \borus\ model, which is effective in characterizing the physical and geometrical properties of the obscuring materials near the SMBH. The main findings of this work are reported as follows.

\begin{itemize} 
\item The best-fit values of the spectral parameters obtained when the sources are fitted with the inclination angle being fixed at the [OIII] measured values are similar to those obtained when inclination angle is left free to vary. Fixing the inclination angle at $\rm \theta_{inc}$ = 60$^\circ$ also gives similar spectral fit results, but incorrect fit results may be obtained for some CT sources out of our sample; fixing the inclination angle at $\rm \theta_{inc}$ = 87$^\circ$ leads to significant different measurements of the torus covering factor and the torus column density even for CT-AGNs, but gives the best constraints on different parameters.

\item In AGN X-ray spectral analysis, one should always let the inclination angle free to vary. If one intends to better constrain the properties of sources when fitting low quality X-ray spectra (i.e., $\leq$300 d.o.f), one should fit the spectra by letting $\theta_{\rm inc}$ free to vary at first, then fix $\theta_{\rm inc}$ at some reasonable values, e.g., $\theta_{\rm inc}$ = 60$^\circ$ or [OIII] measured values. Comparing the best-fit results of the two methods: only when the best-fit values of all parameters fitted when fixing the $\theta_{\rm inc}$ are in good agreement with those obtained when letting $\theta_{\rm inc}$ free to vary in fitting the spectra, one could fix $\theta_{\rm inc}$ at those values, otherwise, fixing $\theta_{\rm inc}$ at some preferred values should always be avoided and $\theta_{\rm inc}$ should be left free to vary in fitting these spectra. 

\item The properties of AGNs in our sample are not dependent on the direction at which they are observed, i.e., the inclination angle.

\item We confirm a strong inverse correlation between the torus covering factor and the Eddington ratio, and a correlation between the torus average column density and the Eddington ratio measured in the sources of our sample, which is in good agreement with the radiative feedback model. We also find an inverse correlation between the torus covering factor and the 2--10\,keV intrinsic luminosity, which has also been measured in previous works. However, we do not find any correlation between the torus average column density and the 2--10\,keV intrinsic luminosity, suggesting that the distribution of the materials in the obscuring torus are regulated by the Eddington ratio rather than the intrinsic luminosity.

\item We found no geometrical correlation between the two components of AGN, i.e., obscuring torus and NLR: the torus covering factors span all values, while the covering factors of NLR do not. The ability to robustly measure the covering factor of the torus in the X-ray band is currently limited by the data quality, sample size, and the lack of sufficiently realistic spectral model, which we expect to improve in future work. However, this result already suggests that AGN geometry might be more complex than what is assumed in the simplistic unified model of AGN.

\end{itemize}

%
%
\acknowledgements
X.Z. thanks the anonymous referee for their detailed and useful comments, which helped significantly improve the paper. X.Z., S.M., and M.A. acknowledge NASA funding under contract 80NSSC17K0635 and 80NSSC19K0531. M.B. acknowledges support from the Black Hole Initiative at Harvard University, which is funded in part by the Gordon and Betty Moore Foundation (grant GBMF8273) and in part by the John Templeton Foundation. \NuSTAR\ is a project led by the California Institute of Technology (Caltech), managed by the Jet Propulsion Laboratory (JPL), and funded by the National Aeronautics and Space Administration (NASA). We thank the NuSTAR Operations, Software and Calibrations teams for support with these observations. This research has made use of the \NuSTAR\ Data Analysis Software (NuSTARDAS) jointly developed by the ASI Science Data Center (ASDC, Italy) and the California Institute of Technology (USA). This research has made use of data and/or software provided by the High Energy Astrophysics Science Archive Research Center (HEASARC), which is a service of the Astrophysics Science Division at NASA/GSFC and the High Energy Astrophysics Division of the Smithsonian Astrophysical Observatory. This work is based on observations obtained with \XMM, an ESA science mission with instruments and contributions directly funded by
ESA Member States and NASA.

\bibliographystyle{aa}
\bibliography{referencezxr}

\appendix
\section{A. \borus\ Model}
\begin{figure*} 
\begin{minipage}[b]{.5\textwidth}
\centering
\includegraphics[width=1\textwidth]{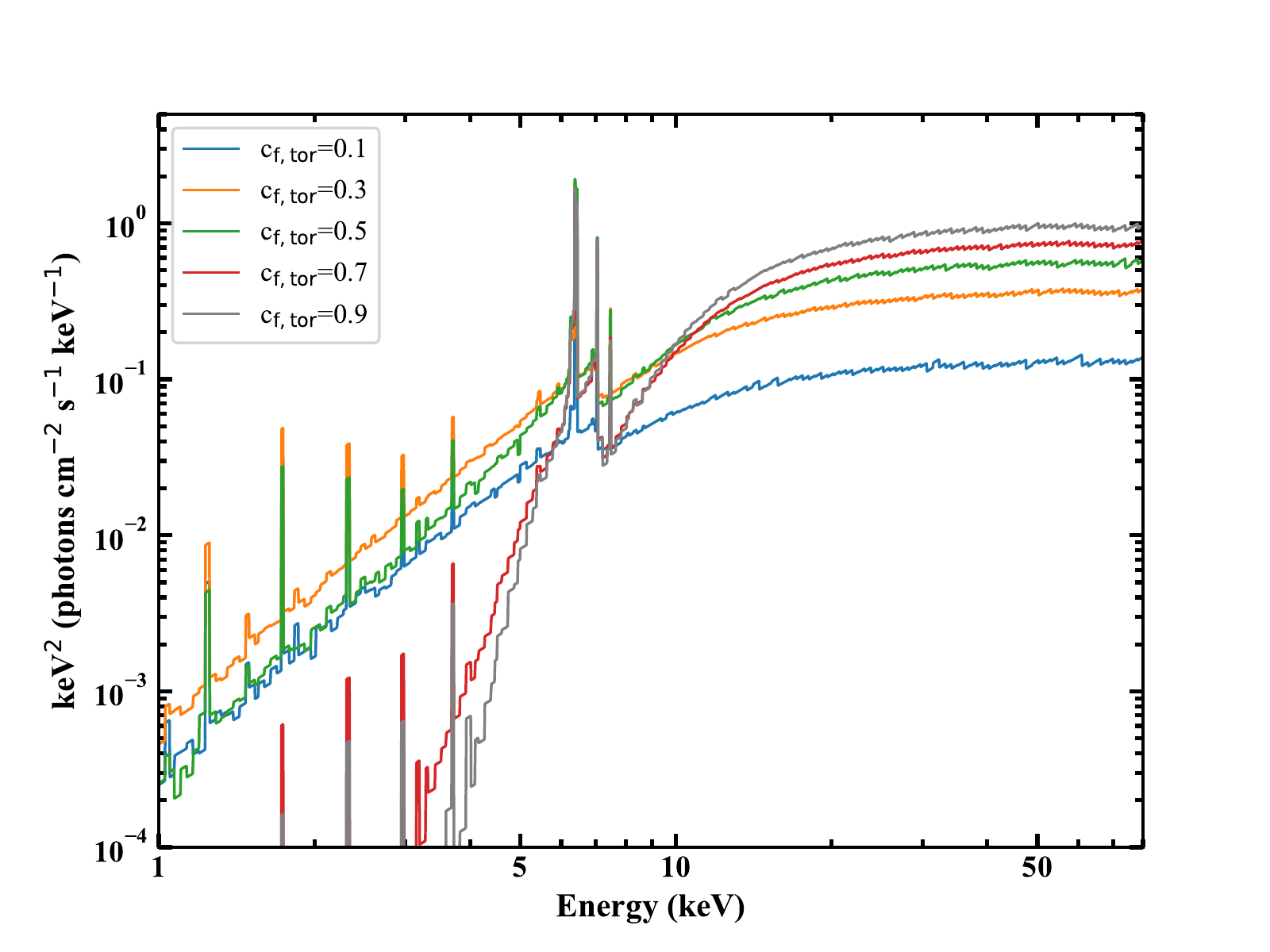}
\includegraphics[width=1\textwidth]{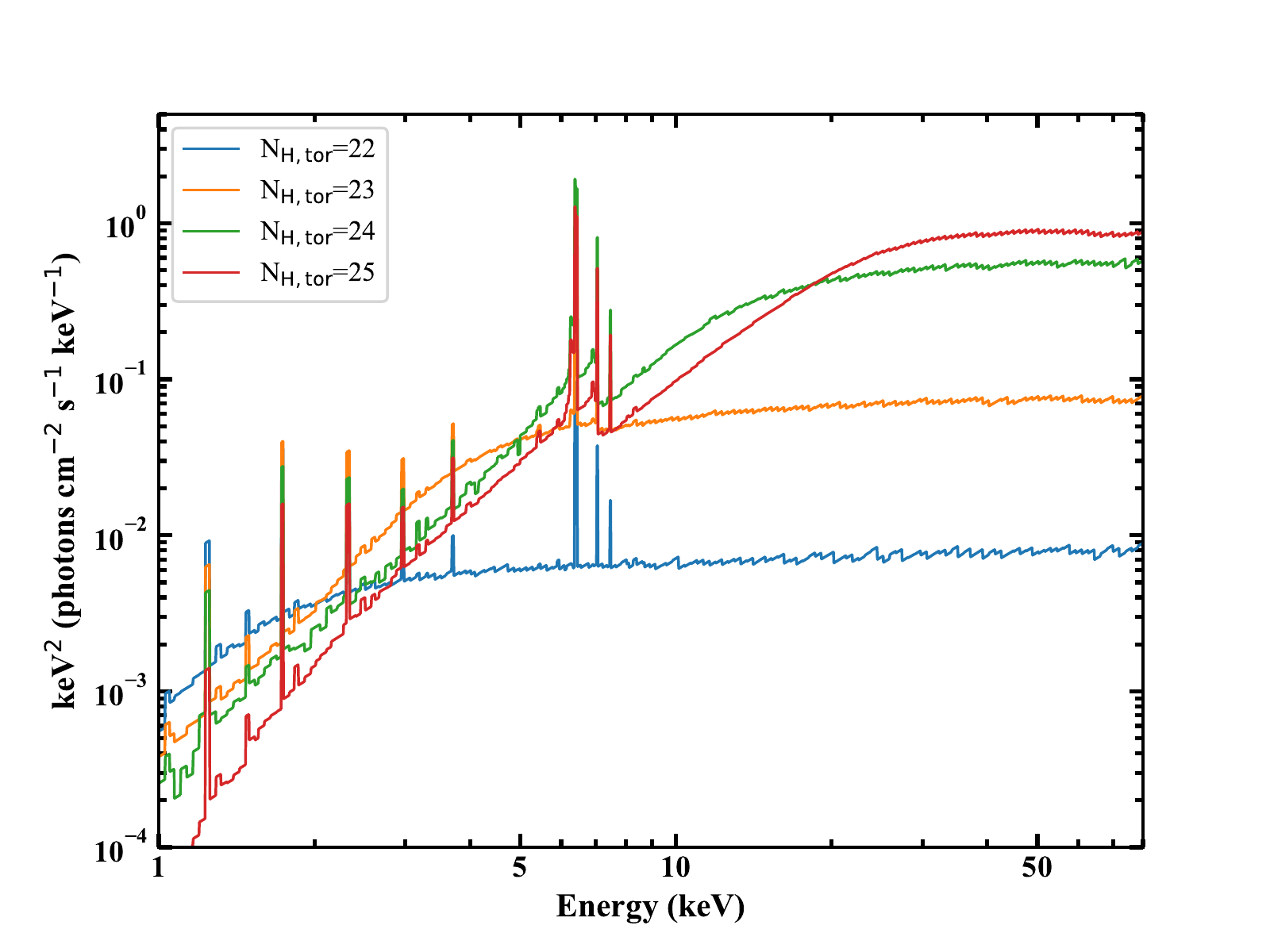}
\end{minipage}
\begin{minipage}[b]{.5\textwidth}
\centering
\includegraphics[width=1\textwidth]{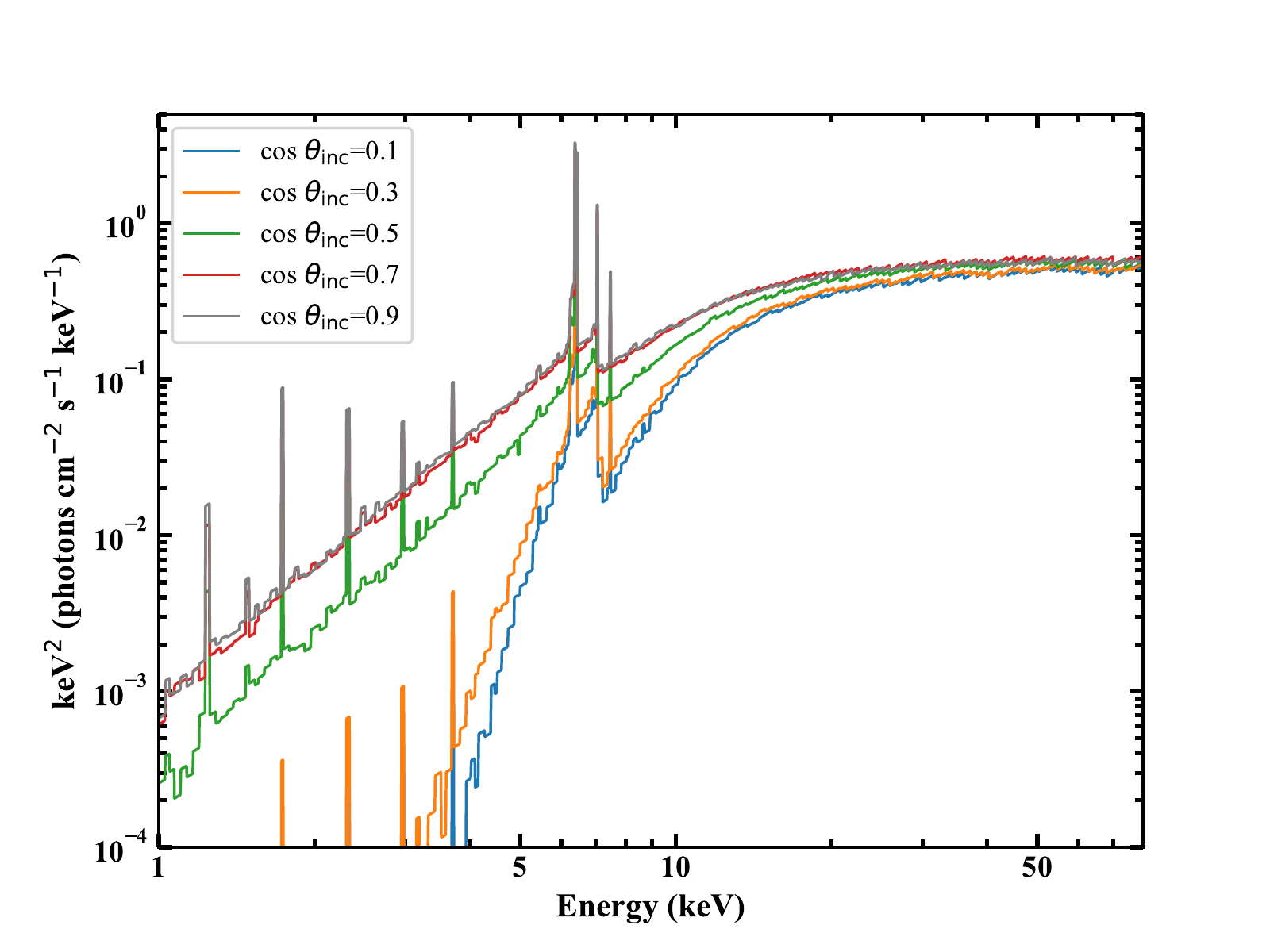}
\includegraphics[width=1\textwidth]{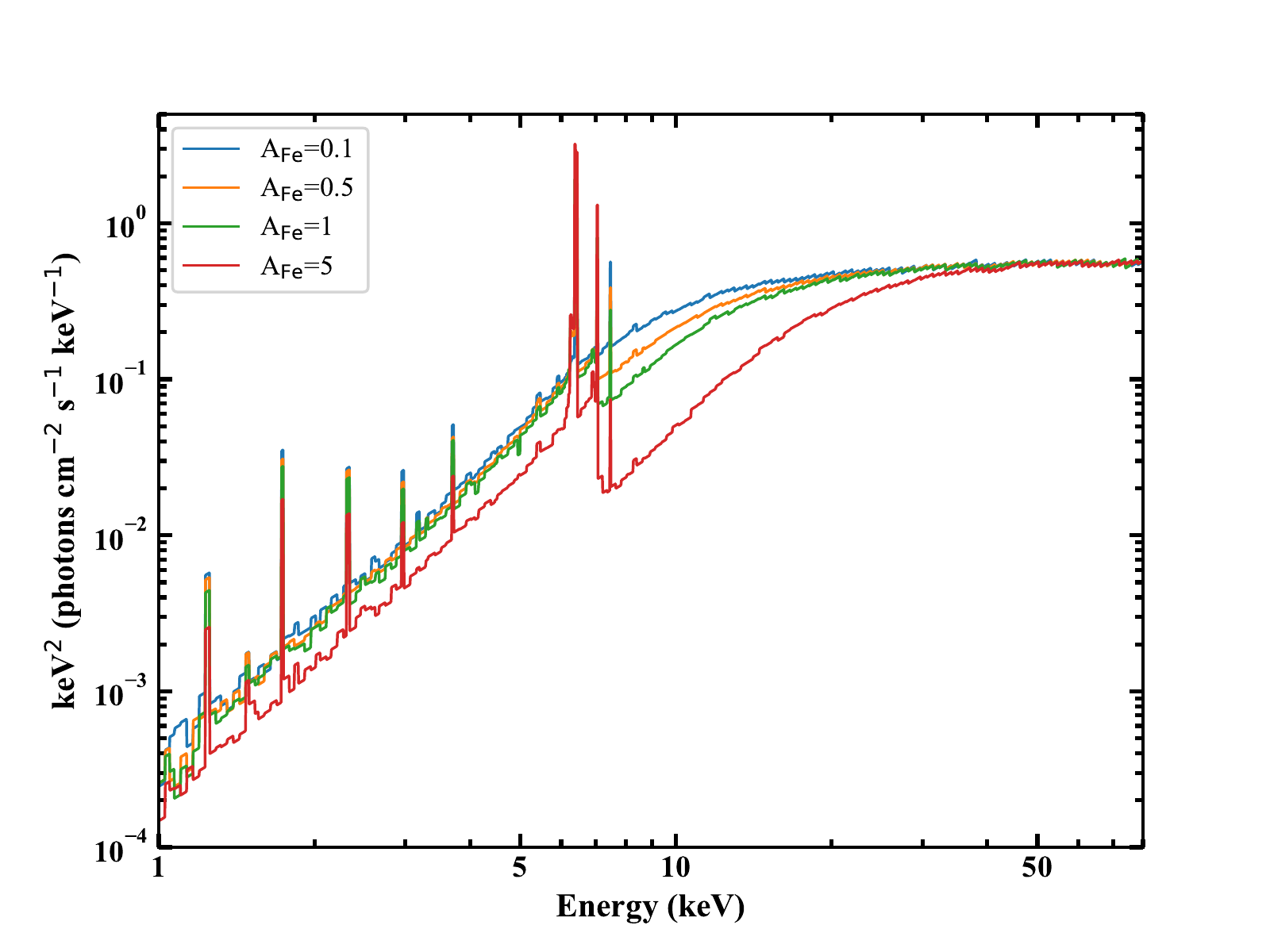}
\end{minipage}
\caption{Top Left: Spectra of the \borus\ \citep{Borus} model prediction of the reprocessed component with varying $c_{\rm f,tor}$, assuming cos($\theta_{\rm inc}$) = 0.5, log($\rm N_{H,tor}$) = 24 and A$_{\rm Fe}$ = 1. Top Right: Spectra of the \borus\ model prediction of the reprocessed component with varying $\theta_{\rm inc}$, assuming $c_{\rm f,tor}$ = 0.5, log($\rm N_{H,tor}$) = 24 and A$_{\rm Fe}$ = 1. Bottom Left: Spectra of the \borus\ model prediction of the reprocessed component with varying $\rm N_{H,tor}$, assuming $c_{\rm f,tor}$ = 0.5, cos($\theta_{\rm inc}$) = 0.5 and A$_{\rm Fe}$ = 1. Bottom Left: Spectra of the \borus\ model prediction of the reprocessed component with varying A$_{\rm Fe}$, assuming $c_{\rm f,tor}$ = 0.5, cos($\theta_{\rm inc}$) = 0.5 and log($\rm N_{H,tor}$) = 24. All spectra assumed a $\Gamma$ = 1.8.}
\label{fig:model}
\end{figure*}

We plot the spectra of \borus\ model prediction of the reprocessed component when varying different parameters, i.e., torus covering factor, $c_{\rm f,tor}$, inclination angle, $\theta_{\rm inc}$, torus column density, $\rm N_{H,tor}$, and relative iron abundance, A$_{\rm Fe}$, in Figure~\ref{fig:model}. 

\section{B. Dependence of the properties of AGNs on their inclinations}
\begin{figure*} 
\begin{minipage}[b]{.5\textwidth}
\centering
\includegraphics[width=1\textwidth]{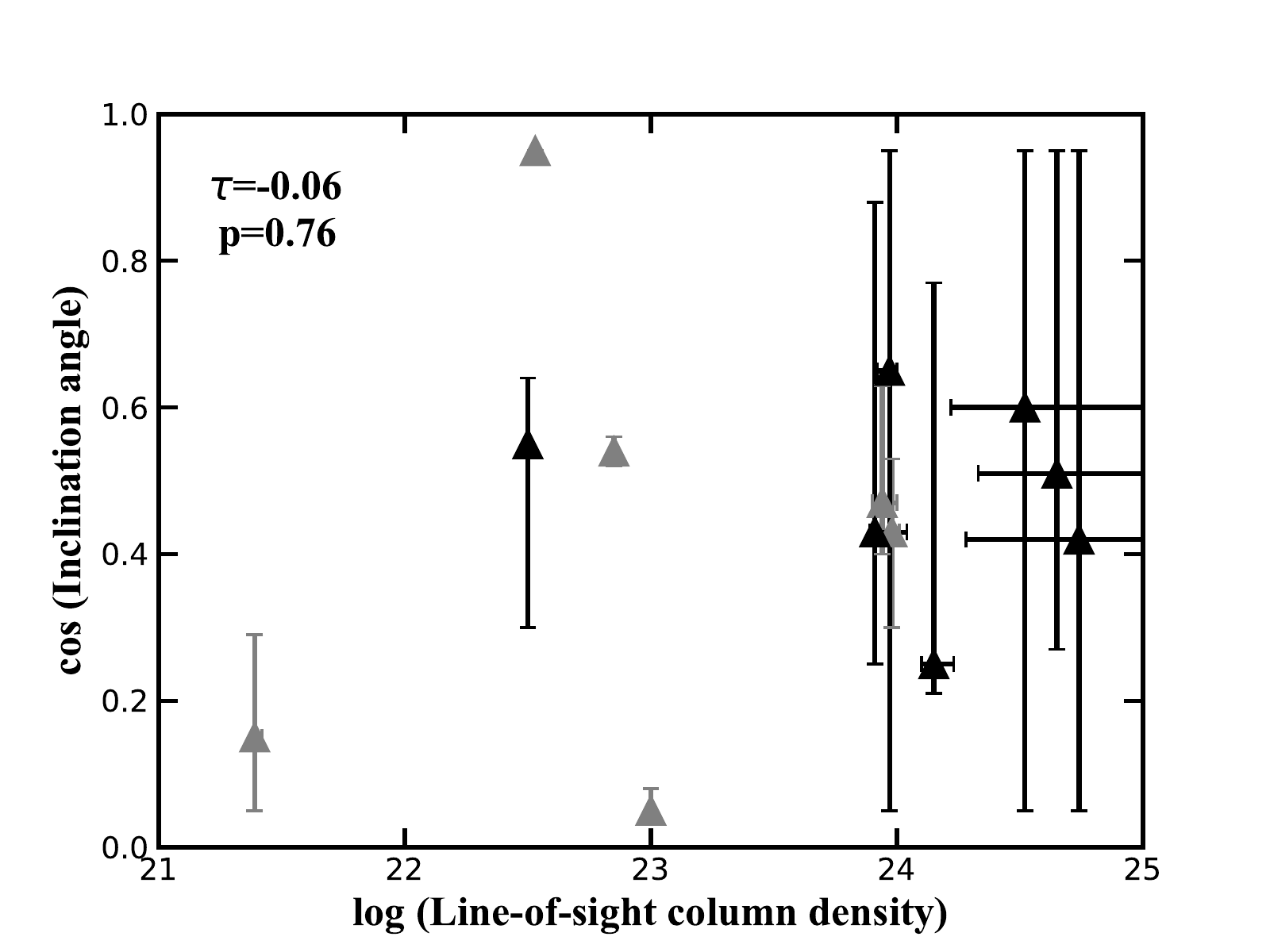}
\includegraphics[width=1\textwidth]{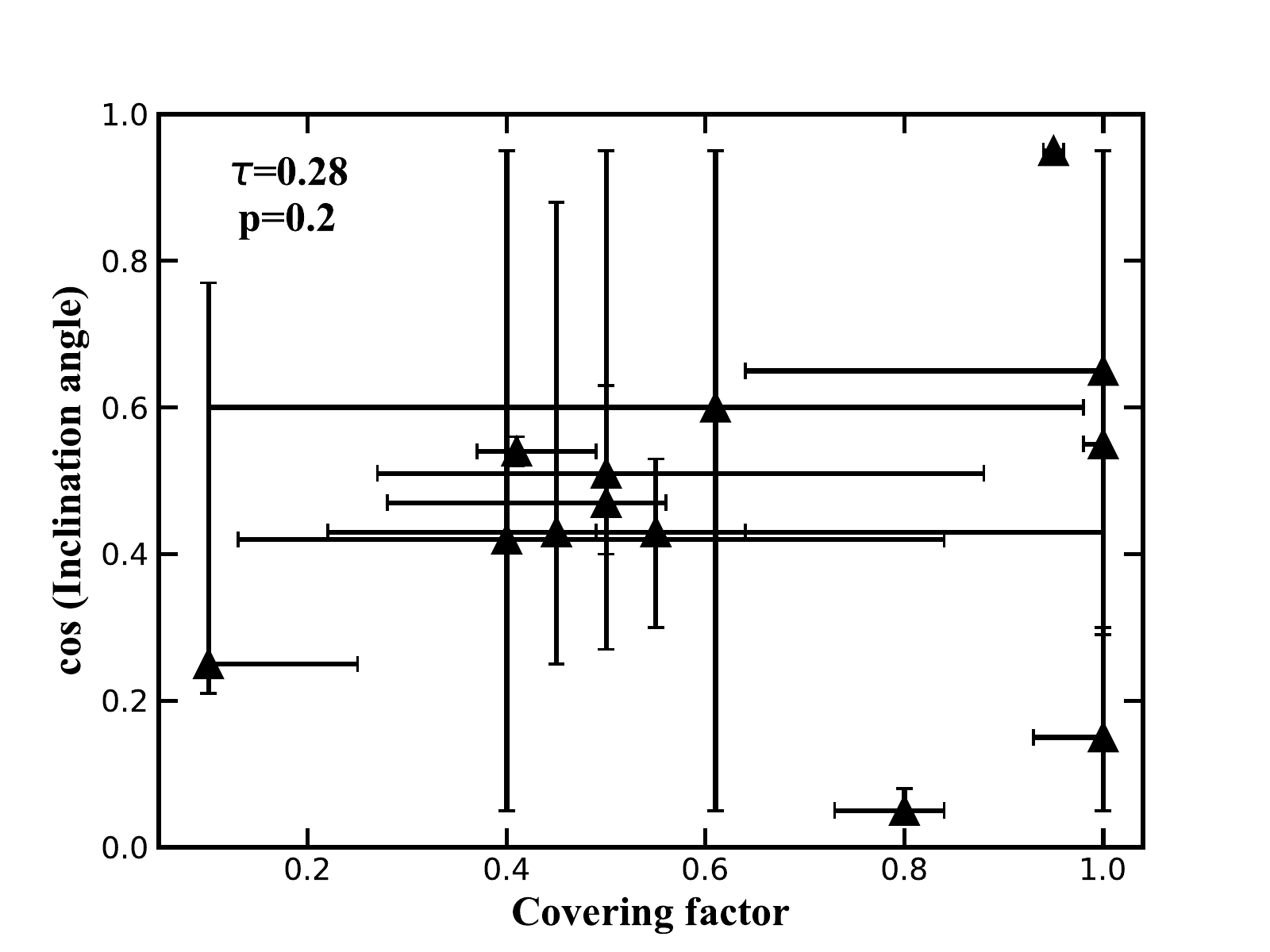}
\includegraphics[width=1\textwidth]{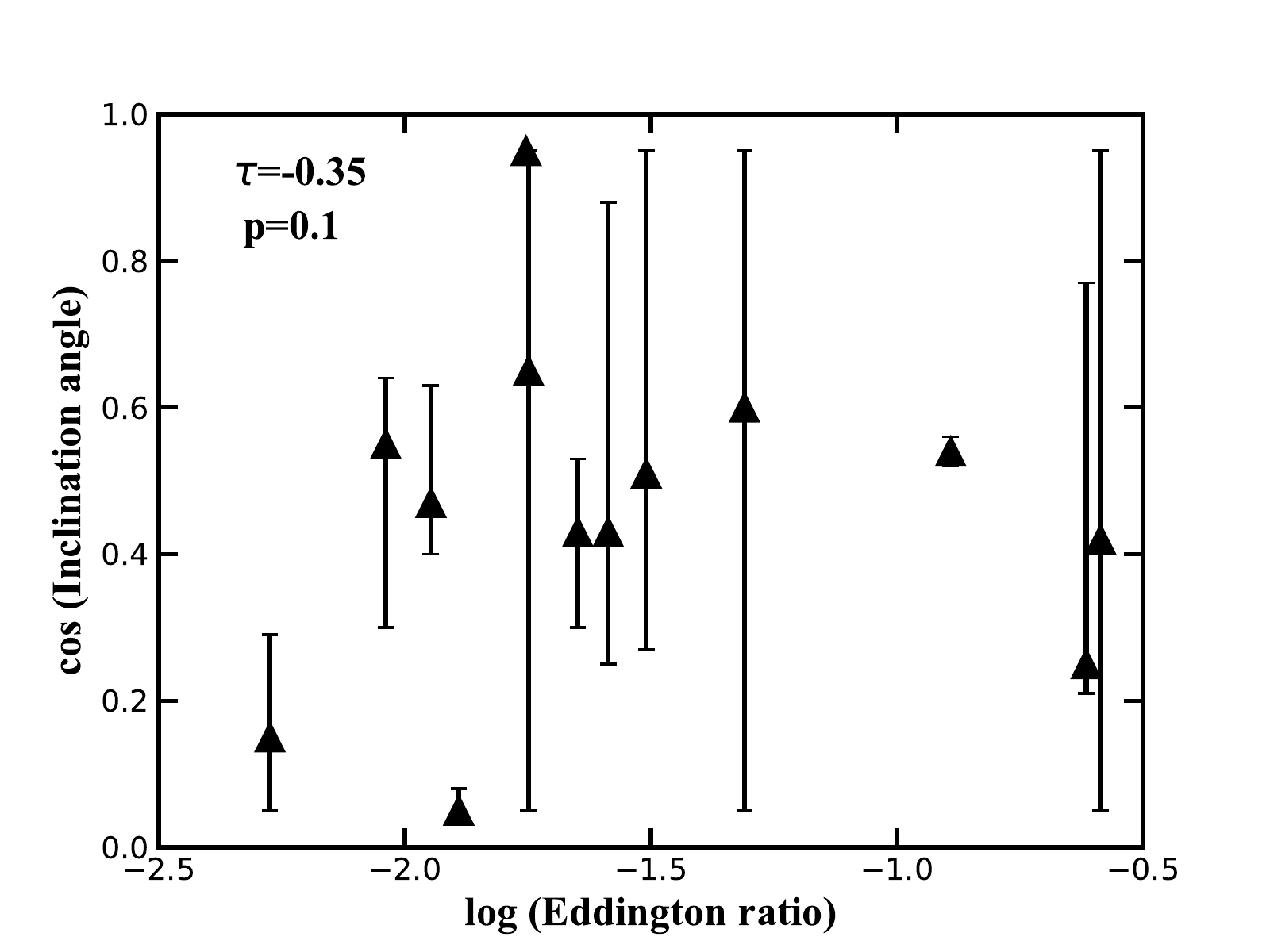}
\end{minipage}
\begin{minipage}[b]{.5\textwidth}
\centering
\includegraphics[width=1\textwidth]{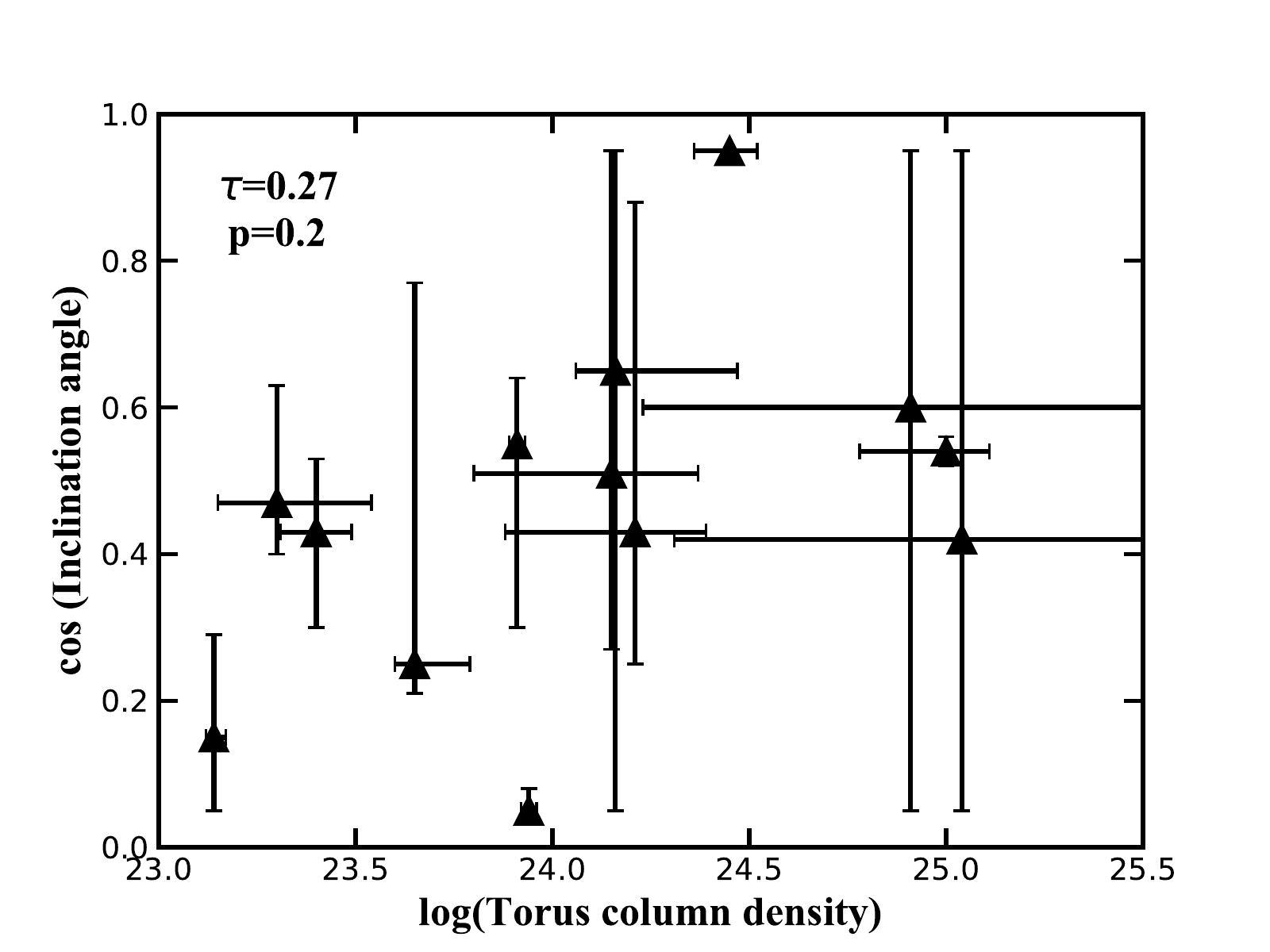}
\includegraphics[width=1\textwidth]{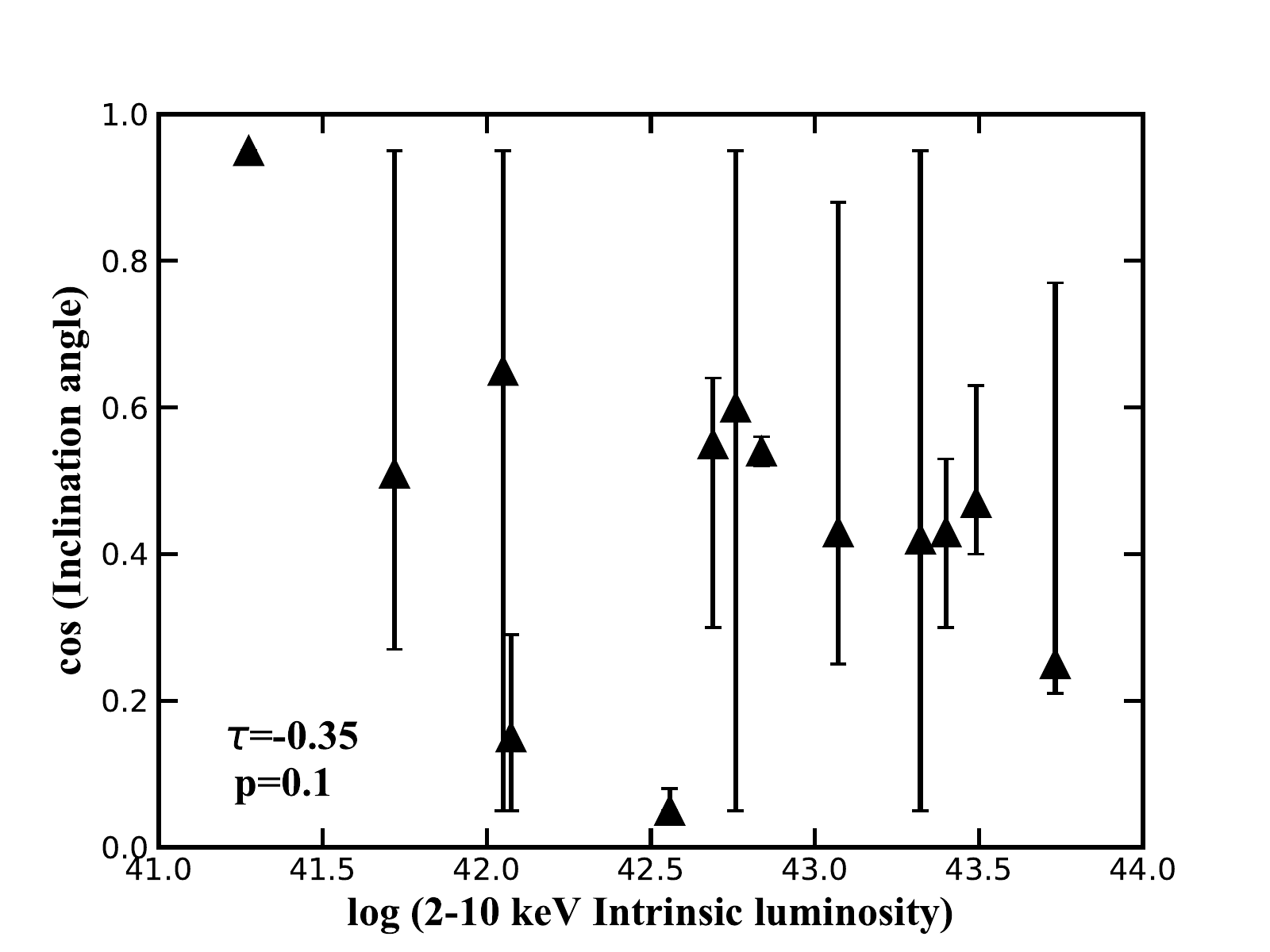}
\end{minipage}
\caption{Best-fit inclination angle of the 13 sources in our sample as functions of their line-of-sight column density, torus covering factor, torus column density, Eddington ratio, 2--10\,keV intrinsic luminosity in Figure~\ref{fig:inc}. The tau and $p$ values for each pair of properties are calculated and reported in each subplot. Sources which have been observed to be variable due to the variability of line-of-sight column density are marked as gray.}
\label{fig:inc}
\end{figure*}

We plot the best-fit inclination angle of the 13 sources in our sample as functions of their line-of-sight column density, torus covering factor, torus column density, Eddington ratio, 2--10\,keV intrinsic luminosity in Figure~\ref{fig:inc}. The tau and $p$ values for each pair of properties are calculated and reported in each subplot. We find no correlation between the measured inclination angle of AGN and the other physical and geometrical properties of AGNs.

\section{C. Fitting details and Spectra of 13 sources}
\subsection{Fitting Details}
Mrk~3: the cut-off energy $E_{\rm cut}$ of the source is found to be much less than the default 500\,keV, thus we let $E_{\rm cut}$ free to vary in fitting the spectra of Mrk~3 and a $E_{\rm cut}$ = 85$_{-20}^{+55}$\,keV is measured. The relative iron abundance is found to be less than the default value, thus the relative iron abundance is left free to vary in fitting the spectrum of Mrk~3 and an A$\rm _{Fe}$ = 0.44$_{-0.08}^{+0.14}$\,A$\rm _{Fe,\odot}$ is measured. The best-fit statistic of the scenario of letting $\theta_{\rm inc}$ free to vary is $\chi^2$/degree of freedom (d.o.f) = 1056/1073 $\approx$0.98. The best-fit photon index for Mrk~3 is $\Gamma$ = 1.48$_{-u}^{+0.11}$, where $u$ means the parameter cannot be constrained at 90\% confidence level within the range of the parameter in the \borus\ model, which is [1.4--2.6] for $\Gamma$. The best-fit line-of-sight column density is log(N$\rm _{H,l.o.s}$) = 23.94$_{-0.04}^{+0.06}$ and the best-fit torus column density is log(N$\rm _{H,tor}$) = 23.30$_{-0.15}^{+0.24}$. The best-fit inclination angle is cos($\theta\rm _{inc}$) = 0.47$_{-0.07}^{+0.16}$ and the best-fit covering factor is $c_{\rm f,tor}$ = 0.50$_{-0.22}^{+0.06}$.

Mrk~34: an energy range of 0.6\,keV--78\,keV rather than 0.5\,keV--78\,keV is used in fitting the spectrum of Mrk~34 since we found that the spectrum between 0.5\,keV--2\,keV cannot be fitted by a single {\texttt{mekal}} model, however, the spectrum between 0.6\,keV--2\,keV can be well fitted by a single {\texttt{mekal}} model. In addition, we added a number of unresolved Gaussian lines as needed to reach a good fit of Mrk~34. The best-fit statistic of Mrk~34 is $\chi^2$/d.o.f = 74/82 $\approx$0.90. The best-fit photon index is $\Gamma$ = 1.45$_{-u}^{+0.67}$. The best-fit line-of-sight column density is log(N$\rm _{H,l.o.s}$) $>$24.28 and the best-fit torus column density is log(N$\rm _{H,tor}$) = 25.04$_{-0.73}^{+u}$, where the parameter range of log(N$\rm _{H,tor}$) in \borus\ table is [22.0--25.5]. The best-fit inclination angle is cos($\theta\rm _{inc}$) = 0.42$_{-u}^{+u}$, suggesting that the inclination angle of Mrk~34 is unconstrained with current data. The best-fit covering factor of Mrk~34 is $c_{\rm f,tor}$ = 0.40$_{-0.27}^{+0.44}$.

Mrk~78: We adopted \NuSTAR\ data from 4\,keV to 78\,keV since the \NuSTAR\ data between 3--4\,keV showed strong discrepancy with \XMM\ and \cha\ data. We found a strong emission line at $\sim$8.265\,keV in the spectrum, which belongs to the Ni K$\beta$ emission line. Thus a gaussian line centered at $E_l$ = 8.265\,keV with zero width is added to better fit the spectrum. We compare the flux of the Ni K$\beta$ emission line and Fe K$\alpha$ line: the measured flux of the Fe K$\alpha$ line of Mrk~78 is F$_{\rm lux,Fe\,K\alpha}$ = 4.5 $\times$ 10$^{-15}$\,erg\,cm$^{-2}$\,s$^{-1}$ between 6.39\,keV and 6.41\,keV and the measured flux of the Ni K$\beta$ line is F$_{\rm lux,Ni\,K\beta}$ = 3.8 $\times$ 10$^{-15}$\,erg\,cm$^{-2}$\,s$^{-1}$ between 8.255\,keV and 8.275\,keV. Strong variability is also found between the \NuSTAR\ observations and the three soft X-ray observations. The 2--10\,keV flux of Mrk~78 observed in \NuSTAR\ increased by $\sim$135\% with respect to the flux observed in \XMM. We found that the variability observed in Mrk~78 is caused by both the variability of intrinsic emission and N$_{\rm H,l.o.s}$ variability. The intrinsic emission of Mrk~78 measured using \NuSTAR\ observation increased by $\sim$38\% with respect to the intrinsic emission measured using \XMM\ observation and the N$_{\rm H,l.o.s}$ observed in \NuSTAR\ decreased by 49\% with respect to the N$_{\rm H,l.o.s}$ observed in \XMM, i.e., log(N$_{\rm H,l.o.s,NuS}$) = 23.62$_{-0.08}^{+0.08}$ from log(N$_{\rm H,l.o.s,XMM}$) = 23.91$_{-0.02}^{+0.13}$. The best-fit statistic of Mrk~78 is $\chi^2$/d.o.f = 276/271 $\approx$ 1.02. The best-fit photon index of Mrk~78 is $\Gamma$ = 1.40$_{-u}^{+0.21}$. The best-fit torus column density is log(N$\rm _{H,tor}$) = 24.21$_{-0.33}^{+0.18}$. The best-fit inclination angle is cos($\theta\rm _{inc}$) = 0.43$_{-0.18}^{+0.45}$ and the best-fit covering factor is found to be $c_{\rm f,tor}$ $>$0.22.

Mrk~573: an energy range of 0.3\,keV--78\,keV rather than 0.5\,keV--78\,keV is used in fitting the spectrum of Mrk~573 because we found the spectrum of Mrk~573 between 0.5\,keV--2\,keV can not be fitted by a single {\texttt{mekal}} model. Although the spectrum between 0.8\,keV--2\,keV can be fitted by a single {\texttt{mekal}} model, to exploit the high-quality \XMM\ data below 0.8\,keV, which provide $\sim$2600 more counts (the total counts in the spectra between 0.8\,keV--78\,keV are $\sim$2400 cts), we fit the spectra of Mrk~573 in the 0.3\,keV--78\,keV energy band and add another {\texttt{mekal}} in our modeling. The best-fit statistic of Mrk~573 is $\chi^2$/d.o.f = 152/194 $\approx$ 0.78. The best-fit photon index is $\Gamma$ = 2.35$_{-0.65}^{+u}$. The best-fit line-of-sight column density is log(N$\rm _{H,l.o.s}$) $>$24.22 and the best-fit average torus column density is log(N$\rm _{H,tor}$) $>$24.23. The best-fit inclination angle is cos($\theta\rm _{inc}$) = 0.60$_{-u}^{+u}$ and the best-fit torus covering factor is $c_{\rm f,tor}$ $<$0.98, suggesting that we are unable to constrain both the inclination and the torus covering factor of Mrk~573.

Mrk~1066: the cut-off energy $E_{\rm cut}$ of Mrk~1066 is found to be $E_{\rm cut}<$28\,keV. The relative iron abundance is found to be A$\rm _{Fe}$ = 3.2$_{-1.0}^{+0.7}$\,A$\rm _{Fe,\odot}$. In addition, we added a number of unresolved Gaussian lines as needed to reach a good fit of Mrk~1066. The best-fit statistic of Mrk~1066 is $\chi^2$/d.o.f = 142/147 $\approx$ 0.97. The best-fit photon index is $\Gamma$ = 1.52$_{-u}^{+0.02}$. The best-fit line-of-sight column density is log(N$\rm _{H,l.o.s}$) = 23.97$_{-0.05}^{+0.03}$ and the best-fit average torus column density is log(N$\rm _{H,tor}$) = 24.16$_{-0.10}^{+0.31}$. The best-fit inclination angle is cos($\theta\rm _{inc}$) = 0.65$_{-u}^{+u}$, suggesting that the inclination angle of Mrk~1066 is fully unconstrained with current data. The best-fit covering factor is $c_{\rm f,tor}$ $>$0.64.

NGC~3227: the source is found to be an unobscured AGN with N$_{\rm H,l.o.s}<$ 10$^{22}$\,cm$^{-2}$. We found that the scattering component contribute marginally to the spectrum and abandon of this component does not worsen the fit. Therefore, we set the fraction of the scattering component to be $f_s$ = 0 to decrease the free parameters in fitting. The spectra below 1\,keV has no obvious emission signature and is hard to be well fitted by {\texttt{mekal}}. We add an phenomenological gaussian and an unabsorbed cut-off power law with a different photon index to fit the soft X-ray spectrum of NGC~3227. The center of the gaussian is at E$_l$ = 0.62\,keV and the width is $\sigma$ = 0.05\,keV. The photon index of this phenomenological cut-off power law is measured as $\Gamma_{\rm soft}$ = 3.9$_{-0.1}^{+0.1}$ compared to the best-fit photon index of the intrinsic cut-off power-law is $\Gamma$ = 1.68$_{-0.01}^{+0.01}$. The best-fit statistic of NGC~3227 is $\chi^2$/d.o.f = 4684/4008 $\approx$ 1.17. The best-fit line-of-sight column density is log(N$\rm _{H,l.o.s}$) = 21.39$_{-0.01}^{+0.03}$ and the best-fit average torus column density is log(N$\rm _{H,tor}$) = 23.14$_{-0.02}^{+0.03}$. The best-fit inclination angle is cos($\theta\rm _{inc}$) $<$0.29 and the best-fit covering factor is $c_{\rm f,tor}$ $>$0.93.

NGC~3783: \NuSTAR\ data above 70\,keV are polluted by background and the data $<$10\,keV show strong discrepancy with \XMM, so we fit the \NuSTAR\ spectrum only between 10--70\,keV following the approach adopted in previous works \citep{Mehdipour2017,Mao2019,Marco_2020}.
The cut-off energy $E_{\rm cut}$ of NGC~3783 is found to be $E_{\rm cut}$ = 37$_{-4}^{+2}$\,keV. In addition, we added a number of unresolved Gaussian lines as needed to reach a good fit of Mrk~3783. The best-fit statistic of NGC~3783 is $\chi^2$/d.o.f = 3349/2929 $\approx$ 1.14. The best-fit photon index is $\Gamma$ = 1.51$_{-0.04}^{+0.02}$. The best-fit line-of-sight column density is log(N$\rm _{H,l.o.s}$) = 22.85$_{-0.01}^{+0.01}$ and the best-fit average torus column density is log(N$\rm _{H,tor}$) = 25.00$_{-0.22}^{+0.11}$. The best-fit inclination angle is cos($\theta\rm _{inc}$) = 0.54$_{-0.02}^{+0.02}$ and the best-fit covering factor is $c_{\rm f,tor}$ = 0.41$_{-0.04}^{+0.08}$.

NGC~4051: the source is known to exhibit strong spectra and flux variation in X-ray \citep{Turner2017}. Variability is also found between the \NuSTAR\ and \XMM\ observations that we analyzed. The 2--10\,keV flux of the \NuSTAR\ observation is $\sim$24\% less than the \XMM\ observation. This flux variability is caused by the variability of the intrinsic emission rather than the N$_{\rm H,l.o.s}$ variability based our analysis. The source is also known to possess a warm absorber outflow \citep{Mizumoto2016}, due to its absorption signature, especially the O $_{\rm VIII}$ absorption feature $\sim$0.65\,keV, which is difficult to be fitted by the default  {\texttt{mekal}} model. We tried to use a complex phenomenological model to fit the spectrum below 1\,keV and found that the best-fit results of the key parameters did not vary compared with those obtained when we fitted the spectra only above 1\,keV. Therefore, we fit the spectra of NGC~4051 with energy only above 1\,keV. The cut-off energy $E_{\rm cut}$ of NGC~4051 is found to be $E_{\rm cut}$ = 44$_{-2}^{+2}$\,keV. The best-fit statistic of NGC~4051 is $\chi^2$/d.o.f = 2686/2389 $\approx$ 1.12. The best-fit photon index is $\Gamma$ = 1.72$_{-0.01}^{+0.01}$. The best-fit line-of-sight column density is log(N$\rm _{H,l.o.s}$) = 22.53$_{-0.01}^{+0.01}$ and the best-fit average torus column density is log(N$\rm _{H,tor}$) = 24.45$_{-0.09}^{+0.07}$. The best-fit inclination angle is cos($\theta\rm _{inc}$) $>$0.94 and the best-fit covering factor is $c_{\rm f,tor}$ = 0.95$_{-0.01}^{+0.01}$.

NGC~4151: the source is known to exhibit spectral and flux variability in X-ray \citep{Beuchert2017}. Strong variability was also found between the \NuSTAR\ and \XMM\ observations in our analysis. The 2--10\,keV flux of the \NuSTAR\ observation is $\sim$103\% larger than the \XMM\ observation. This flux variability is caused by both the variability of the intrinsic emission and the N$_{\rm H,l.o.s}$ variability based on our analysis. The intrinsic emission measured using \NuSTAR\ observation is $\sim$80\% larger than the intrinsic emission  measured using \XMM\ observation. The line-of-sight column density of the two observations are log(N$_{\rm H,l.o.s,NuS}$) = 22.86$_{-0.01}^{+0.02}$ and log(N$_{\rm H,l.o.s,XMM}$) = 23.00$_{-0.01}^{+0.01}$, respectively. The cut-off energy $E_{\rm cut}$ of NGC~4151 is found to be $E_{\rm cut}$ = 112$_{-16}^{+10}$\,keV. The relative iron abundance is found to be A$\rm _{Fe}$ = 0.66$_{-0.05}^{+0.05}$\,A$\rm _{Fe,\odot}$ when the inclination is fixed at [OIII] measured value in fitting. In addition, we added a number of unresolved Gaussian lines as needed to reach a good fit of NGC~4151. The best-fit statistic of NGC~4151 is $\chi^2$/d.o.f = 5200/4664 $\approx$ 1.11. The best-fit photon index of NGC~4151 is $\Gamma$ = 1.67$_{-0.04}^{+0.02}$. The best-fit average torus column density is log(N$\rm _{H,tor}$) = 23.94$_{-0.02}^{+0.02}$. The best-fit inclination angle is cos($\theta\rm _{inc}$) $<$0.08 and the best-fit covering factor is $c_{\rm f,tor}$ = 0.80$_{-0.07}^{+0.04}$.

NGC~4507: the source is known to exhibit spectral and flux variability in X-ray \citep{Braito2012,Marinucci2013}. Variability was also found between the \NuSTAR\ and \XMM\ observations in our analysis. The 2--10\,keV flux of the \NuSTAR\ observation is $\sim$51\% larger than the \XMM\ observation. This flux variability is caused by the variability of the intrinsic emission rather than the N$\rm _{H,l.o.s}$ variability based on our analysis. The intrinsic emission measured using \NuSTAR\ observation is 50\% larger than the intrinsic emission measured using \XMM\ observation. The spectrum below 1\,keV is difficult to model with a single {\texttt{mekal}}, so we add another {\texttt{mekal}} in fitting the spectrum. The relative iron abundance is found to be A$\rm _{Fe}$ = 0.5$_{-0.1}^{+0.1}$\,A$\rm _{Fe,\odot}$. In addition, we added a number of unresolved Gaussian lines as needed to reach a good fit of NGC~4507. The best-fit statistic of NGC~4507 is $\chi^2$/d.o.f = 1601/1614 $\approx$ 0.99. The best-fit photon index is $\Gamma$ = 1.71$_{-0.03}^{+0.05}$. The best-fit line-of-sight column density is log(N$\rm _{H,l.o.s}$) = 23.98$_{-0.05}^{+0.03}$ and the best-fit average torus column density is log(N$\rm _{H,tor}$) = 23.40$_{-0.09}^{+0.09}$. The best-fit inclination angle is cos($\theta\rm _{inc}$) = 0.43$_{-0.13}^{+0.10}$ and the best-fit covering factor is $c_{\rm f,tor}$ = 0.55$_{-0.06}^{+0.09}$.

NGC~5506: the source is known to exhibit spectral and flux variability in X-ray \citep{Sun2018}. Variability was also found between the \NuSTAR\ and \XMM\ observations that we adopted. The 2--10\,keV flux of the \NuSTAR\ observation is $\sim$18\% less than the \XMM\ observation. This flux variability is caused by both the variability of the intrinsic emission and the N$_{\rm H,l.o.s}$ variability based on our analysis. The intrinsic emission measured using \NuSTAR\ is 25\% less than the intrinsic emission measured using \XMM\ observation and the line-of-sight column density measured in two observations are log(N$_{\rm H,l.o.s,NuS}$) = 22.29$_{-0.02}^{+0.03}$ and log(N$_{\rm H,l.o.s,XMM}$) = 22.50$_{-0.01}^{+0.01}$, respectively. The relative iron abundance is found to be A$\rm _{Fe}$ = 5.8$_{-0.2}^{+0.4}$\,A$\rm _{Fe,\odot}$. The cut-off energy $E_{\rm cut}$ of NGC~5506 is found to be $E_{\rm cut}<$ 21\,keV. In addition. we found that the fit was significant improved (from $\chi^2$/d.o.f = 5590/4544 to $\chi^2$/d.o.f = 5378/4543) if we let the photon index of the scattering component free to vary from the one of the intrinsic emission. We measured the photon index of the scattering component as $\Gamma_{\rm soft}$ = 1.02$_{-0.02}^{+0.30}$ and the best-fit photon index of intrinsic emission is $\Gamma$ = 1.72$_{-0.01}^{+0.01}$. The best-fit statistic of NGC~5506 is $\chi^2$/d.o.f = 5378/4543 $\approx$ 1.18. The best-fit average torus column density is log(N$\rm _{H,tor}$) = 23.91$_{-0.02}^{+0.02}$. The best-fit inclination angle is cos($\theta\rm _{inc}$) = 0.55$_{-0.25}^{+0.09}$ and the best-fit covering factor is $c_{\rm f,tor}$ $>$0.98.

NGC~5643: a strong emission line at $\sim$1.836\,keV, which belongs to the Si~K$\beta$ emission line is found. Therefore, a gaussian line with zero width is added to better fit the spectrum. The best-fit statistic of NGC~5643 is $\chi^2$/d.o.f = 198/170 $\approx$1.16. The best-fit photon index of NGC~5643 is $\Gamma$ = 1.77$_{-0.37}^{+0.27}$. The best-fit line-of-sight column density is log(N$\rm _{H,l.o.s}$) $>$24.33 and the best-fit average torus column density is log(N$\rm _{H,tor}$) = 24.15$_{-0.35}^{+0.22}$. The best-fit inclination angle is cos($\theta\rm _{inc}$) $>$0.27 and the best-fit covering factor is $c_{\rm f,tor}$ = 0.50$_{-0.23}^{+0.38}$.

NGC~7674: the source is known to exhibit spectra and flux variation in X-ray \citep{Gandhi2017}. Variability was also found between the \NuSTAR\ and \XMM\ observations that we adopted. The 2--10\,keV flux of the \NuSTAR\ observation is $\sim$29\% larger than the \XMM\ observation. This flux variability is caused by both the variability of the intrinsic emission and the N$_{\rm H,l.o.s}$ variability according to our analysis. The intrinsic emission measured using \NuSTAR\ observation is 113\% larger than the intrinsic emission measured using \XMM\ observation and the line-of-sight column density measured in two observations are log(N$_{\rm H,l.o.s,NuS}$) = 24.45$_{-0.02}^{+0.05}$ and log(N$_{\rm H,l.o.s,XMM}$) = 24.15$_{-0.05}^{+0.08}$, respectively. The relative iron abundance is found to be A$\rm _{Fe}$ = 0.42$_{-0.08}^{+0.09}$\,A$\rm _{Fe,\odot}$. The best-fit statistic of NGC~7674 is $\chi^2$/d.o.f = 240/251 $\approx$ 0.96. The best-fit photon index is $\Gamma$ = 2.21$_{-0.18}^{+0.12}$. The best-fit average torus column density is log(N$\rm _{H,tor}$) = 23.65$_{-0.05}^{+0.14}$. The best-fit inclination angle is cos($\theta\rm _{inc}$) = 0.25$_{-0.04}^{+0.52}$ and the best-fit covering factor is $c_{\rm f,tor}$ $<$0.25.

\subsection{Spectra of 13 Sources}
\begin{figure*} 
\begin{minipage}[b]{.5\textwidth}
\centering
\includegraphics[width=1\textwidth]{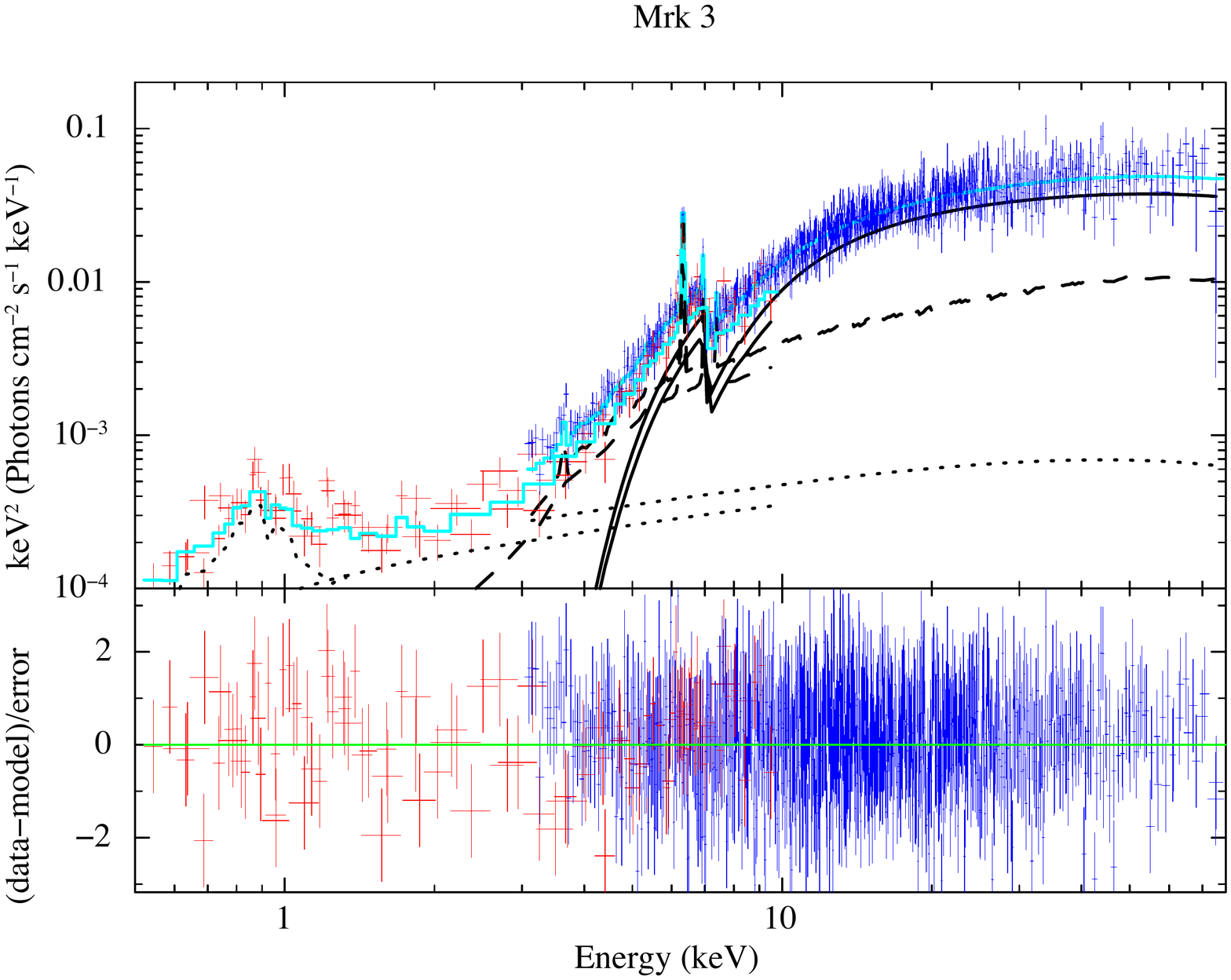}
\includegraphics[width=1\textwidth]{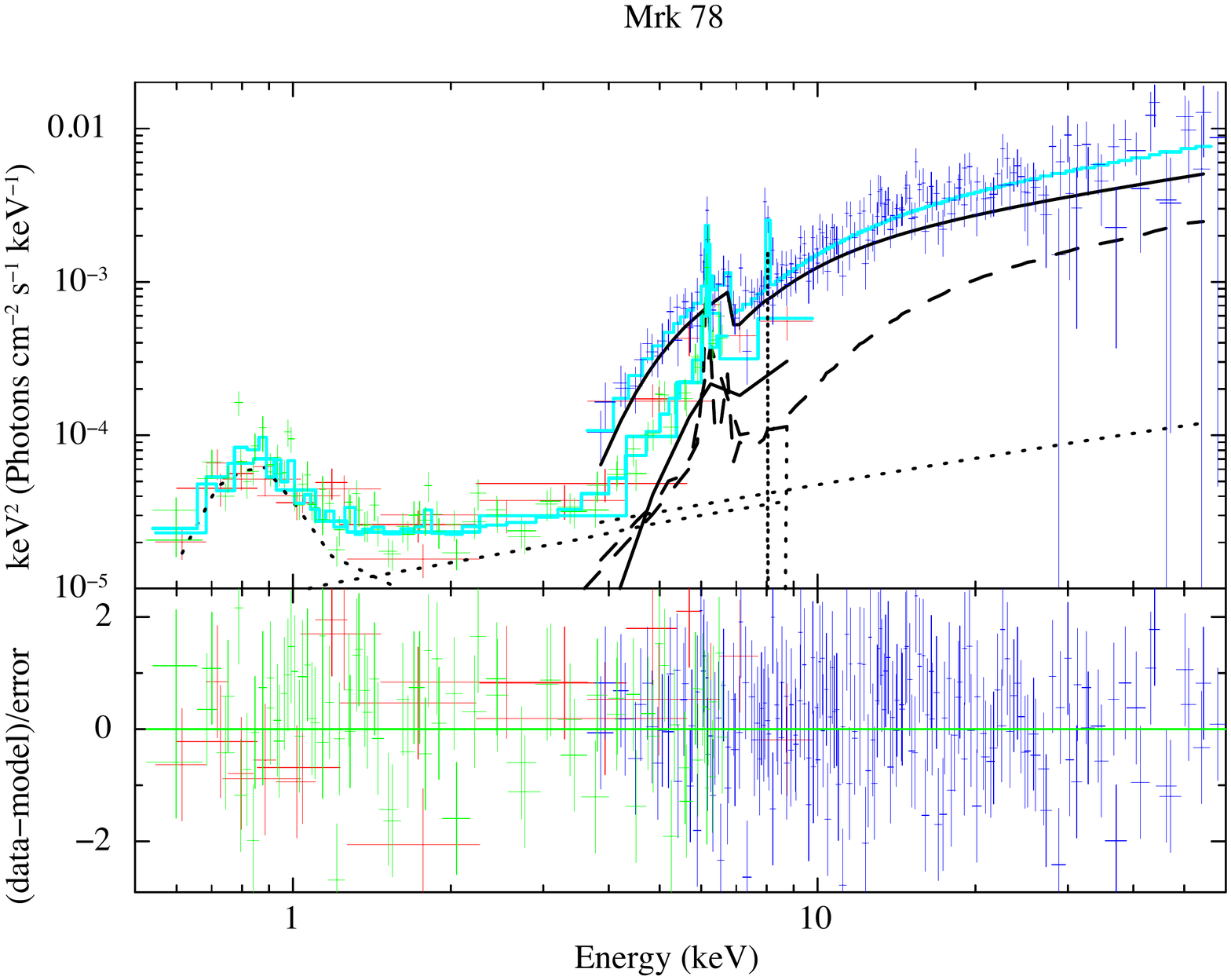}
\includegraphics[width=1\textwidth]{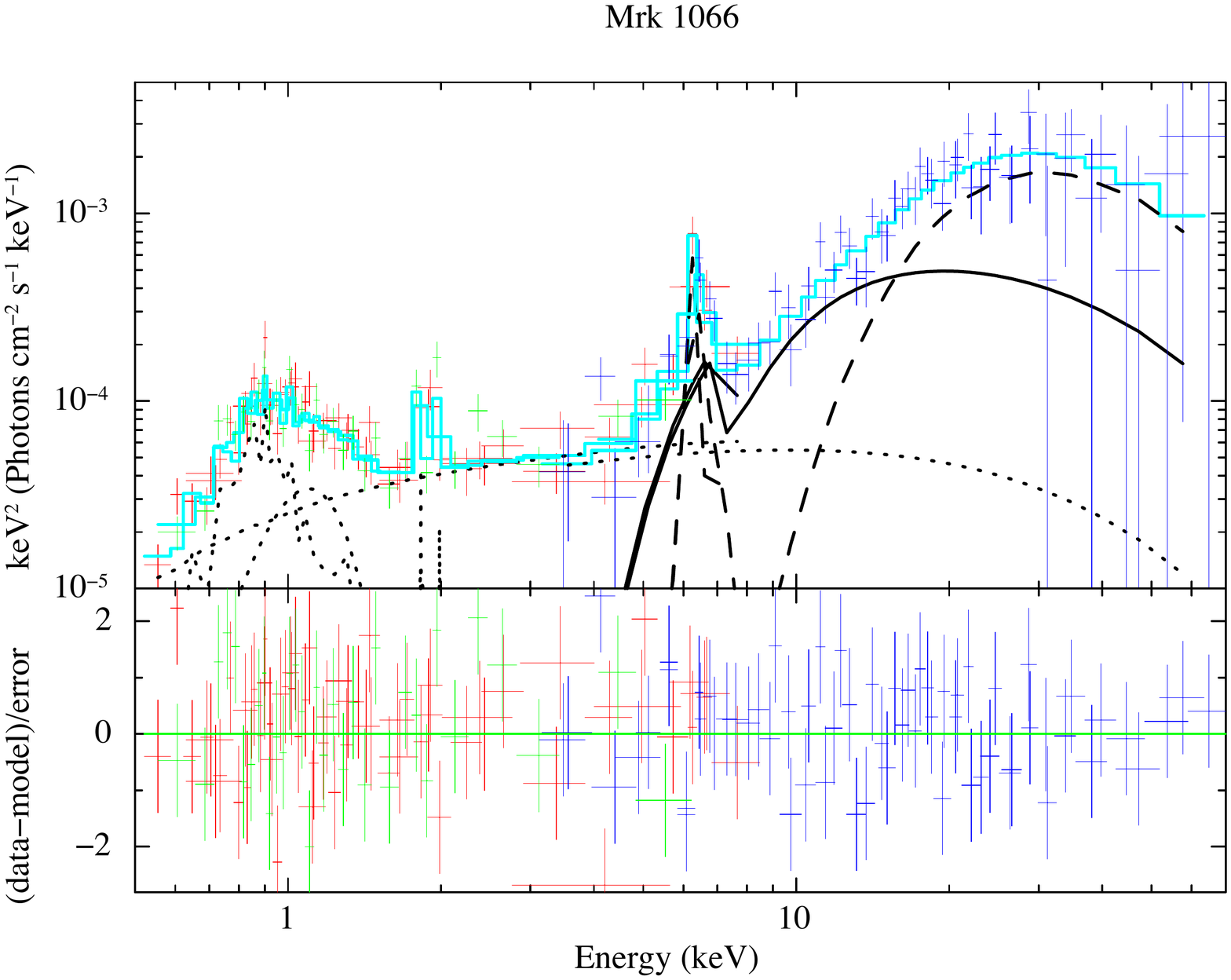}
\end{minipage}
\begin{minipage}[b]{.5\textwidth}
\centering
\includegraphics[width=1\textwidth]{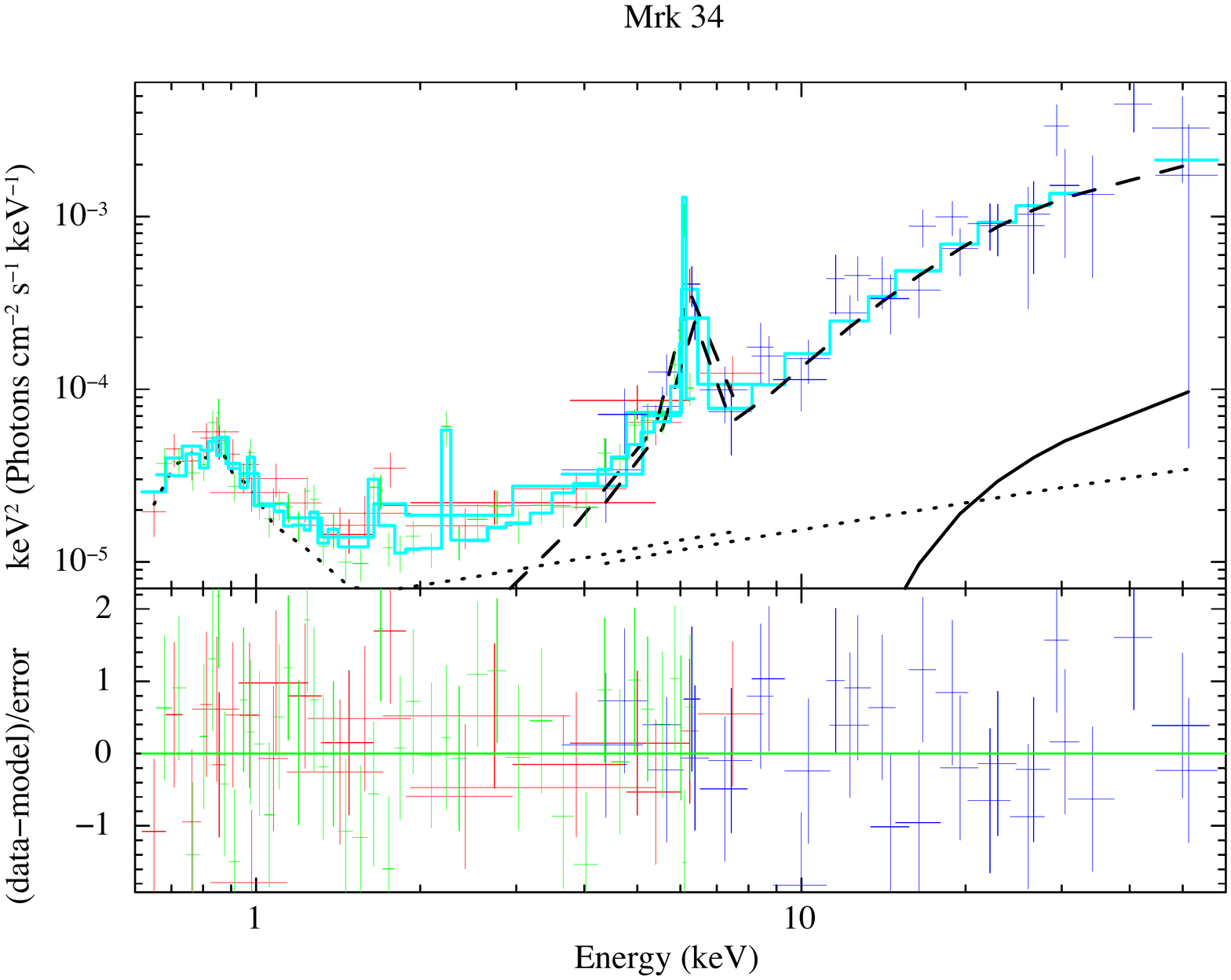}
\includegraphics[width=1\textwidth]{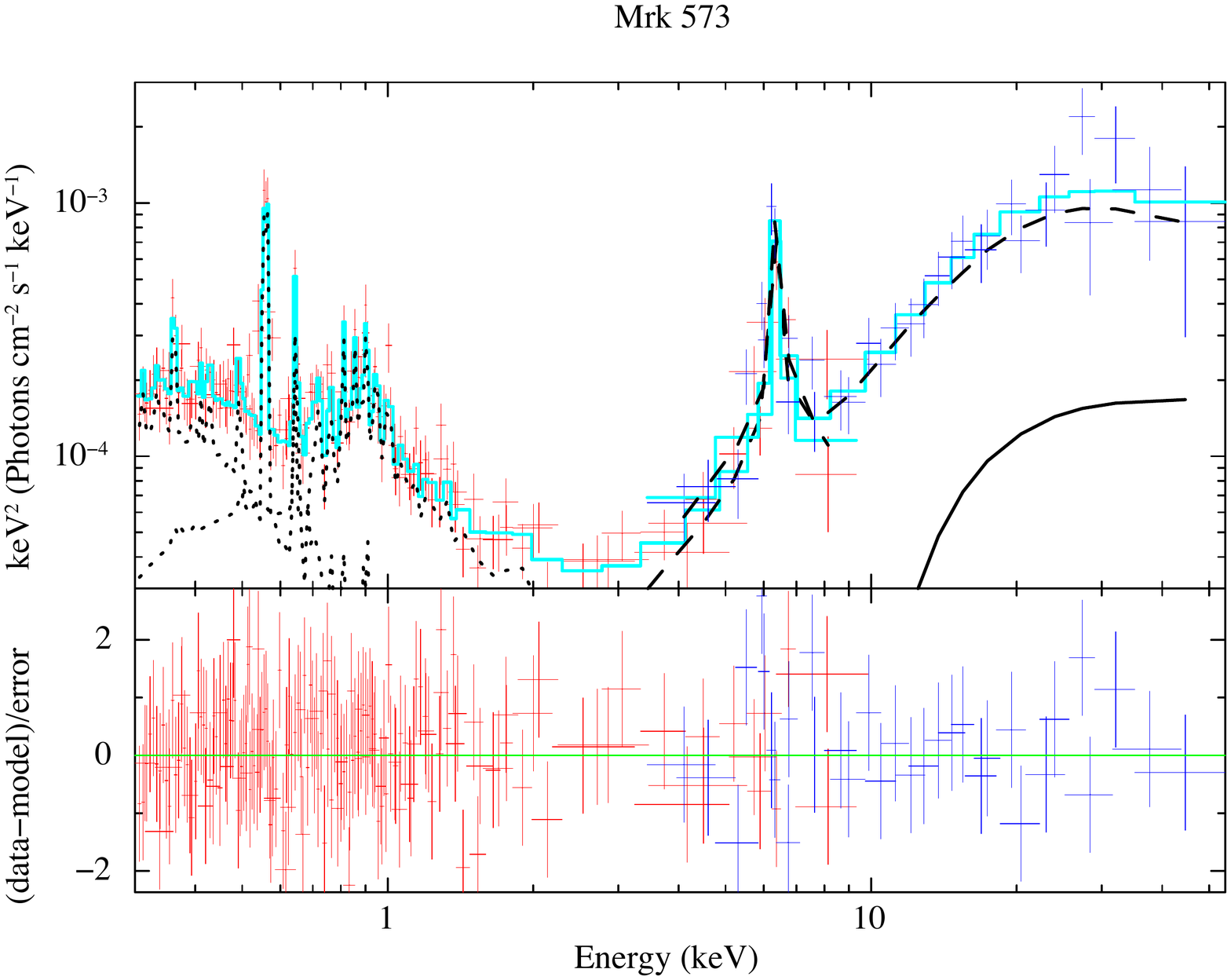}
\includegraphics[width=1\textwidth]{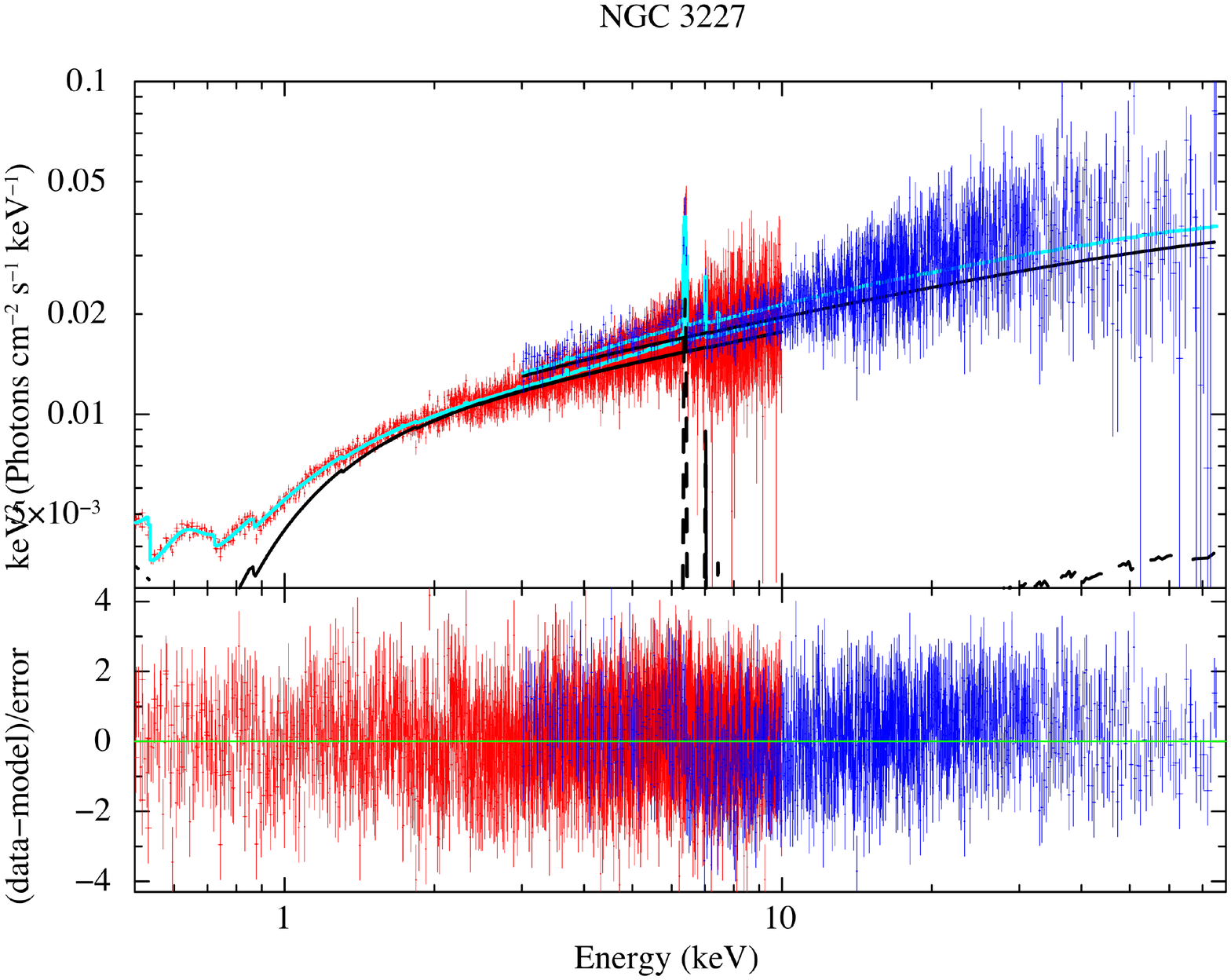}
\end{minipage}
\caption{Unfolded \NuSTAR, \XMM\ and \cha\ spectra of different sources fitted with \borus\ model when the inclination angle is left free to vary and the residuals between the data and best-fit predictions of the model. The \NuSTAR\ data are plotted in blue, the \XMM\ data are plotted in red and the \cha\ data are plotted in green. The best-fit model prediction is plotted as cyan solid lines. The single components of the model are plotted in black with different line styles, i.e., the absorbed intrinsic continuum with solid lines, the reflection component with dashed lines, the scattered component, the {\tt mekal} component and emission lines with dotted lines.}
\label{fig:spectra1}
\end{figure*}

\begin{figure*} 
\begin{minipage}[b]{.5\textwidth}
\centering
\includegraphics[width=1\textwidth]{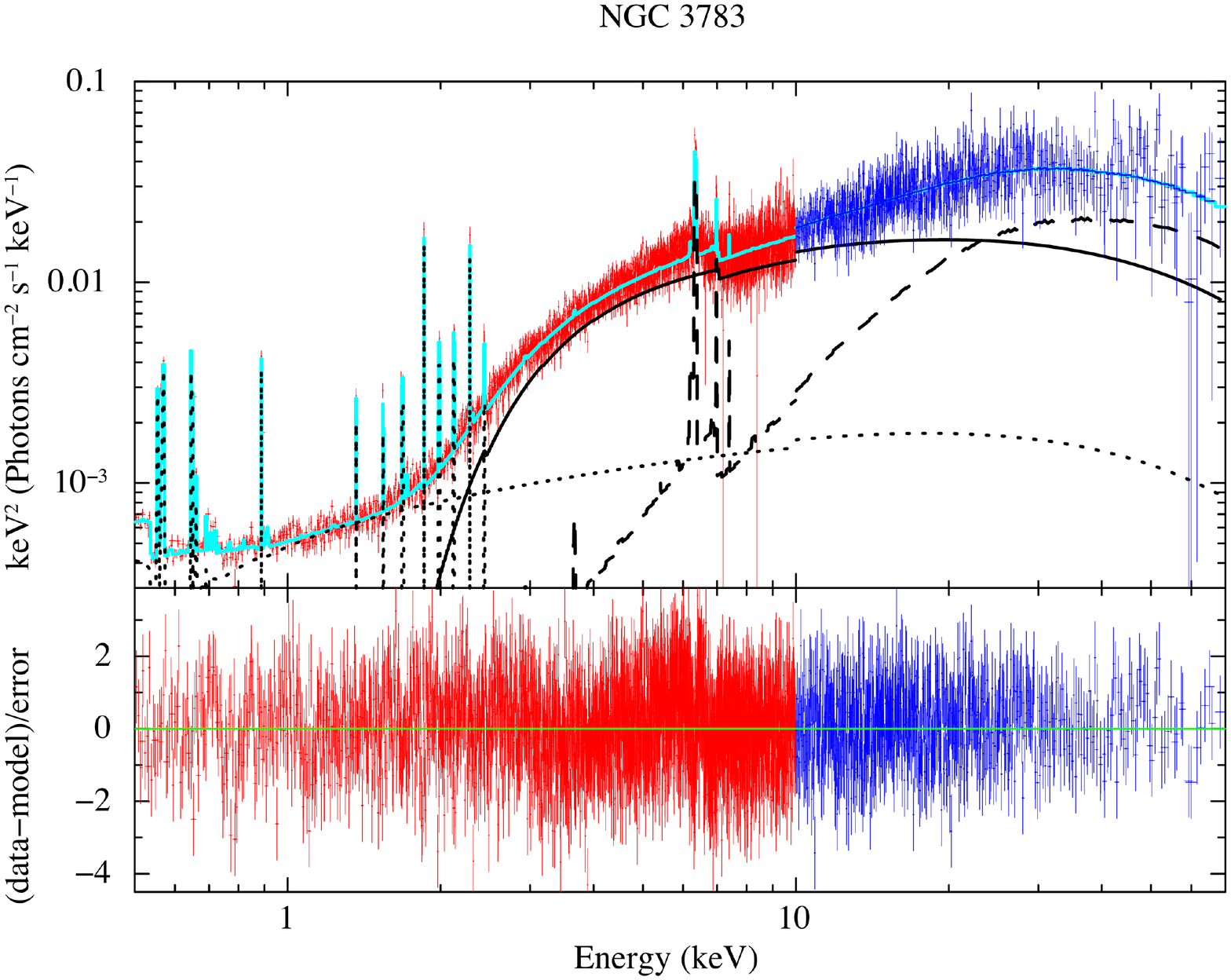}
\includegraphics[width=1\textwidth]{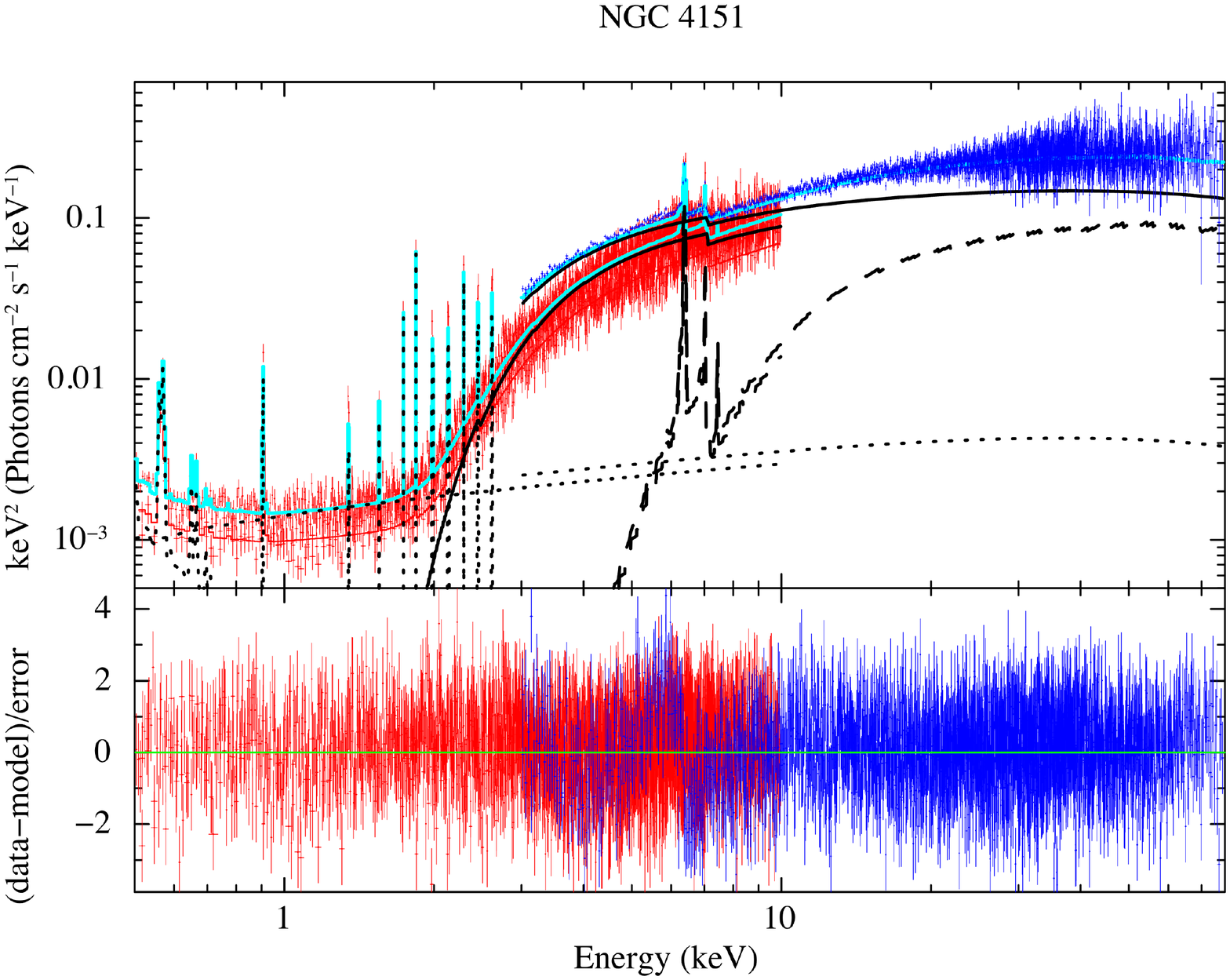}
\includegraphics[width=1\textwidth]{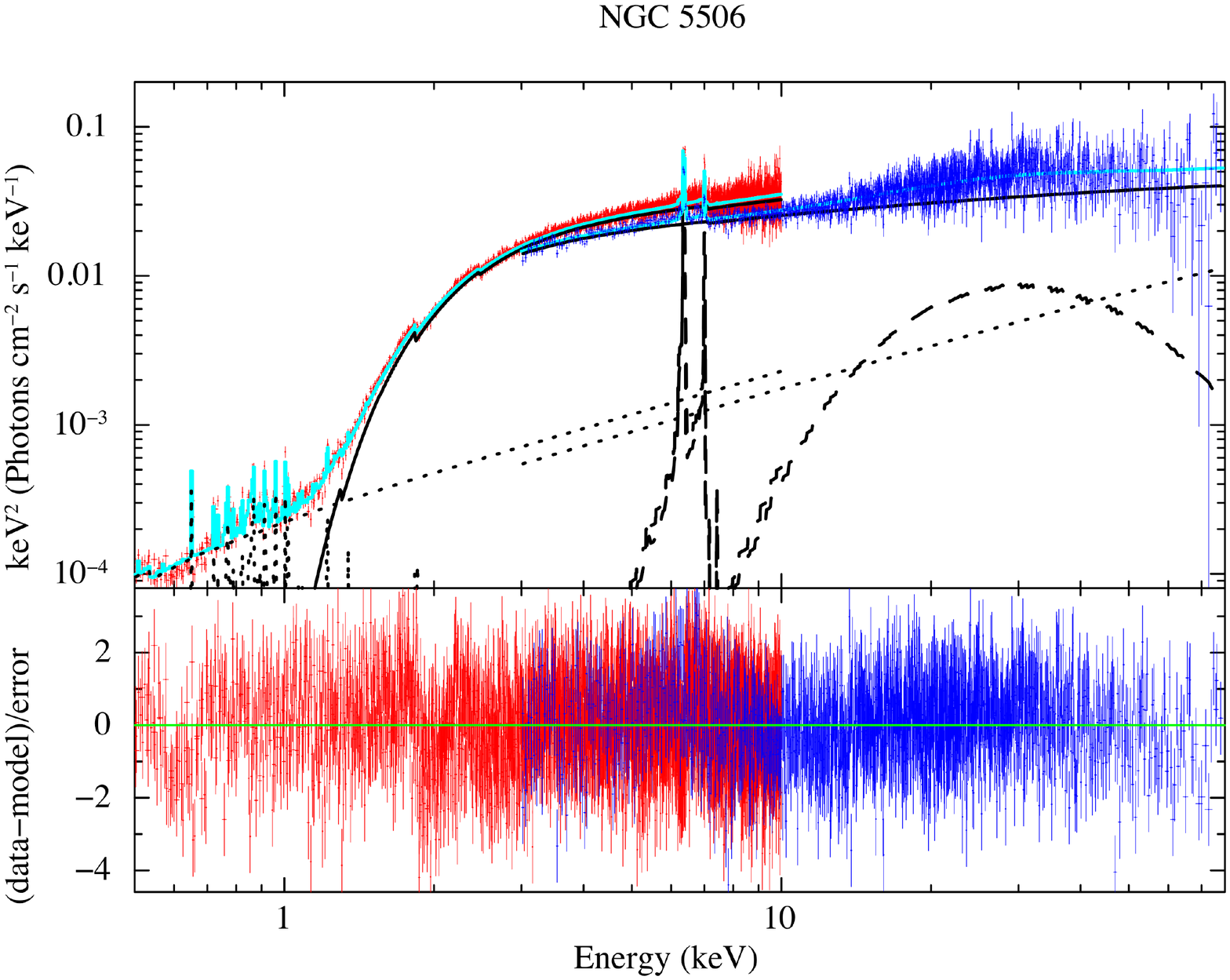}
\end{minipage}
\begin{minipage}[b]{.5\textwidth}
\centering
\includegraphics[width=1\textwidth]{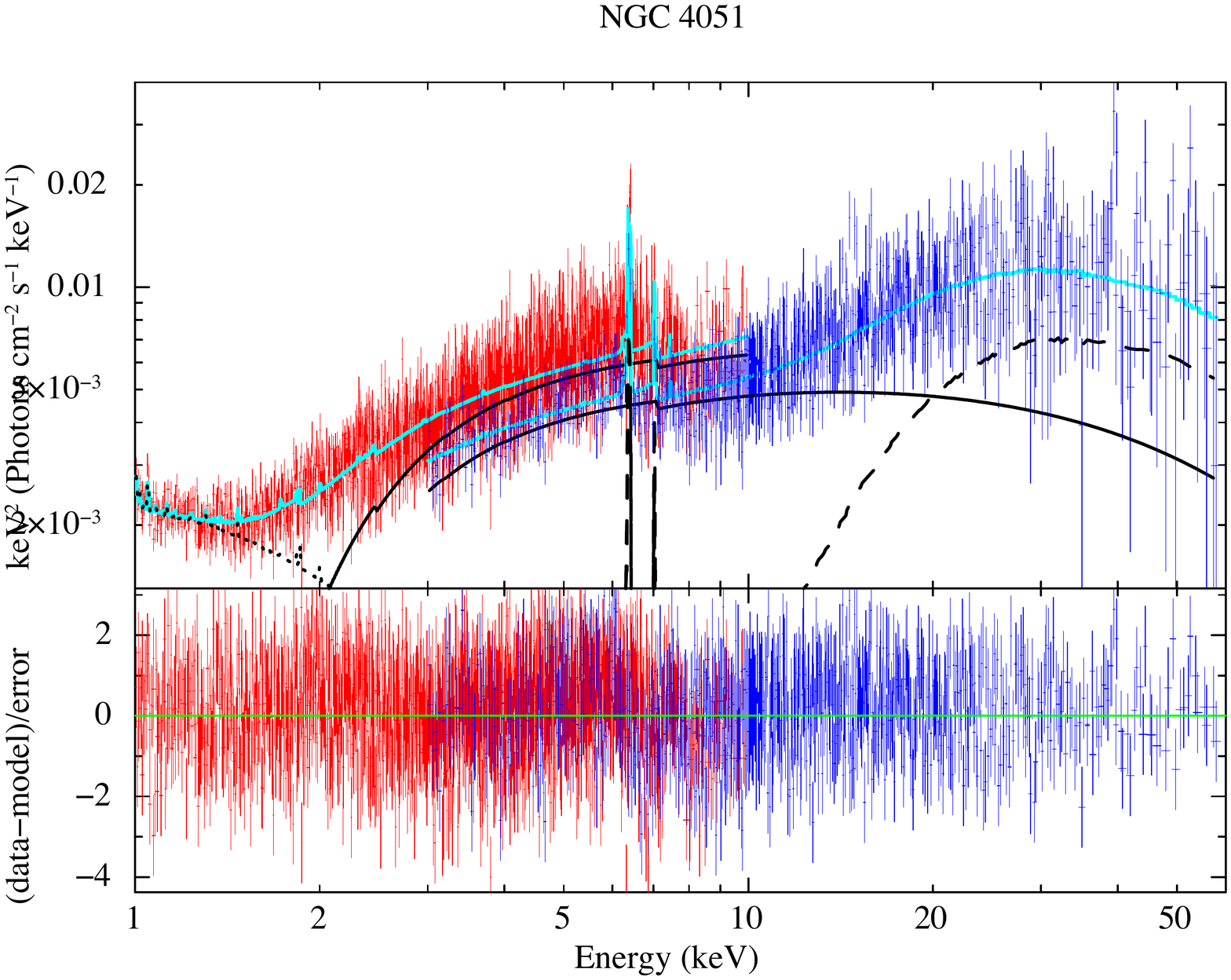}
\includegraphics[width=1\textwidth]{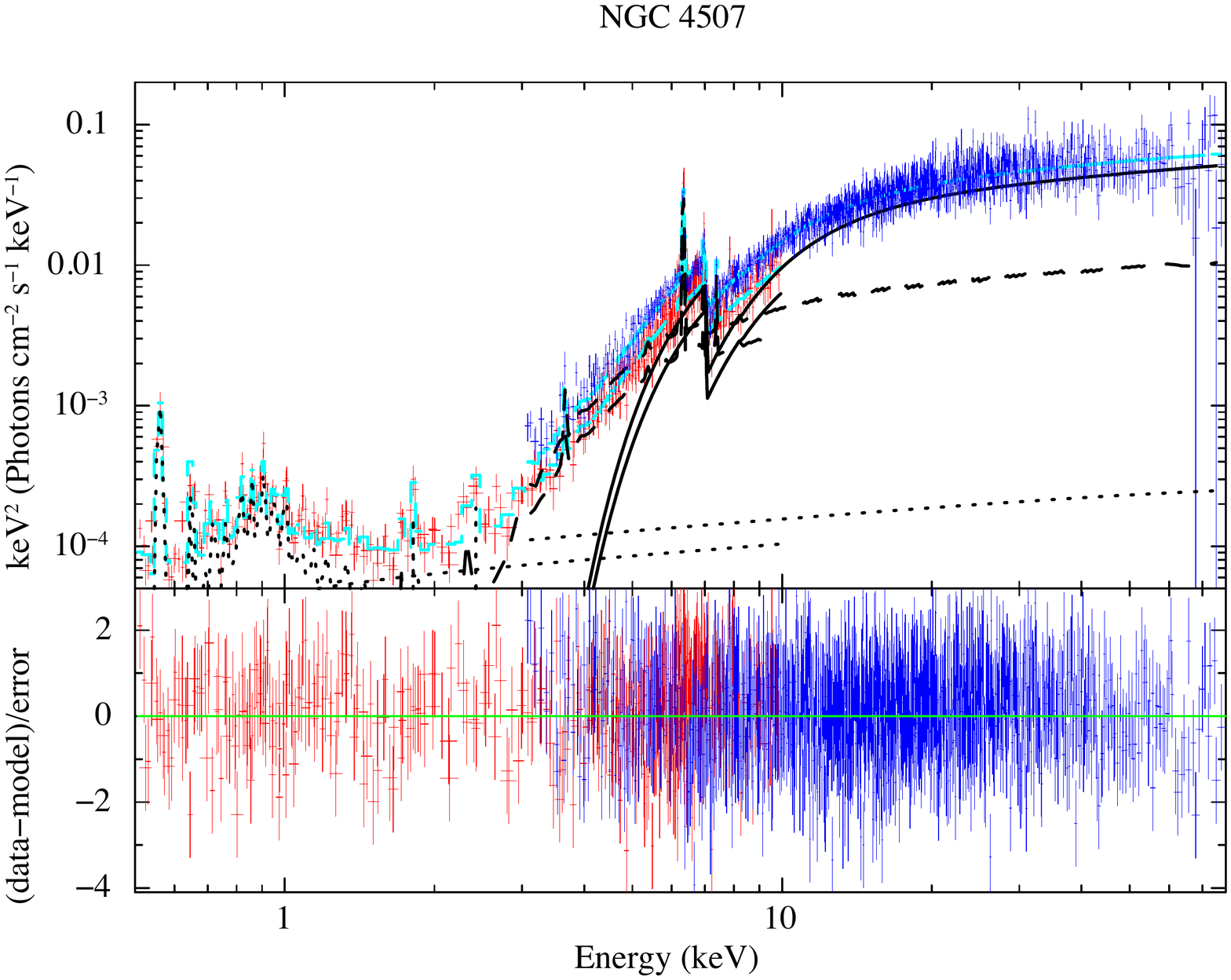}
\includegraphics[width=\textwidth]{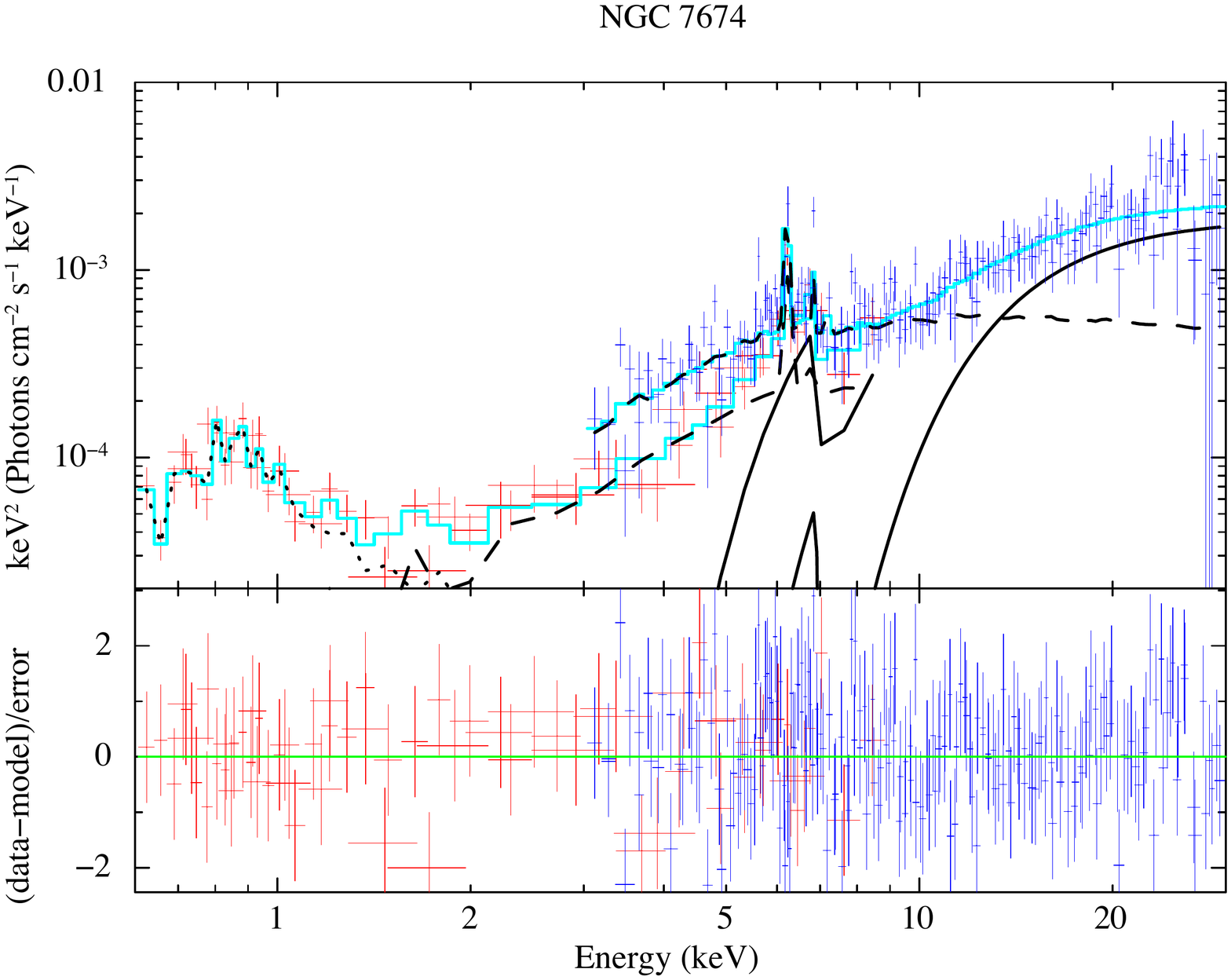}
\end{minipage}
\caption{Unfolded \NuSTAR, \XMM\ and \cha\ spectra of different sources fitted with \borus\ model when the inclination angle is left free to vary and the residuals between the data and best-fit predictions of the model. The \NuSTAR\ data are plotted in blue, the \XMM\ data are plotted in red and the \cha\ data are plotted in green. The best-fit model prediction is plotted as cyan solid lines. The single components of the model are plotted in black with different line styles, i.e., the absorbed intrinsic continuum with solid lines, the reflection component with dashed lines, the scattered component, the {\tt mekal} component and emission lines with dotted lines.}
\label{fig:spectra2}
\end{figure*}

Unfolded \NuSTAR, \XMM\ and \cha\ spectra of different sources fitted with \borus\ model when the inclination angle is left free to vary and the residuals between the data and best-fit predictions of the model are plotted in Fig.~\ref{fig:spectra1} and Fig.~\ref{fig:spectra2}. The \NuSTAR\ data are plotted in blue, the \XMM\ data are plotted in red and the \cha\ data are plotted in green. The best-fit model prediction is plotted as cyan solid lines. The single components of the model are plotted in black with different line styles, i.e., the absorbed intrinsic continuum with solid lines, the reflection component and Fe K$\alpha$ line with dashed lines, the scattered component, the {\tt mekal} component and gaussian lines with dotted lines.

\end{document}